\documentclass[12pt]{article}

\usepackage{latexsym} % Gets \Box etc
\usepackage{amssymb}  % \gtrsim, \geqslant, etc etc: 
\usepackage{epsfig}       % For PostScript figures
\newcommand\eqn[1]{(\ref{#1})}      % parentheses around the LaTex "ref" macro
\newcommand\Eqn[1]{Eq.~(\ref{#1})}  % includes ``Eq.'' in front
\newcommand{\e}{ {\rm e} }
\newcommand{\beq}{\begin{equation}}
\newcommand{\eeq}{\end{equation}}
\newcommand{\ba}{\begin{array}}
\newcommand{\bea}{\begin{eqnarray}}
\newcommand{\ea}{\end{array}}
\newcommand{\eea}{\end{eqnarray}}

\newcommand\comment[1]{ \hbox{[{\it Comment suppressed here.}\/]} }
\newcommand\hide[1]{}

\newcommand{\Tr}{\hbox{Tr }}
\newcommand{\inter}{{\mbox{\scriptsize int}}}
\newcommand{\Gammatpi}{\Gamma_{\mbox{\scriptsize 2PI}}}
\newcommand{\bG}{{\bar{G}}}
\newcommand{\bV}{\bar{V}}

\newcommand{\bphi}{\bar{\phi}}
\newcommand{\bSigma}{\bar{\Sigma}}

\newcommand{\bLambda}{\bar{\Lambda}}

\newcommand{\tp}{\tilde{p}}

\newcommand{\dl}{\delta\lambda}
\newcommand{\Dl}{\Delta\lambda}
\newcommand{\dZ}{\delta Z}
\newcommand{\dG}{\delta G}
\newcommand{\dmsq}{\delta m^2}

\newcommand{\lamNLO}{\lambda^{({\rm NLO})}}

\newcommand{\superA}{\mbox{\scriptsize A}}
\newcommand{\superB}{\mbox{\scriptsize B}}
\newcommand{\Lama}{\Lambda^{(\superA)}}
\newcommand{\Lamb}{\Lambda^{(\superB)}}
\newcommand{\bLama}{\bLambda^{(\superA)}}
\newcommand{\bLamb}{\bLambda^{(\superB)}}
\newcommand{\Dla}{\Dl^{(\superA)}}
\newcommand{\Dlb}{\Dl^{(\superB)}}

\newcommand{\Va}{V^{(\superA)}}
\newcommand{\Vb}{V^{(\superB)}}
\newcommand{\bVa}{\bV^{(\superA)}}
\newcommand{\bVb}{\bV^{(\superB)}}
\newcommand{\bva}{\bar{\cal V}^{(\superA)}}
\newcommand{\bvb}{\bar{\cal V}^{(\superB)}}

\newcommand{\bbD}{\bar{\bf D}}
\newcommand{\bPi}{\bar\Pi}
\newcommand{\skipover}[1]{}
\newcommand{\nn}{\nonumber \\}

\makeatletter %\catcode`\@=11

\def\appendix{\par                              % Have \appendix say
    \setcounter{section}{0}                     % `Appendix A', not just `A'
    \setcounter{subsection}{0}
    \renewcommand{\theequation}{\Alph{section}.\arabic{equation}}
    \renewcommand{\thesection}{Appendix \Alph{section}
                \setcounter{equation}{0}  } %Have eqns numbered (A.1) etc
}

\def\applabel#1{\@bsphack
  \protected@write\@auxout{}%
         {\string\newlabel{#1}{{\Alph{section}}{\thepage}}}%
  \@esphack}
\def\section{
\setcounter{equation}{0}        % Reset eqn numbers at start of section
\@startsection {section}{1}{\z@}{-3.5ex plus -1ex minus 
 -.2ex}{2.3ex plus .2ex}{\large\bf}}
\renewcommand{\theequation}{\arabic{section}.\arabic{equation}}

\def\subsection{\@startsection{subsection}{2}{\z@}{-3.25ex plus -1ex minus 
 -.2ex}{1.5ex plus .2ex}{\normalsize\bf}}

\def\subsubsection{\@startsection{subsubsection}{3}{\z@}{-3.25ex plus
 -1ex minus -.2ex}{1.5ex plus .2ex}{\normalsize}}

\makeatother   %\catcode`\@=12

\newsavebox{\eqlabel}
\makeatletter  %\catcode`\@=11
\newlength{\numblen}
\newsavebox{\eqnumb}
\def\@eqnnum{\savebox{\eqnumb}{\rm (\theequation)}%
\settowidth{\numblen}{\usebox{\eqnumb}}%
\makebox[\numblen][l]{\usebox{\eqnumb}~~~\usebox{\eqlabel}}}
\makeatother   %\catcode`\@=12

\newenvironment{equationwithlabel}[1]{ %
  \begin{equation}\label{#1} }{\end{equation}} %\savebox{\eqlabel}{~}}
\newcommand{\beql}[1]{\begin{equationwithlabel}{#1}}
\newcommand{\eeql}{\end{equationwithlabel}}
\newenvironment{equationarraywithlabel}[1]{ %
  \begin{eqnarray}\label{#1} }{\end{eqnarray}} %\savebox{\eqlabel}{~}}
\newcommand{\beal}[1]{\begin{equationarraywithlabel}{#1}}
\newcommand{\eeal}{\end{equationarraywithlabel}}
\begin{document}

\title{\vspace*{-1cm}
\bf Nonperturbative renormalization for\\ 2PI effective action 
techniques}

\author{
J. Berges\thanks{email: j.berges@thphys.uni-heidelberg.de} $\,^a$,
Sz. Bors\'anyi\thanks{email: borsanyi@thphys.uni-heidelberg.de} $\,^a$,
U. Reinosa\thanks{email: reinosa@hep.itp.tuwien.ac.at} $\,^{ab}$,
J. Serreau\thanks{email: julien.serreau@th.u-psud.fr} $\,^{c}$
\\
\normalsize{$^a$ Institut f{\"u}r Theoretische Physik, Universit{\"a}t 
Heidelberg,}\\[-0.1cm]
\normalsize{Philosophenweg 16, 69120 Heidelberg, Germany.}\\
\normalsize{$^b$ Institut f\"ur Theoretische Physik, Technische 
Universit\"at Wien}\\[-0.1cm]
\normalsize{Wiedner Hauptstrasse 8-10/136, A-1040 Wien, Austria.}\\
\normalsize{$^c$ Astro-Particule et Cosmologie, 11, place Marcelin 
Berthelot,}\\[-0.1cm] 
\normalsize{75231 Paris Cedex 05, France \& 
Laboratoire de Physique Th\'eorique,}\\[-0.1cm]
\normalsize{B\^at. 210, Univ. Paris-Sud 11, 91405 Orsay Cedex, France.}
\vspace*{-0.4cm}
}

\date{}
%\\[1ex] 
%\preprintno}

\begin{titlepage}
\maketitle
\def\thepage{}          % No page number on title page

\begin{abstract}
Nonperturbative approximation schemes based on two-particle 
irreducible (2PI) effective actions provide an important means
for our current understanding of (non-)equilibrium quantum field
theory. A remarkable property is their renormalizability, 
since these approximations involve selective summations to infinite 
perturbative orders. In this paper we show how to renormalize
all $n$-point functions of the theory, which are given by derivatives 
of the 2PI-resummed effective action $\Gamma[\phi]$ for scalar
fields $\phi$. This provides a complete description in terms of the 
generating functional for renormalized proper vertices, 
which extends previous prescriptions in the literature on the 
renormalization for 2PI effective actions. 
The importance of the 2PI-resummed generating functional 
for proper vertices stems from the fact that the latter
respect all symmetry properties of the theory and, in particular, 
Goldstone's theorem in the phase with spontaneous symmetry breaking. 
This is important in view of the application of these techniques to 
gauge theories, where Ward identities play a crucial role.
\end{abstract}

\end{titlepage}

\renewcommand{\thepage}{\arabic{page}}

%============================================================================
%------------------------ Body of paper begins ------------------------------
%============================================================================

\section{Introduction}
\label{sec:intro}

Selective summations to infinite order in perturbation theory
often play an important role in vacuum, thermal equilibrium
or nonequilibrium quantum field theory.
A prominent example concerns theories with bosonic field
content such as QCD, QED or scalar theories at high temperature, 
where standard perturbative expansions are plagued by infrared 
singularities. Screening effects can be taken into account e.g.\ by resumming 
so-called hard thermal loops \cite{Braaten:1989mz}. The resulting 
resummed perturbation theory, however, can reveal a poor convergence behavior 
and typically requires further summations~\cite{Blaizot:2003tw}.
It has been pointed out that improved behavior may be based on
expansions of the two-particle irreducible (2PI) effective 
action~\cite{Blaizot:2000fc}.
This has recently been demonstrated for scalar theories, where a dramatically 
improved convergence behavior is observed once systematic loop-
or coupling-expansions of the 2PI effective
action are employed without further assumptions~\cite{Berges:2004hn}.  
These 2PI approximation schemes also play a 
crucial role for our understanding of quantum fields 
out-of-equilibrium~\cite{Berges:2003pc,Berges:2004yj}. 
There, infinite summations 
are needed to obtain approximations which are uniform in time. 2PI
techniques provide for the first time the link between far-from-equilibrium
dynamics at early times and late-time thermalization in quantum field 
theory~\cite{Berges:2000ur,Berges:2001fi,Cooper:2002qd,Berges:2002wr,Berges:2002cz}. 
The good convergence properties of the 
approach have also been observed in the context of classical statistical 
field theories, where comparisons with exact results are 
possible~\cite{Aarts:2001yn}, or in applications to universal
properties of critical phenomena~\cite{Alford:2004jj}.

The 2PI expansions are known to be consistent with global 
symmetries~\cite{Baym,Knoll:2001jx,vanHees:2001ik}. However, 
Ward identities may not be manifest at intermediate calculational
steps of a 2PI resummation scheme. To obtain the
correct symmetry properties for physical results requires
a consistent renormalization procedure for the 2PI-resummed 
effective action $\Gamma[\phi]$. The latter is the 
generating functional for the proper vertices by derivatives
with respect to the field $\phi$. Very important progress
in this direction has been made in 
Refs.~\cite{vanHees:2001ik,Blaizot:2003br}\footnote{Renormalization 
for the special case of Hartree--type 
approximations, which can be related to the two-loop 2PI
effective action~\cite{Cornwall:1974vz}, has a long history~\cite{Hartree}. 
Loop approximations of the 2PI effective
action are also called ``$\Phi$-derivable'' approximations.}.
It has been shown, in particular, that the resummation equation for 
the two-point function in the context of scalar $\varphi^4$-theories
can be made finite by adjusting a finite number of local counterterms. 
A similar analysis~\cite{Cooper:2004rs} has recently been 
performed for the 2PI $1/N$-expansion~\cite{Berges:2001fi,Aarts:2002dj}. 
Although this allows one to compute various physical quantities such as 
the thermodynamic pressure, we emphasize that the renormalization 
presented in these works 
is not sufficient to fix all counterterms needed to compute the
proper vertices encoded in $\Gamma[\phi]$. The aim of this work  
is to provide this missing prescription for general approximations schemes,
including systematic loop-, coupling- or $1/N$-expansions.
We exemplify these techniques using scalar field theories for
simplicity. Very similar techniques as those employed here can be 
used for more complicated systems, including fermionic or gauge 
degrees of freedom, or for higher $n$PI effective actions.
The importance of the 2PI-resummed generating functional 
$\Gamma[\phi]$ for proper vertices stems from the fact that the latter
respect all symmetry properties of the theory and, in particular, Goldstone's
theorem in the phase with spontaneous symmetry breaking. 
This is a crucial step towards the renormalization
for 2PI-resummed effective actions in gauge theories, where
Ward identities play a particularly important role.
The application of our techniques to the latter 
will be presented in a separate work~\cite{BBRS}.

It is a remarkable property that 2PI approximations schemes are 
renormalizable, since they involve selective summations to infinite 
perturbative orders. The proof requires to show that all $n$-point 
functions of the theory, which are given by derivatives of the
effective action, can be renormalized. This is exemplified in the
present work for scalar field theories with quartic self-interactions.
The non-trivial character of this result is two-fold: firstly, 
the divergences due to the infinite selective summation can be absorbed in a 
finite number of counterterms and, secondly, the procedure involves
only standard ``local'' counterterms despite the non-local character 
of the 2PI summation. Furthermore, the renormalization procedure is 
medium-independent. Thus no temperature, chemical potential or even 
time-dependence is hidden in the counterterms.
We have published previously a first {\it application} of our methods 
to resolve the convergence problems of perturbative expansions at 
high temperature in Ref.~\cite{Berges:2004hn}. The aim was to demonstrate the
applicability for practical calculations, before
presenting in detail the discussion of the somewhat contrived foundations.
We mention that a summary of the present work was presented in~\cite{Urko:SEWM04}.

The paper is organized as follows: Sec.~\ref{sec:scheme} describes basics
of the 2PI approach required for the present discussion. In 
Sec.~\ref{sec:npoint}, we explicitly 
construct the $n$-point functions of the theory in terms of functional 
derivatives of the 2PI effective action. These derivatives involve
summations of infinite series of diagrams, which can be all conveniently
generated by a single four-point vertex-function through an
integral, Bethe-Salpeter--type equation.\footnote{This equation is the
same as that introduced in Refs.~\cite{vanHees:2001ik,Blaizot:2003br}
in discussing the renormalization of the propagator equation.} Once expressed
in terms of the solution of this equation, the renormalization analysis
for $n$-point functions is considerably simplified. Sec.~\ref{sec:renorm} presents
a comprehensive discussion of the nonperturbative renormalization 
to be performed here and will be of interest for readers who are not 
interested in going through the technical details of the proof, which we 
present in Sec.~\ref{sec:finite}. We discuss in 
Sec.~\ref{sec:finite2} that once the theory has been renormalized in 
the symmetric phase, no new divergences appear for a non-vanishing field
expectation value.
Various aspects concerning symmetries and renormalization for the case of 
multiple scalar field components and the application to the 2PI
$1/N$-expansion are discussed in Sec.~\ref{sec:sym}. 
We end with a summary and conclusions in Sec. \ref{Sec:summaryconclusions}.
Appendix~\ref{sec:appcorrel} is devoted to the derivation of useful 
identities between the various functional derivatives of the 2PI 
effective action, and Appendix~\ref{2PItop} 
discusses some of their topological 
properties for diagrammatics.

%%%%%%%%%%%%%%%%%%%%%%%%%%%%%%%%%%%%%%%%%%%%%%%%%%%%%%%%%%%%%%%%%%%%%%%%%%%%%%%%%%%%%%%%%%%%%%%%%%%%%%%%%%%%%%%%%%%%%%%%%%%%%%%%%%%%%%%%%%   The 2PI self-consistent resummation scheme   %%%%%%%%%%%%%%%%%%%%%%%%%%%%%%%%%%%%%%%%%%%%%%%%%%%%%%%%%%%%%%%%%%%%%%%%%%%%%%%%%%%%%%%%%%%%%%%%%%%%%%%%%%%%%%%%%%%%%%%%%%%%%%%%%%%%%%%%%%%%%%%%%%%%%%%%%%%%%%%%%%%%%%%%%%%
\section{The 2PI-resummed effective action}
\label{sec:scheme}

%%%%%%%%%%%%%%%%%%%%%%%%%%%%%%%%%%%%%%%%%%%%%%%%%%%%%%%%%%%%%%%%%%%%%%%%%%%%%%%%%%%%%%%%%%%%%%%%%%%%%%%%%%%%%%%%%%%%%%%%%%%%%%%%%%%%%%%%%%   Generalities   %%%%%%%%%%%%%%%%%%%%%%%%%%%%%%%%%%%%%%%%%%%%%%%%%%%%%%%%%%%%%%%%%%%%%%%%%%%%%%%%%%%%%%%%%%%%%%%%%%%%%%%%%%%%%%%%%%%%%%%%%%%%%%%%%%%%%%%%%%%%%%%%%%%%%%%%%%%%%%%%%%%%%%%%%%%%%%%%%%%%%%%%%%%%%%%%%%%%%%%%%
\subsection{Definition}

All information about the quantum theory can be obtained from the (1PI) 
effective action $\Gamma[\phi]$, which is the generating functional
for proper vertex functions. This effective action is represented as a 
functional of the field expectation value or one-point function $\phi$ 
only. In contrast, the 2PI 
effective action $\Gamma_{\rm 2PI}[\phi,G]$ is written as a functional
of $\phi$ and the connected two-point 
function $G$ by introducing an additional bilinear 
source in the defining functional integral~\cite{Baym,Cornwall:1974vz}. 
Higher functional representations,
so-called $n$PI effective actions $\Gamma_{\rm nPI}[\phi,G,V_3,\ldots,V_n]$,
are constructed accordingly and  
include in addition the three-point, four-point, \ldots, and 
$n$-point functions or the corresponding proper vertices $V_3,\ldots,V_n$ 
\cite{Baym,Dominicis,Berges:2004pu}.
The different functional representations of the
effective action are equivalent in the sense that they agree 
for the exact theory in the absence of additional sources: 
\beq
\label{nPI}
 \Gamma[\phi] = \Gamma_{\rm 2PI}[\phi,G=\bar{G}(\phi)] = \cdots =
 \Gamma_{\rm nPI}[\phi,G=\bar{G}(\phi), \ldots, V_n = \bar{V}_n(\phi)] \,.
\eeq
The absence of additional sources corresponds to the 
stationarity conditions~\cite{Baym,Dominicis}
\beq
\frac{\delta \Gamma_{\rm nPI}}{\delta G}\Big|_{G=\bar{G}} = 0 
\quad , \ldots, \quad
\frac{\delta \Gamma_{\rm nPI}}{\delta V_n}\Big|_{V_n = \bar{V}_n} = 0 
\, ,
\eeq
by which the $n$-point functions are determined self-consistently and
become implicit functions of the field:
$G = \bar{G}(\phi)$, $V_3 = \bar{V}_3(\phi)$, \ldots, 
$V_n = \bar{V}_n(\phi)$. 

The importance of different functional representations for the
effective action stems from the fact that one typically cannot solve the
theory exactly. Higher $n$PI effective actions turn out to provide a very 
efficient tool to build systematic nonperturbative approximation schemes. 
For instance, a loop expansion of the 1PI effective action to a given order  
differs in general from an expansion of $\Gamma_{\rm 2PI}[\phi,G]$ to the same 
number of loops.\footnote{Cf., however, the equivalence hierarchy 
relating the loop expansions of the various $n$PI effective 
actions to certain orders in the loops~\cite{Berges:2004pu}.}
The 2PI loop-expansion truncated at a given order can be seen to 
resum an infinite series of contributions for the 1PI loop-expansion.
More generally, any systematic 
e.g.\ loop-, coupling-, or $1/N$-expansion of higher 
effective actions at a given order resums infinitely many contributions in
the corresponding expansion of lower effective actions.
Higher effective actions, therefore, provide a powerful tool to devise 
systematic nonperturbative approximation schemes for the calculation of
$\Gamma[\phi]$. In the following we concentrate on the 2PI effective action
since this has been the most frequently used so far.\footnote{For
applications of higher than 2PI effective actions in QCD
see Ref.~\cite{Berges:2004pu}.} 

In the following we use the abbreviated notation
$\phi_1\equiv\phi(x_1)$ and $G_{12}\equiv G(x_1,x_2)$, where
indices denote the space-time arguments. For theories with more
than one field these labels can denote the various field-components 
as well. We consider a real scalar field theory with classical action
for the fluctuating field $\varphi$
\beq
 S[\varphi]=S_0[\varphi]+S_\inter[\varphi]\,,
\eeq
with the free (quadratic) part
\beq
 S_0[\varphi]=\frac{1}{2}\varphi_a\, iG_{0,ab}^{-1}\, \varphi_b \,,
\label{eq:S0}
\eeq
and where $S_\inter[\varphi]$ represents the interaction part. 
Here $G_0$ is the free field theory propagator and
the expectation value of $\varphi$ is denoted by $\phi$.
Summation/integration over repeated indices is 
implied.\footnote{For non-equilibrium systems, 
this includes an integration over a closed time-path \cite{Schwinger:1960qe}.} 
The corresponding 2PI generating functional can be written as 
\beq
\label{eq:2PI}
 \Gammatpi[\phi,G] = S_0[\phi]+\frac{i}{2}\,\Tr\! \ln G^{-1}
 +\frac{i}{2}\,\Tr G_0^{-1}G +\Gamma_\inter[\phi,G]\,,
\eeq
where $\Gamma_\inter$ is given by all closed two-particle-irreducible graphs 
times an overall factor $(-i)$, with lines representing the 
(auxiliary) two-point function $G$ \cite{Baym,Cornwall:1974vz}.\footnote{In terms of the 
usual Corwall-Jackiw-Tomboulis parametrization \cite{Cornwall:1974vz}, 
one has
$$
 \Gamma_\inter[\phi,G]=S_\inter[\phi]
 +\frac{1}{2}\,\Tr \frac{\delta^2S_\inter}{\delta\phi\delta\phi} G
 +\Gamma_2[\phi,G]\,.
$$\label{fo:CJT}} 
The vertices are obtained in the same way as for the
1PI effective action from the shifted action 
$S_\inter[\phi+\varphi]$ by collecting all terms higher than 
quadratic in the fluctuating field $\varphi$ \cite{Cornwall:1974vz}.
This provides a functional representation of the theory in terms of the 
one-point and two-point fields, $\phi$ and $G$. 
The effective action $\Gamma[\phi]$ is obtained from
(\ref{eq:2PI}) by taking $G$ to be at its 
physical value, given by the stationarity condition
\beq
\label{eq:statG}
 \frac{\delta\Gammatpi}{\delta G_{12}}\Big|_{\bG}=0\,.
\eeq
Using Eq.\ (\ref{eq:2PI}), this condition can be rewritten 
as
\beq
\label{eq:gap}
 \bG_{12}^{-1}(\phi)=G_{0,12}^{-1}-\bSigma_{12}(\phi)\,,
\eeq
with the self-energy\footnote{Note that this definition
differs from the conventional one~\cite{Cornwall:1974vz}, given by 
$2i \delta \Gamma_2/\delta G$ with the $\Gamma_2$
of footnote \ref{fo:CJT}.}
\beq
\label{eq:se}
 \bSigma_{12}(\phi) \equiv 
 2i\frac{\delta\Gamma_\inter}{\delta G_{21}}\Big|_{\bG}\,.
\eeq
The equation for the two-point function (\ref{eq:gap}) 
can be rewritten as an infinite series in terms of the 
free field theory propagator $G_0$:
\beq
\bG_{12}(\phi) = G_{0,12} + G_{0,1a}\, \bSigma_{ab}(\phi)\, G_{0,b2} 
+ G_{0,1a}\, \bSigma_{ab}(\phi)\, G_{0,bc}\, 
\bSigma_{cd}(\phi)\, G_{0,d2} + \ldots
\eeq
Each 2PI diagram with propagator lines associated to $G$ that contributes
to $\Gamma_{\rm 2PI}[\phi,G]$, therefore, encodes a selective summation of
an infinite series of perturbative diagrams. The effective action $\Gamma[\phi]$
is obtained from a given approximation of the 2PI effective action according
to
\beq
\label{eq:effac}
 \Gamma[\phi]=\Gammatpi[\phi,\bG(\phi)]\,.
\eeq
This defines the 2PI-resummed effective action. Equation \eqn{eq:effac} yields
an efficient starting point for systematic nonperturbative approximations 
of the (1PI) effective action. The resummation  
enters through the solution of the stationarity equation (\ref{eq:statG}),
which leads to the self-consistent equation (\ref{eq:gap}) for $\bG(\phi)$. 
We will use the symbols $\Gamma[\phi]$ and $\Gammatpi[\phi,\bG(\phi)]$
synonymously in the following.

In order to be able to renormalize all $n$-point 
functions it is important that all approximations are 
done on the level of the effective action. Once an approximate
$\Gammatpi[\phi,G]$ is 
specified, there are no further approximations involved on the level of 
the equation of motion (\ref{eq:gap}) for $\bG(\phi)$. We emphasize that 
this last property is 
crucial for the program of renormalization in these self-consistent
resummation schemes.\footnote{It turns out that recent 
claims~\cite{Braaten:2001en} 
about the non-renormalizability of 2PI expansions are
an artefact of additional approximations employed for the 
equations of motion. Cf.~also the discussion in Ref.~\cite{Berges:2004hn}.}

%%%%%%%%%%%%%%%%%%%%%%%%%%%%%%%%%%%%%%%%%%%%%%%%%%%%%%%%%%%%%%%%%%%%%%%%%%%%%%%%%%%%%%%%%%%%%%%%%%%%%%%%%%%%%%%%%%%%%%%%%%%%%%%%%%%%%%%%%%%%%%%%   Physical $n$-point functions   %%%%%%%%%%%%%%%%%%%%%%%%%%%%%%%%%%%%%%%%%%%%%%%%%%%%%%%%%%%%%%%%%%%%%%%%%%%%%%%%%%%%%%%%%%%%%%%%%%%%%%%%%%%%%%%%%%%%%%%%%%%%%%%%%%%%%%%%%%%%%%%%%%%%%%%%%%%%%%%%%%%%%%%%%%%%%%%%%%%%%
\subsection{2PI-resummed $n$-point functions}\label{sec:npoint}

The functional derivatives of the 2PI-resummed effective action 
(\ref{eq:effac}),
\beq
\label{eq:npoint}
 \Gamma^{(n)}_{12\ldots n}\equiv
 \frac{\delta^n\Gamma[\phi]}{\delta\phi_1\cdots\delta\phi_n}\Big|_{\bphi}\,,
\eeq
provide a complete description of the theory in terms of
$n$-point functions $\Gamma^{(n)}_{12\ldots n}$ for a given
approximation. Here $\phi=\bphi$ is determined by the stationarity 
condition
\beq
\label{eq:statphi}
\Gamma^{(1)}_{1} \equiv
 \frac{\delta \Gamma [\phi]}{\delta\phi_1}\Big|_{\bphi}=0\,.
\eeq
We emphasize that the functions $\Gamma^{(n)}_{12\ldots n}$ reflect all 
symmetry properties of the theory. In particular, aspects
such as Goldstone's theorem in the phase with spontaneous 
symmetry breaking are fulfilled to any order in a systematic
2PI expansion such as loop-, coupling- or $1/N$-expansions
\cite{Aarts:2002dj,vanHees:2001ik}. 

The functions (\ref{eq:npoint}) are related to derivatives of the 
2PI effective action, $\Gamma_{\rm 2PI}[\phi,G]$, through 
Eq.~(\ref{eq:effac}). One has, for instance,
\beq
 \frac{\delta\Gamma}{\delta\phi_1} = 
\frac{\delta \Gammatpi}{\delta \phi_1}\Big|_{\bG}
 +\frac{\delta \Gammatpi}{\delta G_{ab}}\Big|_{\bG}
\frac{\delta\bG_{ab}}{\delta\phi_1}
 =\frac{\delta \Gammatpi}{\delta \phi_1}\Big|_{\bG}\,.
\eeq
The last equality arises from Eq.\ (\ref{eq:statG}). 
Using the representation 
(\ref{eq:2PI}) one finds:
\beq
 \frac{\delta\Gamma}{\delta\phi_1} = iG_{0,1a}^{-1}\, \phi_a +
 \frac{\delta\Gamma_\inter}{\delta \phi_1}\Big|_{\bG}\,.
\eeq
The functional derivatives of $\Gamma_\inter$, which we call
2PI kernels, play a central role in what follows.
In particular, an important feature concerns their two-particle-irreducibility
properties. This is discussed in Appendix~\ref{2PItop}. 
It is useful to express the function $\delta \bG/\delta \phi$ 
in terms of 2PI kernels. One has
\beq
\label{eq:link}
 \frac{\delta \bG_{12}}{\delta \phi_3}=-\bG_{1a}\bG_{b2}
 \frac{\delta \bG^{-1}_{ab}}{\delta \phi_3}
 =\bG_{1a}\bG_{b2}\frac{\delta\bSigma_{ab}}{\delta\phi_3}\,,
\eeq
where we have used Eq.\ (\ref{eq:gap}) for the second equality.
With this, one obtains for the second derivative of the 2PI
resummed effective action:
\bea
 \frac{\delta^2 \Gamma}{\delta\phi_1\delta\phi_2} &=& 
 iG_{0,12}^{-1}
 +\frac{\delta^2\Gamma_\inter}{\delta\phi_1\delta\phi_2}\Big|_{\bG}
 +\frac{\delta^2\Gamma_\inter}{\delta\phi_1\delta G_{ab}}\Big|_{\bG}
 \bG_{ac}\bG_{db}
 \frac{\delta \bSigma_{cd}}{\delta \phi_2}\,.
\label{eq:2pt}
\eea
Taking the field derivative of Eq.\ (\ref{eq:se}) and using Eq.\ 
(\ref{eq:link}), one obtains the following integral equation for 
$\delta\bSigma/\delta\phi$:
\bea
 \frac{\delta\bSigma_{12}}{\delta\phi_3}
 &=&\frac{2i\delta^2\Gamma_\inter}{\delta G_{12}\delta\phi_3}\Big|_{\bG}
 +\frac{2i\delta^2\Gamma_\inter}{\delta G_{12}\delta G_{ab}}\Big|_{\bG}
 \bG_{ac}\bG_{db}\frac{\delta\bSigma_{cd}}{\delta \phi_3}\,.
\label{eq:resum1}
\eea
The above equation resums the infinite series of ``ladder'' graphs 
made of ``rungs'' 
$2i\delta^2\Gamma_\inter/\delta G^2|_\bG$ connected by lines $\bG$ and ``ended''
by the kernel $2i\,\delta^2\Gamma_\inter/\delta G\delta\phi|_\bG$. 
In the following, we will explain the diagrammatic representation of
this equation.
\begin{figure}
\begin{center}
\includegraphics[width=11cm]{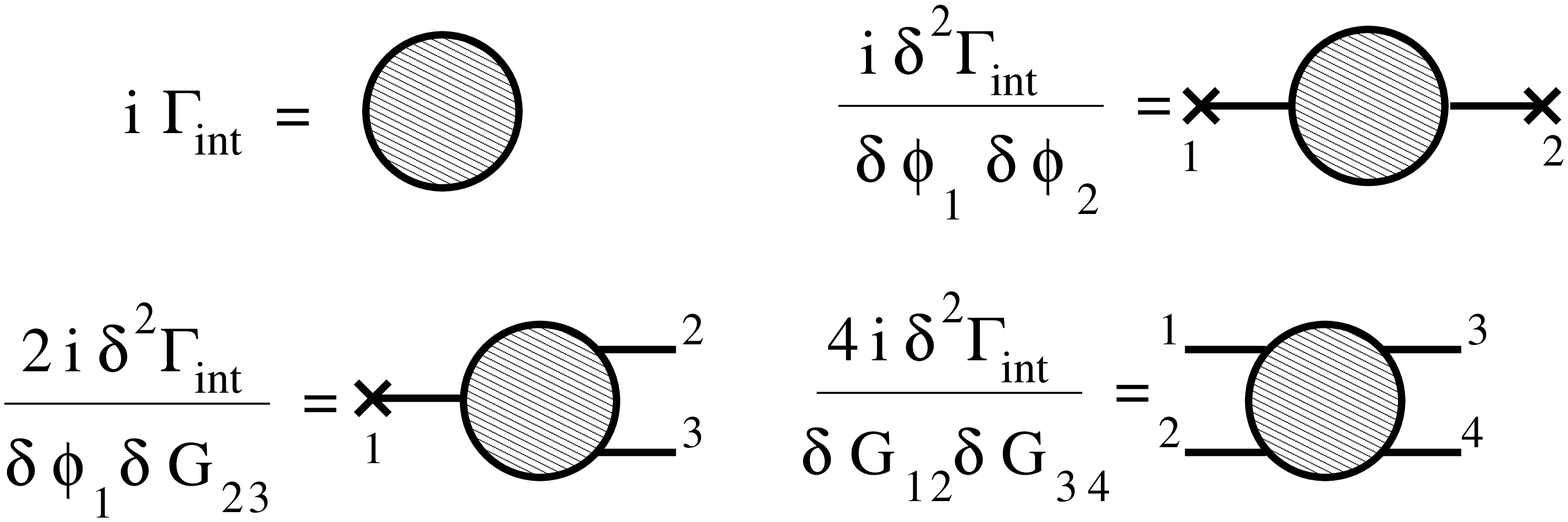}
\caption{\small \label{fig:Gamma_int_der} Diagrammatic representation of 
$\Gamma_\inter$ and its derivatives. Note that external legs do not 
represent propagators. An overall $i$ and a factor $2$ per each derivative 
with respect to $G$ are factored out such that the diagrams contributing 
to the kernels correspond to the standard Feynman rules.}
\end{center}
\end{figure}
\begin{figure}
\begin{center}
\includegraphics[width=10cm]{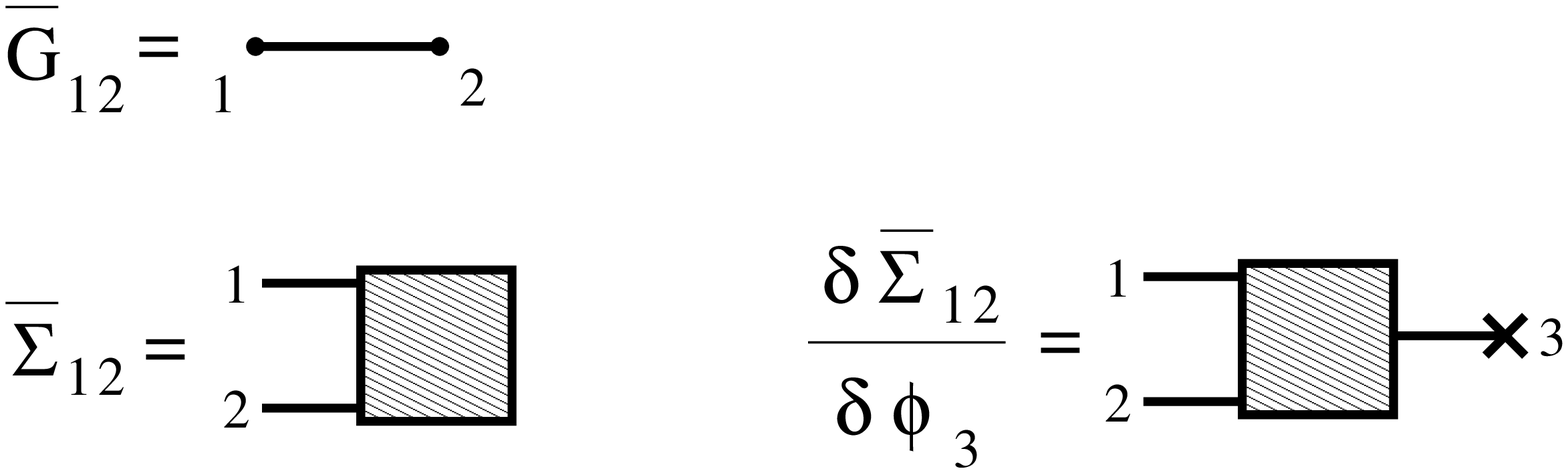}
\caption{\small \label{fig:G_Sigma} 
Diagrammatic representations of the functions $\bG_{12}$, $\bSigma_{12}$ and 
$\delta\bSigma_{12}/\delta\phi_3$.}
\end{center}
\end{figure}
\begin{figure}
\begin{center}
\includegraphics[width=11cm]{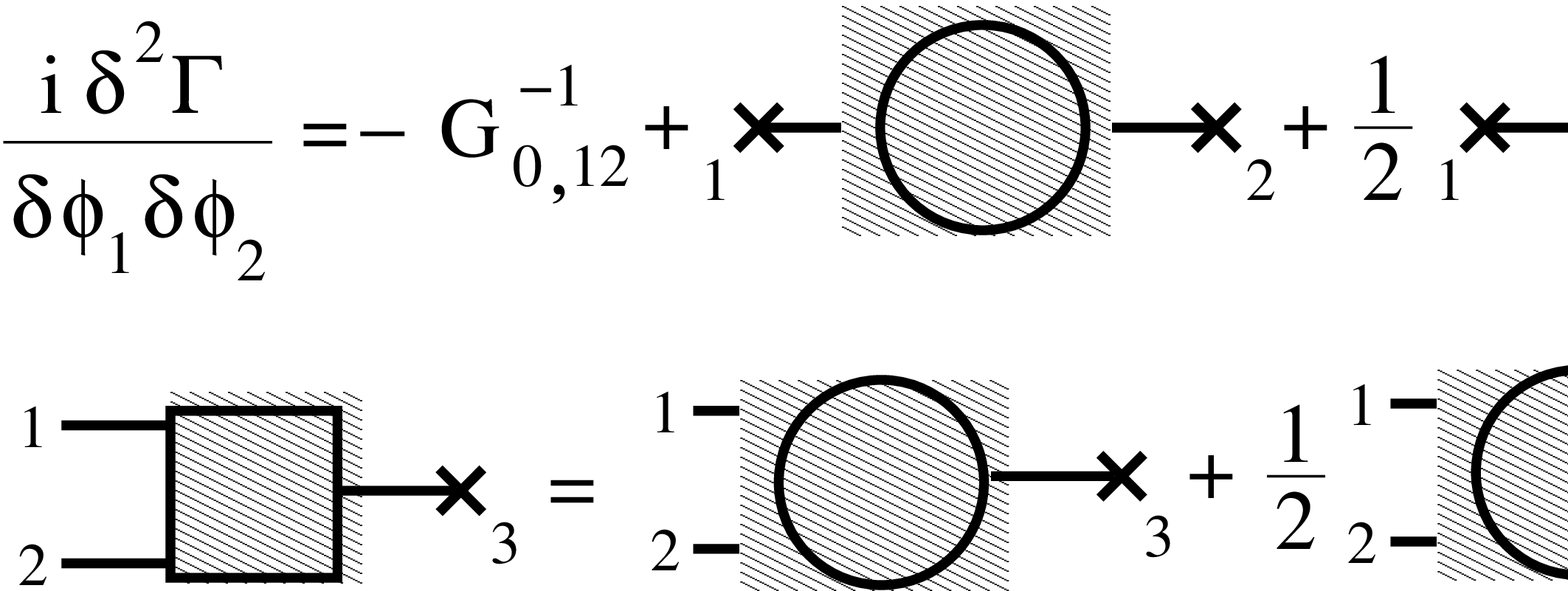}
\caption{\small Diagrammatic representation of equations (\ref{eq:2pt}) 
and (\ref{eq:resum1}).
\label{fig:Sigma}}
\end{center}
\end{figure}

All higher order derivatives can be obtained by differentiating the two
equations (\ref{eq:2pt}) and (\ref{eq:resum1}).
For this purpose, it is useful to introduce the following diagrammatic 
notation: We represent the 2PI kernels 
\mbox{$\sim \delta^{m+n}\Gamma_\inter/\delta\phi^m\delta G^n\big|_{\bG}$}
having $m+2n$ external legs by dashed circles. Legs ended by crosses 
represent derivatives with respect to the field $\phi$ and 
pairs of legs (without crosses) represent derivatives with respect to $G$.
For each derivative with respect to $G$ we add a symmetry factor of two. 
This is illustrated in Fig.~\ref{fig:Gamma_int_der}. Kernels can connect to 
other structures with lines representing the two-point function $\bG$ 
(cf.~Fig.~\ref{fig:G_Sigma}). 
It is useful to introduce a notation for the self-energy $\bSigma_{12}$, defined
in \Eqn{eq:se}: We use a square box with a pair of legs, which follows
the above notation since
$\bSigma_{12}$ is obtained as a derivative with respect to $G_{12}$.
A leg with a cross is added to this box for each derivative with respect to the 
field $\phi$, as illustrated in Fig.~\ref{fig:G_Sigma}. The diagrammatic 
representation 
of Eqs.\ (\ref{eq:2pt}) and (\ref{eq:resum1}) is shown in Fig.\ \ref{fig:Sigma}.
\begin{figure}
\begin{center}
\includegraphics[width=12cm]{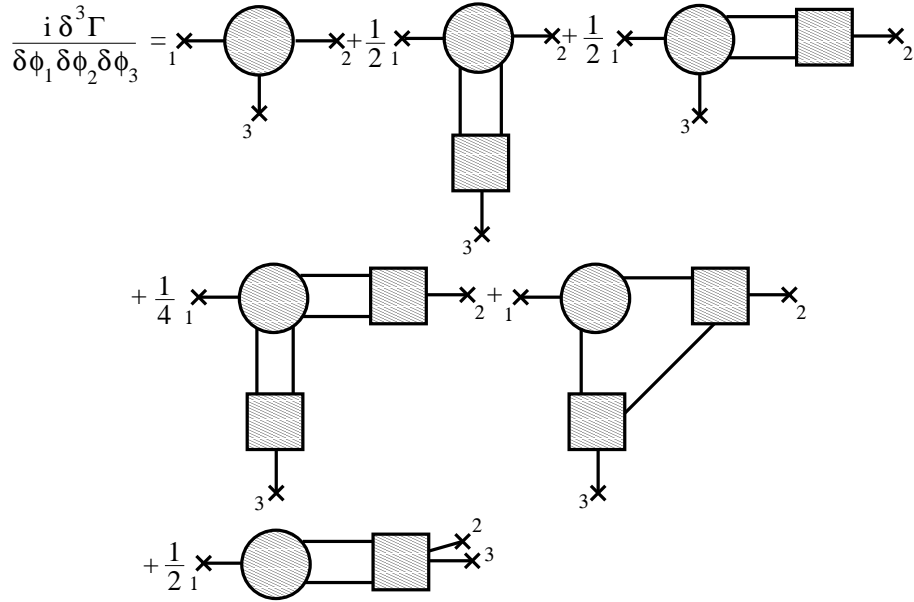}
\caption{\small Diagrammatic representation of the third derivative 
$\delta^3 \Gamma/\delta\phi_1\delta\phi_2\delta\phi_3$
of the 2PI-resummed effective action.
The last graph contains the function $\delta^2\bSigma/\delta\phi^2$,
which sums an infinite series of terms described by the
integral equation shown in Fig.~\ref{fig:sigma_der2}.
\label{fig:third_derivative}}
\end{center}
\end{figure}
\begin{figure}
\begin{center}
\includegraphics[width=14cm]{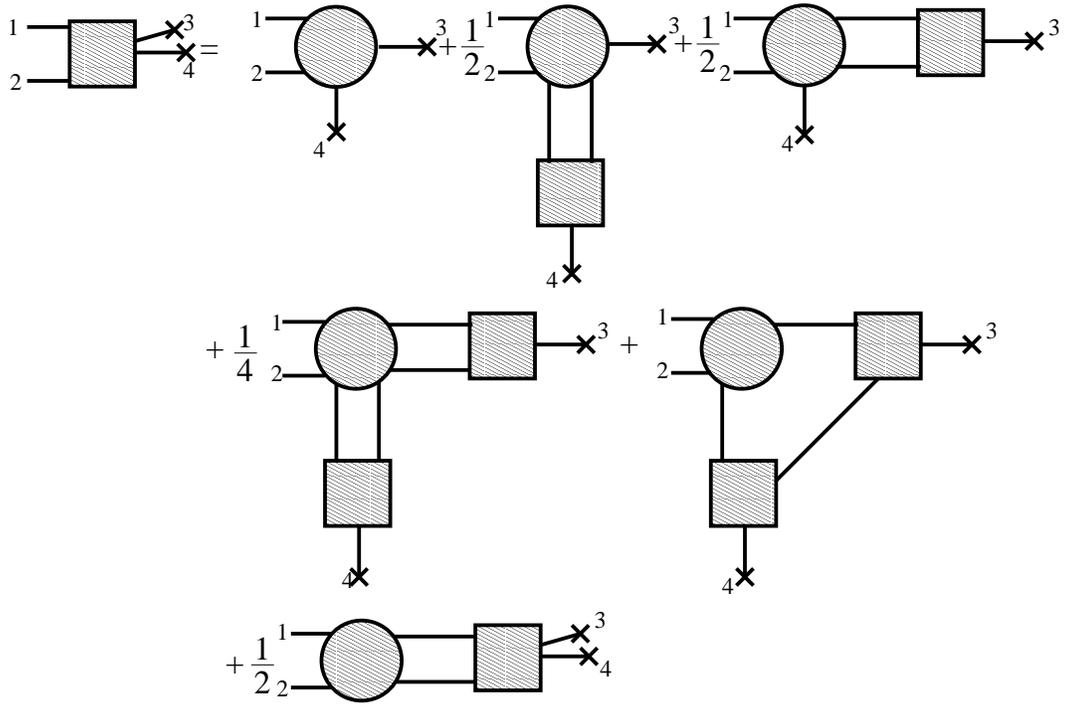}
\caption{\small
Diagrammatic representation of the integral equation 
for $\delta^2\bSigma_{12}/\delta\phi_3\delta\phi_4$.
\label{fig:sigma_der2}}
\end{center}
\end{figure}

\begin{figure}[tb]
\begin{center}
\includegraphics[width=10cm]{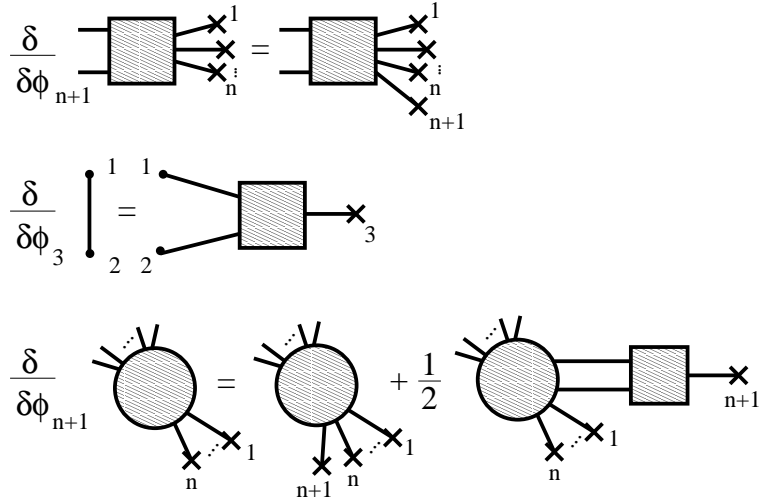}
\caption{\small Rules for obtaining the diagrammatic representation of proper vertices: 
  A field derivative can act on three different objects, 
  the self-energy $\bSigma(\phi)$, 
  the propagator $\bG(\phi)$ and the various derivatives of 
  $i\Gamma_{\rm int}$ represented by circles. The first rule represents 
  the fact that, by definition, adding a leg with a cross
  to a square box corresponds to taking 
  a derivative with respect to the field. The second rule essentially corresponds 
  to \Eqn{eq:link}. Finally the third rule arises from the fact that a given 
  kernel has both an explicit field-dependence and an implicit one through the 
  function $\bG(\phi)$. 
  Thus to the usual explicit derivative with respect to the field, one must add an explicit 
  derivative with respect to $G$ convoluted with a total derivative of the function $\bG(\phi)$, 
  hence the additional square box.\label{fig:derivatives}}
\end{center}
\end{figure}

Using this diagrammatic notation we show the three-point function 
$\delta^3 \Gamma/\delta \phi_1\delta \phi_2\delta \phi_3$ in 
Fig.~\ref{fig:third_derivative}. One observes the appearance of the 
function $\delta^2\bSigma/\delta\phi^2$ in the last graph of 
that figure. By differentiating Eq.~(\ref{eq:resum1}) once, one sees 
that the latter satisfies an integral equation
represented in Fig.\ \ref{fig:sigma_der2}.
Note that these expressions have a very similar structure
as those appearing already for the two-point function:
According to Eq.~(\ref{eq:2pt}), the expression $\Gamma^{\rm (2)}$
involves a finite number of terms and a term 
containing the derivative $\delta \bSigma/\delta\phi$.
The latter is the solution of the linear integral equation
\eqn{eq:resum1}, represented in Fig.~\ref{fig:Sigma}. By taking 
further derivatives of Eqs.~(\ref{eq:2pt}) and (\ref{eq:resum1})
one observes that a similar structure appears for all higher
$n$-point functions: $\Gamma^{(n)}$ involves derivatives 
$\delta^p\bSigma/\delta\phi^p$ with $1\le p \le n-1$ and the latter 
satisfy linear integral equations of the form:
\beq
\label{eq:resumgen}
 \frac{\delta^p\bSigma_{ab}}{\delta\phi_1\cdots\delta\phi_p}=
 \mathcal{A}^{ab}_{1\ldots p}
 +\frac{2i\delta^2\Gamma_\inter}{\delta G_{ab}\delta G_{cd}}\Big|_{\bG}
 \bG_{ce}\bG_{fd}\frac{\delta^p\bSigma_{ef}}{\delta\phi_1\cdots\delta\phi_p}\,.
\eeq
Here the function $\mathcal{A}$ involves a finite number of 
2PI kernels including lower derivatives $\delta^k\bSigma/\delta\phi^k$ 
with $1\le k \le p-1$.
Similarly to Eq.\ (\ref{eq:resum1}), the above equation resums an infinite
series of ladder graphs with rungs given by $2i\,\delta^2\Gamma_\inter/
\delta G^2|_\bG$ and where each ladder is ended by the function 
$\mathcal{A}$. This general structure is most easily seen by a diagrammatic
analysis, using the rules for derivation with respect to the field depicted
in Fig.~\ref{fig:derivatives} (see the comments in the caption).

As emphasized in Eq.~(\ref{eq:resumgen}), all these ladder 
resummations are generated by the kernel
\beq
\label{eq:BSkernel}
 \bLambda_{12,34}\equiv
 4\frac{\delta^2\Gamma_\inter}{\delta G_{12}\delta G_{34}}\Big|_{\bG}\,.
\eeq
It is therefore useful to introduce the infinite series of ladder graphs,
denoted as $\bV_{12,34}$, which satisfies the following integral, 
Bethe-Salpeter--like equation:\footnote{This equation plays a central role
in the work of Refs.~\cite{vanHees:2001ik,Blaizot:2003br}. Its importance has
also been emphasized in the context of transport coefficient calculations at high 
temperature \cite{Aarts:2003bk}, or when discussing the so-called 
Landau-Pomeranchuk-Migdal effect in the context of photon-production in 
high-energy heavy-ion collisions~\cite{Serreau:2003wr}.}
\beq
\label{eq:BS}
 \bV_{12,34}=\bLambda_{12,34}
 +\frac{i}{2}\,\bLambda_{12,ab}\,\bG_{ac}\bG_{db}\,\bV_{cd,34}\,.
\eeq
In terms of this
vertex-function the set of integral equations (\ref{eq:resumgen}) 
(cf.~also Eq.~(\ref{eq:resum1})) can be solved explicitly:
\beq
\label{eq:resumsol}
 \frac{\delta^p\bSigma_{ab}}{\delta\phi_1\cdots\delta\phi_p}=
 \mathcal{A}^{ab}_{1\ldots p}
 +\frac{i}{2}\,\bV_{ab,cd}\,\bG_{ce}\bG_{fd}\,\mathcal{A}^{ef}_{1\ldots p}\,.
\eeq
When expressed in terms of $\bV$, the $n$-point functions (\ref{eq:npoint}) 
only involve a finite number of contributions. This considerably simplifies 
the analysis of UV singularities.

%%%%%%%%%%%%%%%%%%%%%%%%%%%%%%%%%%%%%%%%%%%%%%%%%%%%%%%%%%%%%%%%%%%%%%%%%%%%%%%%%%%%%%%%%%%%%%%%%%%%%%%%%%%%%%%%%%%%%%%   Explicit example   %%%%%%%%%%%%%%%%%%%%%%%%%%%%%%%%%%%%%%%%%%%%%%%%%%%%%%%%%%%%%%%%%%%%%%%%%%%%%%%%%%%%%%%%%%%%%%%%%%%%%%%%%%%%%%%%%%%%%%%%%%%%%%%%%%%%%%%%%%%%%%%%%%%%%%%%%%%%%%%%%%%%%%%%%%%%%%%%%%%%%
\subsection{Vertex functions for vanishing field expectation value}
\label{sec:explicit}

In the following we present some relevant equations for $\bphi \equiv 0$,
since for the purpose of renormalization it is sufficient 
to consider the symmetric phase (see Sec.~\ref{sec:finite2}). 
To be explicit, we specify the above general 
considerations to a four-dimensional $Z_2$-symmetric theory with 
$\varphi^4$-interaction. As a consequence all functions made of an odd 
number of field derivatives of the 2PI effective action vanish.
We will make frequent use of the symmetry properties
$G_{12} = G_{21}$, $\bLambda_{12,34}=\bLambda_{21,34}
=\bLambda_{12,43}=\bLambda_{34,12}$ (cf. \Eqn{eq:BSkernel})
and equivalently for the vertex-function $\bV_{12,34}$ given in 
\Eqn{eq:BS}. The two- and four-point functions then read:
\bea
\label{eq:phi4_2pf1}
 \Gamma^{(2)}_{12}&=&
 iG_{0,12}^{-1}
 +\frac{\delta^2\Gamma_\inter}{\delta\phi_1\delta\phi_2}\Big|_{\bG}\,,\\
 \nonumber\\
\label{eq:phi4_4pf1}
 \Gamma^{(4)}_{1234}&=& 
 \frac{\delta^4\Gamma_\inter}{\delta\phi_1\cdots\delta\phi_4}\Big|_{\bG}
 +\left(\frac{\delta^3\Gamma_\inter}{\delta\phi_1\delta\phi_2\delta G_{ab}}\Big|_{\bG}
 \,\bG_{ac}\bG_{db}\,\frac{\delta^2\bSigma_{cd}}{\delta\phi_3\delta\phi_4}+{\rm perm.}\right)\,,\nn
\eea
where 'perm.' denotes possibe permutations of the indices $(2,3,4)$.
The second derivative of the self-energy appearing on the RHS of \Eqn{eq:phi4_4pf1}
satisfies the integral equation
\begin{figure}
\begin{center}
\includegraphics[width=13cm]{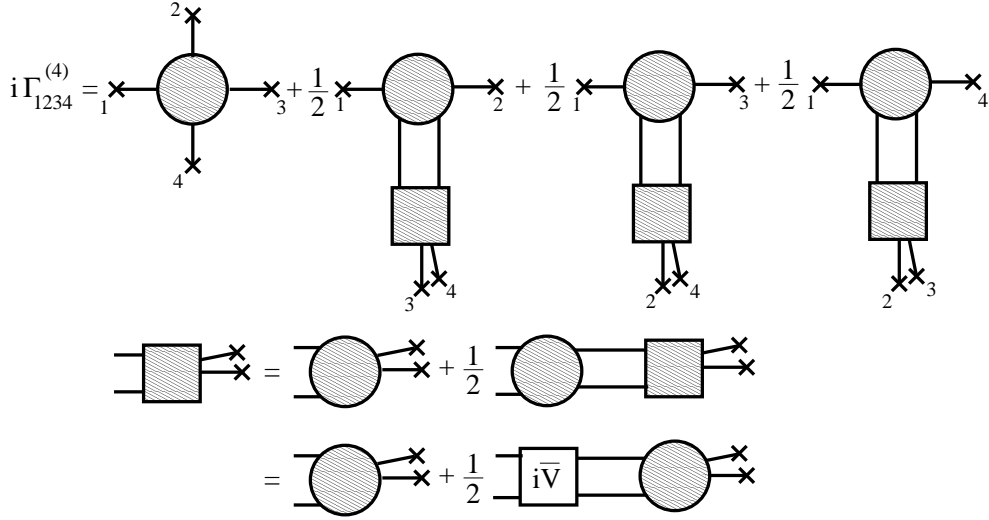}
\caption{\small \label{fig:fourth_der} Diagrammatic representation of the fourth 
derivative of the resummed 2PI effective action in the symmetric phase. It involves the function $\delta^2\bSigma/\delta\phi^2$ which equation can be solved in terms of $\bV$.}
\end{center}
\end{figure}
\beq
\label{eq:phi4_4pf2}
 \frac{\delta^2\bSigma_{12}}{\delta\phi_3\delta\phi_4}=
 \frac{2i\delta^3\Gamma_\inter}
 {\delta G_{12}\delta\phi_3\delta\phi_4}\Big|_{\bG}
 +\frac{i}{2} \bLambda_{12,ab}\,
 \bG_{ac}\bG_{db}\,\frac{\delta^2\bSigma_{cd}}{\delta\phi_3\delta\phi_4}\,.
\eeq
This equation is the one depicted in Fig.~\ref{fig:sigma_der2} 
for the symmetric phase, where $\bphi = 0$. Equations \eqn{eq:phi4_4pf1} 
and \eqn{eq:phi4_4pf2} are represented in Fig.\ \ref{fig:fourth_der}. As 
described above, the integral equation (\ref{eq:phi4_4pf2}) can be solved 
explicitely in terms of the vertex-function $\bV$ (cf.\ 
Eq.~(\ref{eq:BS}) and Fig. \ref{fig:fourth_der}):
\beq
\label{eq:phi4_4pf3}
 \frac{\delta^2\bSigma_{12}}{\delta\phi_3\delta\phi_4}=
 \frac{2i\delta^3\Gamma_\inter}
 {\delta G_{12}\delta\phi_3\delta\phi_4}\Big|_{\bG}
 +\frac{i}{2}\,\bV_{12,ab}\,\bG_{ac}\bG_{db}\,
 \frac{2i\delta^3\Gamma_\inter}
 {\delta G_{cd}\delta\phi_3\delta\phi_4}\Big|_{\bG}\,.
\eeq
Inserting this expression in Eq.\ (\ref{eq:phi4_4pf1}) for the four-point 
function, one obtains a closed expression in terms of $\bV$ and 2PI kernels. 
This is represented in Fig.~\ref{fig:fourth_der2} 
(cf.\ also Eq.\ (\ref{eq:4pt2symb}) below).
\begin{figure}
\begin{center}
\includegraphics[width=14cm]{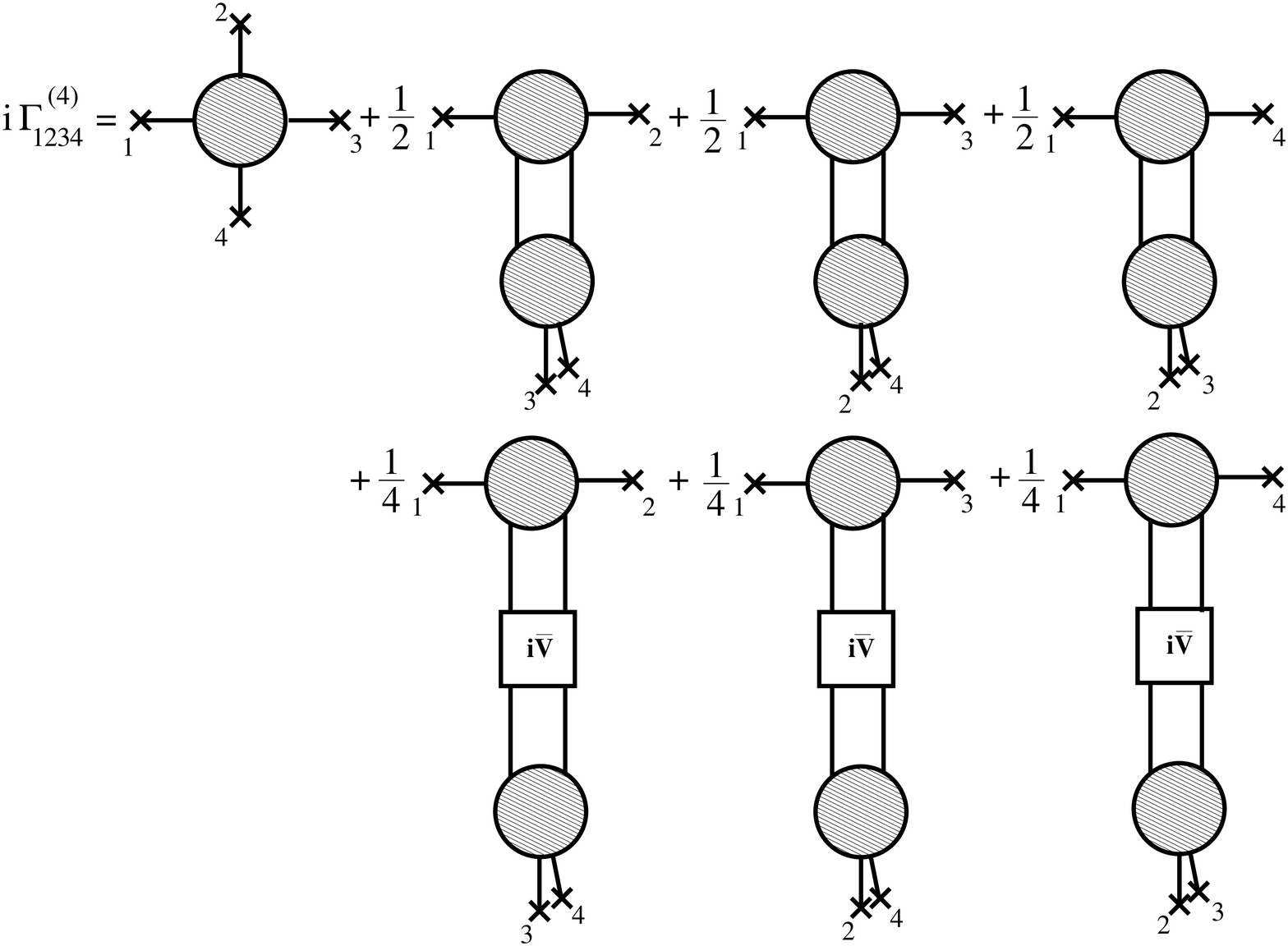}
\caption{\small \label{fig:fourth_der2} Diagrammatic representation 
of the fourth derivative of the resummed 2PI effective 
action in the symmetric phase.}
\end{center}
\end{figure}

In order to summarize the set of relevant equations to be used in the 
following, we define:
\beq
\label{eq:2kernel2res}
 \Sigma_{12}\equiv 
 i\frac{\delta^2\Gamma_\inter}{\delta\phi_1\delta\phi_2}\Big|_{\bG}\,,
\eeq
as well as
\beq
\label{eq:4kernel2res}
 \Lambda_{12,34}\equiv 2\frac{\delta^3\Gamma_\inter}
 {\delta\phi_1\delta\phi_2\delta G_{34}}\Big|_{\bG}\,,
\eeq
with symmetry properties $\Lambda_{12,34}=\Lambda_{21,34}
=\Lambda_{12,43}$. In analogy with the vertex function $\bV$, 
we introduce the notation:
\beq 
\label{eq:V}
 V_{12,34}\equiv -i\frac{\delta^2\bSigma_{34}}{\delta\phi_1\delta\phi_2}\,.
\eeq
Leaving space-time indices implicit, the integral equation (\ref{eq:BS})
takes the following compact form:
\beq
\label{eq:BSsymb}
 \bV=\bLambda+\frac{i}{2}\,\bLambda\,\bG^2\,\bV
 =\bLambda+\frac{i}{2}\,\bV\,\bG^2\,\bLambda\,,
\eeq
where the second equality follows from the symmetry properties of the functions
$\bLambda$ and $\bV$. Similarly, the integral equation (\ref{eq:phi4_4pf2}) 
and its solution (\ref{eq:phi4_4pf3}) in terms of $\bV$ now read:\footnote{Notice 
the different ordering of the functions 
as compared to Eqs.\ (\ref{eq:phi4_4pf2}) and (\ref{eq:phi4_4pf3}). This is a mere 
consequence of the ordering of space-time indices in our definition (\ref{eq:V}). 
This choice proves the most convenient for later use 
(cf.~Sec.~\ref{sec:finite}).}
\beq
\label{eq:intVsymb}
 V=\Lambda+\frac{i}{2}\,V\,\bG^2\,\bLambda
 =\Lambda+\frac{i}{2}\,\Lambda\,\bG^2\,\bV\,.
\eeq
It is important to realize that all the quantities we define are, up to an 
overall factor, made of a resummation of perturbative diagrams, with no extra 
factors other than the standard symmetry factors. 
This is useful in order to discuss renormalization in a diagrammatic way.
Finally, it is useful to define the functions $\Lambda^\dagger$ and $V^\dagger$,
such that $\Lambda^\dagger_{12,34}\equiv\Lambda_{34,12}$ and similarly for $V^\dagger$. 

With these notations, the two- and four-point functions read:
\bea
\label{eq:2ptsymb}
 \Gamma^{(2)}&=&iG_{0}^{-1}-i\Sigma\,,\\
\label{eq:4ptsymb}
 \Gamma^{(4)}&=&\frac{\delta^4\Gamma_\inter}{\delta\phi^4}\Big|_{\bG}
 +\frac{i}{2}\,\Big(\Lambda\,\bG^2\,V^\dagger+{\rm perm.}\Big)\,.
\eea
Using the explicit expression of the function $V$, Eq.\ (\ref{eq:intVsymb}), the 
four-point function is given by (cf.~Fig.\ \ref{fig:fourth_der2}):
\beq
\label{eq:4pt2symb}
 \Gamma^{(4)}=\frac{\delta^4\Gamma_\inter}{\delta\phi^4}\Big|_{\bG} 
 +\frac{i}{2}\,\Big(\Lambda\,\bG^2\,\Lambda^\dagger
 +\frac{i}{2}\,\Lambda\,\bG^2\,\bV\,\bG^2\,\Lambda^\dagger
 +{\rm perm.}\Big)\,.\nonumber\\
\eeq
As emphasized previously, we are left with closed expressions involving only 
2PI kernels and the vertex function $\bV$, appearing in a finite number of 
loop integrals. The same is true for higher $n$-point functions as well.
This property will simplify considerably the discussion of
renormalization in the next sections.

We point out that a simplification occurs for approximations where the
following relation between two-point 2PI kernels is satisfied:\footnote{This 
is for instance the case for the 2PI $1/N$-expansion at NLO \cite{Aarts:2002dj} 
(see also Sec.~\ref{sec:appN}), or for the approximation discussed in 
Ref.~\cite{Berges:2004hn}. Notice that this relation is actually fulfilled
in the exact theory, see Appendix \ref{sec:appcorrel}.}
\beq
\label{eq:relation}
 \frac{\delta^2\Gamma_\inter}{\delta\phi_1\delta\phi_2}\Big|_{\phi=0}=
 2\frac{\delta\Gamma_\inter}{\delta G_{12}}\Big|_{\phi=0}\,.
\eeq
As shown in Appendix \ref{sec:appcorrel}, this implies, in particular, that 
$\bSigma=\Sigma$, $\bLambda=\Lambda$ and, consequently, $\bV=V$. Using the 
integral equation \eqn{eq:BSsymb}, the expressions \eqn{eq:2ptsymb} and 
\eqn{eq:4ptsymb} for the two- and four-point functions simplify to:
\bea
 \Gamma^{(2)}&=&i\bG^{-1} \, ,\\
\label{simpler}
 \Gamma^{(4)}&=&\frac{\delta^4\Gamma_\inter}{\delta\phi^4}\Big|_{\bG} 
 +\Big(\bV-\bLambda+{\rm perm.}\Big)\,.
\eea
Obviously, the renormalization of $\Gamma^{(2)}$ and $\Gamma^{(4)}$ is 
greatly simplified in that case. 

Finally, we mention that in the exact theory, the various two- and four-point 
functions introduced above satisfy the relations: $\Gamma^{(2)}=i\bG^{-1}$ and 
$\bV=V=\Gamma^{(4)}$, as shown in Appendix \ref{sec:appcorrel}. Although such 
relations are generally not respected anymore once approximations are introduced,
they are important when imposing renormalization conditions as is discussed
below.

%%%%%%%%%%%%%%%%%%%%%%%%%%%%%%%%%%%%%%%%%%%%%%%%%%%%%%%%%%%%%%%%%%%%%%%%%%%%%%%%%%%%%%%%%%%%%%%%%%%%%%%%%%%%%%%%%%%%%%%%%%%%%%%%%%%%%%%%%%%%%%%%%%%%%%   Renormalizability   %%%%%%%%%%%%%%%%%%%%%%%%%%%%%%%%%%%%%%%%%%%%%%%%%%%%%%%%%%%%%%%%%%%%%%%%%%%%%%%%%%%%%%%%%%%%%%%%%%%%%%%%%%%%%%%%%%%%%%%%%%%%%%%%%%%%%%%%%%%%%%%%%%%%%%%%%%%%%%%%%%%%%%%%%%%%%%%%%%%%%
\section{Renormalization}
\label{sec:renorm}

The diagrammatic tools introduced in the previous section allow one to analyze
the origin of UV divergences in a very efficient way. As 
exemplified in Fig.~\ref{fig:third_derivative} for the case of the three-point 
function, all diagrammatic contributions to a given proper vertex share a common
generic structure: They involve particular 2PI kernels (represented by 
circles) and field derivatives of the self-energy $\bSigma$ (represented by
boxes), which enter loops
with lines associated to the two-point function $\bG$. The 2PI kernels, self-energy
derivatives and $\bG$ generally contain divergences. Furthermore, the loops 
they enter may also be divergent. Our renormalization program thus aims at 
making kernels, self-energy derivatives and $\bG$
finite and at removing the remaining divergences in the loop integrals
involving these objects. 
In order to do so, we make extensive use of the techniques put forward 
in Refs.~\cite{vanHees:2001ik,Blaizot:2003br}, where the BPHZ subtraction 
procedure is applied to 2PI diagrams with lines associated to the 
resummed two-point function $\bG$. As a consequence of the two-particle 
irreducibility of the diagrams, this ``2PI'' BPHZ analysis automatically gives the 
correct mass and field-strength counterterms. Moreover, 
it enables one to identify the 
counterterms needed to renormalize all kernels 
$\delta^{m+2n}\Gamma_\inter/\delta G^m
\delta\phi^{2n}|_\bG$ with $2(m+n)\ge 4$ external legs. The two-point kernels
$\delta\Gamma_\inter/\delta G|_\bG$ and $\delta^2\Gamma_\inter/\delta\phi^2|_\bG$, 
however, require a more careful analysis: In contrast to standard perturbation
theory, the BPHZ procedure applied to resummed diagrams actually misses an infinite 
number of coupling sub-divergences hidden in the resummed two-point function $\bG$.\footnote{These 
sub-divergences arise from an interplay between the asymptotic logarithmic behavior of the various 
two-point 2PI kernels. The fact that these subtleties do not show up in higher 
2PI kernels follows from simple power counting arguments.}  This problem was 
investigated in detail in Refs.~\cite{vanHees:2001ik,Blaizot:2003br} for the 
case of the kernel $\delta\Gamma_\inter/\delta G|_\bG$ in the symmetric phase. 
There, it has been shown that the hidden sub-divergences are all generated by 
successive iterations of the Bethe-Salpeter--type equation (\ref{eq:BSsymb}) 
and that they can in fact be absorbed in the renormalization of the function $\bV$.
This in turn amounts to a single subtraction, which corresponds to an infinite shift 
of the tree-level contribution to $\delta^2\Gamma_\inter/\delta G^2|_\bG$, i.e.\ to a 
simple coupling counterterm.\footnote{We note that, as a consequence, the kernel 
$\delta^2\Gamma_\inter/\delta G^2|_\bG$ is not finite. However, as a consequence
of the previous BPHZ procedure, its divergent part is local and is actually given by the
one counterterm needed to renormalize the function $\bV$.} 
Here, we apply a similar analysis to the kernel $\delta^2\Gamma_\inter/
\delta\phi^2|_\bG$. We show that the corresponding coupling sub-divergences
are in fact generated by successive iterations of the integral equation
(\ref{eq:intVsymb}) and can be absorbed by a shift of the tree-level 
contribution to the four-point kernel $\delta^3\Gamma_\inter/\delta G\delta\phi^2|_\bG$. 
This nonperturbative shift renormalizes the function $V$.

Once the two-point kernels have been renormalized, a careful analysis of the expressions derived 
in the previous section for proper vertices reveals that all potential sub-divergences 
in the latter can be absorbed in the renormalization of the four-point functions 
$\bV$ and $V$ and of 2PI kernels with two, six, or more than six legs. In fact, after 
the latter have been made finite, there remains only an overall divergence in the 
four-point function $\Gamma^{(4)}$, which can be eliminated by a standard 
counterterm. The latter corresponds to a shift of the tree-level contribution 
to the third four-point 2PI kernel $\delta^4\Gamma_\inter/\delta\phi^4|_\bG$. 
The remarkable result is that the only modifications to the 2PI BPHZ analysis described 
above reduce to a redefinition of the tree-level contributions to each of the 
four-point 2PI kernels $\delta^2\Gamma_\inter/\delta G^2|_\bG$, 
$\delta^3\Gamma_\inter/\delta G\delta\phi^2|_\bG$ and $\delta^4
\Gamma_\inter/\delta\phi^4|_\bG$. This is the main result of 
the present paper. With these counterterms being determined, all proper vertices 
$\Gamma^{(n)}$ are finite. Furthermore, the zero-point function is finite up to 
an irrelevant field-independent constant \cite{vanHees:2001ik}.

%%%%%%%%%%%%%%%%%%%%%%%%%%%%%%%%%%%%%%%%%%%%%%%%%%%%%%%%%%%%%%%%%%%%%%%%%%%%%%%%%%%%%%%%%%%%%%%%%%%%%%%%%%%%%%%%%%%%%%%%%%%%%%%%%%%%%%%%%%%%%%%%%%%%%%   Counterterms   %%%%%%%%%%%%%%%%%%%%%%%%%%%%%%%%%%%%%%%%%%%%%%%%%%%%%%%%%%%%%%%%%%%%%%%%%%%%%%%%%%%%%%%%%%%%%%%%%%%%%%%%%%%%%%%%%%%%%%%%%%%%%%%%%%%%%%%%%%%%%%%%%%%%%%%%%%%%%%%%%%%%%%%%%%%%%%%%%%%%%%%%%%%%%
\subsection{Counterterms}\label{sec:ct}
\begin{figure}
\begin{center}
\includegraphics[width=4cm]{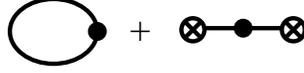}
\caption{\small \label{fig:ct} Diagrammatic representation of the mass and field-strength 
counterterms. The black dots denote indistinctly 
the mass counterterm $\dmsq$ or field-strength counterterm
$\dZ$ appearing in the first two terms on the RHS of 
Eq.~(\ref{eq:gammaint_ren}). The line of the closed loop represents
$G_R$, while a circled cross denotes $\phi_R$.}
\end{center}
\end{figure}

We consider a quantum field theory with classical action
\beq
\label{eq:action}
 S[\varphi] = -\int_x \,
 \left\{\frac{1}{2}\,\varphi\left(\square+m^2\right)\varphi
 +\frac{\lambda}{4!}\varphi^4\right\}\,,
\eeq
employing 
the notation $\int_x\equiv\int_\Lambda d^4x$, where the subscript $\Lambda$ 
refers to some given regularization procedure
such as, for instance, cutoff or dimensional regularization. 
Introducing the renormalized field 
\beq
\label{wfr}
 \varphi_R=Z^{-1/2}\varphi\,,
\eeq
the action reads:\footnote{To prevent a proliferation of symbols, we will
distinguish the action $S$ (and the effective action $\Gamma$) in terms of 
renormalized fields by its arguments.}
\bea
 S[\varphi_R] &=&-\int_x\,
 \left\{\frac{1}{2}\,\varphi_R \left(\square+m_R^2\right)\varphi_R
 +\frac{\lambda_R}{4!}\varphi_R^4\right.\nonumber\\
\label{eq:action_ren}
 && \left.\hspace{1.cm}+\,\frac{1}{2}\,
\varphi_R\left(\dZ\,\square+\dmsq\right)\varphi_R
 +\frac{\dl}{4!}\varphi_R^4\right\},
\eea
with the standard definitions
\bea
 \dZ &=& Z-1\,,\nonumber\\
 Z m^2 & = & m_R^2+\dmsq\,,\\
 Z^2 \lambda & = & \lambda_R +\dl\,.\nonumber
\eea
Including the counterterms $\sim \dZ$, $\dmsq$ and $\dl$ into 
the interaction part of the action, we define
\beq
 S_{0R}[\varphi_R]=-\frac{1}{2}\,\int_x\,
 \varphi_R \left(\square+m_R^2\right)\varphi_R
 \equiv\frac{1}{2}\,\int_{xy}\,
 \varphi_R(x) iG_{0R}^{-1}(x,y) \varphi_R(y)\,,
\eeq
where $G_{0R}$ denotes the renormalized free propagator. 
It follows with \Eqn{eq:S0} that
\beq
 ZG_{0}^{-1}=G_{0R}^{-1}+\dG_0^{-1}\,,
\eeq
where we introduced the notation
\beq
 i\dG_{0}^{-1} (x,y) =-(\delta Z \,\square+\delta m^2)\,\delta^4(x-y)\,.
\eeq
The corresponding 2PI effective action in terms of the fields 
$\phi$ and $G$ is given by Eq.~(\ref{eq:2PI}).
In terms of the renormalized fields
\beq
 \phi_R = Z^{-1/2}\phi \quad, \quad G_R = Z^{-1}G \,,
\eeq
it can be written, up to an irrelevant constant, as
\bea
\label{eq:2PI_ren}
 \Gammatpi[\phi_R,G_R] &=& 
 S_{0R}[\phi_R]+\frac{i}{2}\,\Tr \ln G_R^{-1}+\frac{i}{2}\,\Tr G_{0R}^{-1} G_R
 \nonumber\\
 &&+\Gamma_\inter^R[\phi_R,G_R;\lambda_R+\dl,\dZ,\dmsq]\,,
\eea
where we defined the ``interaction'' functional $\Gamma_\inter^R$ in terms of 
renormalized fields as (note that $\delta G_0^{-1}$ is treated as part
of the interaction):
\bea\label{eq:2PI_ren_int}
 \Gamma_\inter^R[\phi_R,G_R;\lambda_R+\dl,\dZ,\dmsq] &=&
 \frac{i}{2}\, \int_{xy}\,
 \phi_R(x) \dG_{0}^{-1}(x,y) \phi_R(y) 
 \nn
 &&+\frac{i}{2}\, \Tr\! \dG_{0}^{-1} G_R
 +\Gamma_\inter[\phi_R,G_R;\lambda_R+\dl]\,. \nn
\label{eq:gammaint_ren}
\eea
Here, we have used the fact that
\beq
 \Gamma_\inter[\phi,G;\lambda]=
 \Gamma_\inter[\phi_R,G_R;\lambda_R+\delta\lambda]\,,
\eeq
which follows from the standard relation between the number of vertices $v$
and the number of external and internal lines, $e$ and $i$, of a given diagram:
$4v=e+2i$. Alternatively,
one can construct the 2PI effective action in terms of renormalized fields
directly from the defining functional integral with action
(\ref{eq:action_ren}), treating the counterterms 
as part of the interaction. The coupling counterterm merely shifts the coupling 
constant to be used in the vertices of 2PI diagrams whereas the quadratic
counterterms $\dZ$ and $\dmsq$ give new (two-legs) vertices. There are only two 
2PI diagrams one can construct with such vertices, which are represented in 
Fig.~\ref{fig:ct}. These correspond to the first two terms on the RHS of 
Eq.~(\ref{eq:gammaint_ren}).

%%%%%%%%%%%%%%%%%%%%%%%%%%%%%%%%%%%%%%%%%%%%%%%%%%%%%%%%%%%%%%%%%%%%%%%%%%%%%%%%%%%%%%%%%%%%%%%%%%%%%%%%%%%%%%%%%%%%%%%%%%%%%%%%%%%%%%%%%%%%%%%%%%%%%%   Necessary conditions   %%%%%%%%%%%%%%%%%%%%%%%%%%%%%%%%%%%%%%%%%%%%%%%%%%%%%%%%%%%%%%%%%%%%%%%%%%%%%%%%%%%%%%%%%%%%%%%%%%%%%%%%%%%%%%%%%%%%%%%%%%%%%%%%%%%%%%%%%%%%%%%%%%%%%%%%%%%%%%%%%%%%%%%%%%%%%%%%%%%%%
\subsection{Conditions for renormalizability}

As for perturbative renormalizability, 2PI approximations are typically
only renormalizable for systematic expansions. In such cases, each new order
of the expansion involves a new selective summation of an infinite series of 
perturbative contributions and it is non-trivial that such approximations turn 
out to be renormalizable order by order. There may also be cases where a 
suitable expansion parameter is missing and one would like to retain only 
specific 2PI diagrams while dropping others.
In this subsection, we give a set of necessary conditions which any 2PI 
approximation has to fulfill in order to be renormalizable. We will 
see in the next subsection that these necessary conditions are
actually sufficient. 

For this it is useful to write\footnote{We omit in the notation the 
explicit dependence 
on the counterterms (cf. \Eqn{eq:2PI_ren_int}) for simplicity.}  
$\Gamma_\inter^R[\phi_R,G_R]$ as a power series in $\phi_R$:\footnote{Note that
this does not imply that the 2PI-resummed effective action, which includes
$\Gamma_{\rm int}^R[\phi_R,\bG_R(\phi_R)]$, can be written as a power series
in $\phi_R$.}
\bea
\label{eq:fieldexp}
 \Gamma_\inter^R[\phi_R,G_R] &=& \Gamma_\inter^{(0)}[G_R]+\sum_{n=1}^\infty
 \frac{1}{(2n)!}
 \int_{x_1,\ldots,x_{2n}}\,\Gamma_\inter^{(2n)}(x_1,\ldots,x_{2n};G_R)\,
 \nonumber\\
 && \times\, \phi_R(x_1)\cdots\phi_R(x_{2n})\,,
\eea
with zero-field part $\Gamma_\inter^{(0)}[G_R]\equiv\Gamma_\inter^R[\phi_R=0,G_R]$
and where
\beq
\label{eq:fieldexpcoeff}
 \Gamma_\inter^{(2n)}(x_1,\ldots,x_{2n};G_R)\equiv
 \left.\frac{\delta^{2n}\Gamma_\inter^R[\phi_R,G_R]}
 {\delta\phi_R(x_1)\cdots\delta\phi_R(x_{2n})}\right|_{\phi_R=0}\,.
\eeq
The number of fields is even because of the $Z_2$-symmetry of
the $\varphi^4$-theory.

\begin{figure}
\begin{center}
\includegraphics[width=14cm]{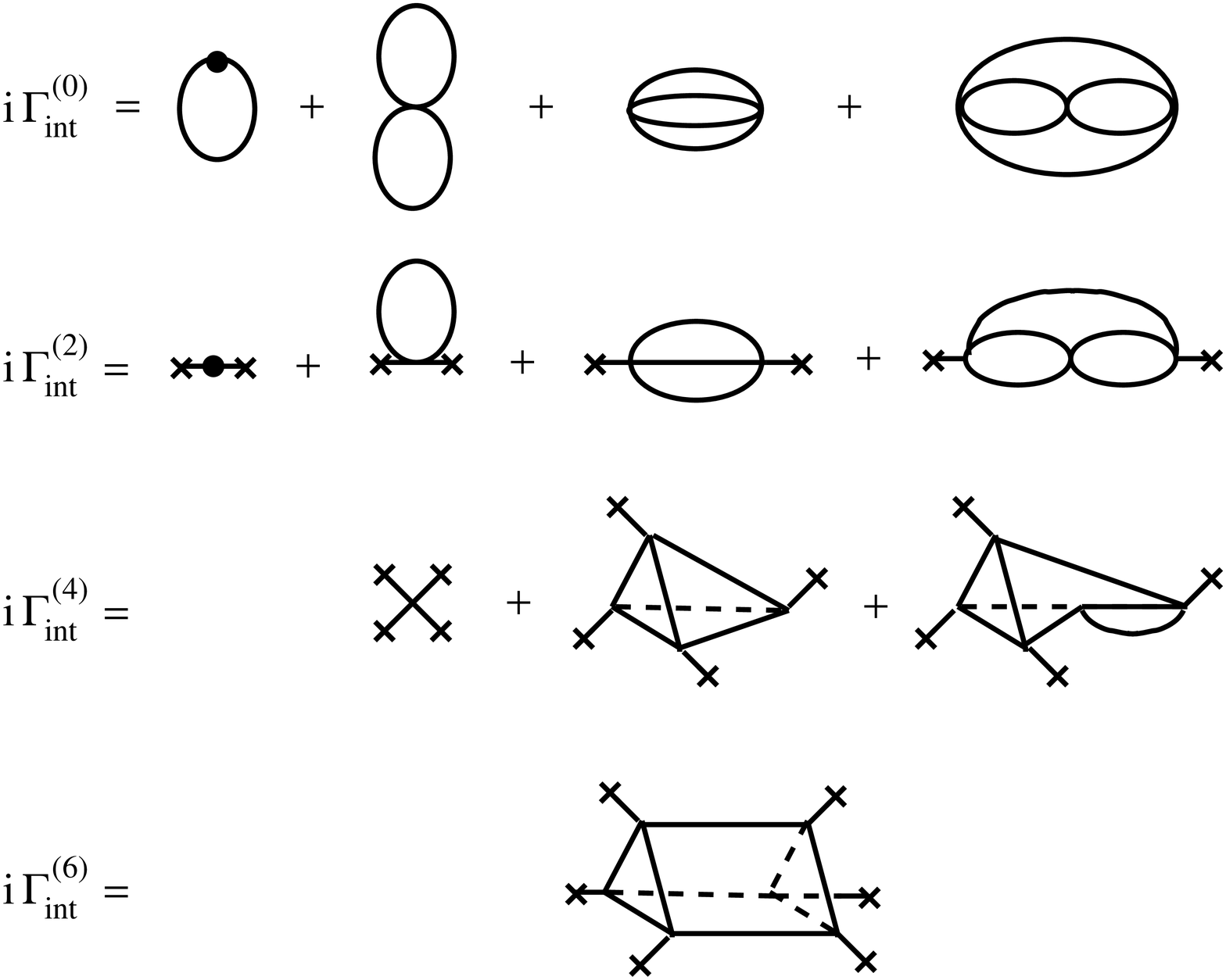}
\caption{\small \label{fig:exemples} Examples of approximations for 
$\Gamma_\inter^{(2n)}(G_R)$, see \Eqn{eq:fieldexp}, for $n=0,1,2,3$.}
\end{center}
\end{figure}

If $\Gamma_\inter^{(2n)}(G_R)$ is finite, all the 
kernels generated from it by taking derivatives 
with respect to $G_R$ are automatically finite. In Fig.~\ref{fig:exemples} 
we display possible contribution to each of the $\Gamma_\inter^{(2n)}$ for illustration, 
with the notation introduced in the caption of Fig.~\ref{fig:ct} for mass 
and field-strength counterterms.  We will consider in the following
the minimal requirements for a given approximation (ensemble of graphs) to be renormalizable. 
A necessary condition is that, whenever a graph contains a potentially divergent sub-diagram, 
there must be a corresponding graph in the truncation where the divergent sub-diagram has been 
replaced by a point. The latter corresponds to the counterterm (or BPHZ subtraction)
needed to cancel the sub-divergence. 

\begin{figure}
\begin{center}
\includegraphics[width=11cm]{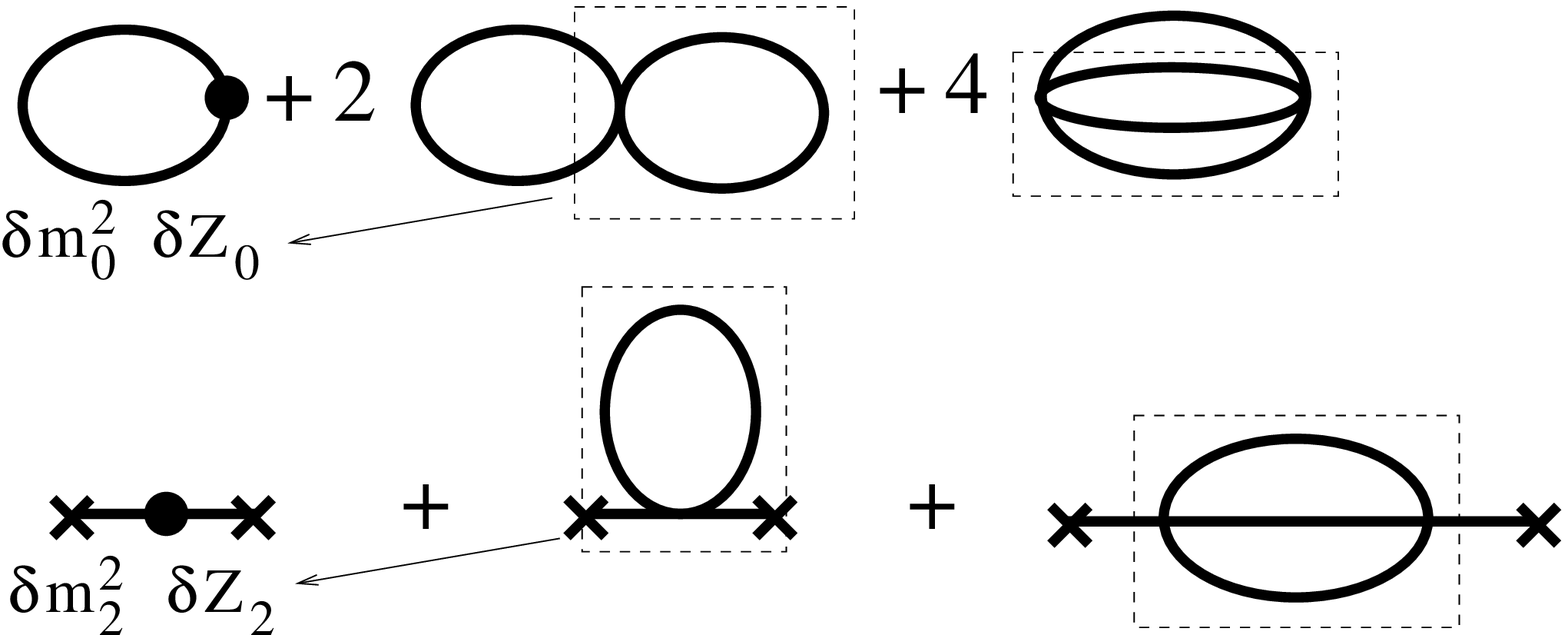}
 \caption{\small Illustration for the analysis of two-point singularities. 
   The factors arise from the different possibilities 
   to draw a two-point box.\label{fig:2point}}
\end{center}
\end{figure}

We first discuss two-point singularities. 
Following the standard procedure, we graphically represent sub-divergences by boxes 
surrounding the corresponding sub-diagrams.\footnote{More precisely, our boxes only 
represent the overall divergence of the considered sub-diagram. This assumes that 
all possible sub-divergences of the latter have been subtracted according to the
usual recursive BPHZ procedure.}
Because of the two-particle irreducibility of the diagrams (see Appendix~\ref{2PItop})
for the zero-field part $\Gamma_\inter^{(0)}$ 
the only two-point boxes one can draw 
are those which contain all the lines of the diagram but one. 
As illustrated in Fig.~\ref{fig:2point}, there is one diagram to absorb 
all these structures. It is given by the 
first diagram in Fig.~\ref{fig:ct}, which precisely corresponds to mass and field-strength 
counterterms. We denote the latter by $\dmsq_0$ and $\dZ_0$ to emphasize 
that they arise from the analysis of the zero-field part $\Gamma_\inter^{(0)}$. 
Similarly, for a given diagram with two external fields, which contributes 
to $\Gamma_\inter^{(2)}$, 
the only possible two-point box one can draw is the one containing 
the whole graph. 
The associated divergences can be absorbed in the mass and field-strength counterterms 
represented by the second diagram in Fig.~\ref{fig:ct}. We denote the latter by
$\dmsq_2$ and $\dZ_2$. We emphasize that for a given truncation these are not 
necessarily the same as $\dmsq_0$ and $\dZ_0$, which just reflects the fact
that these are different approximations of the same counterterm (cf.~below). 
Finally, there are no two-point boxes in 2PI diagrams with more than two 
external fields ($\Gamma_\inter^{(2n)}$ with $n\ge2$) for precisely the same topological 
reasons as there can be no other graphs than those of Fig.~\ref{fig:ct} with
mass and field-strength counterterms. This in turn is related to the fact that the 
latter only arise in the renormalization of the two-point 2PI kernels 
$\delta\Gamma_\inter^R/\delta G_R|_{\bG_R}$ and $\delta^2\Gamma_\inter^R/
\delta\phi_R^2|_{\bG_R}$.

\begin{figure}
\begin{center}
\includegraphics[width=7cm]{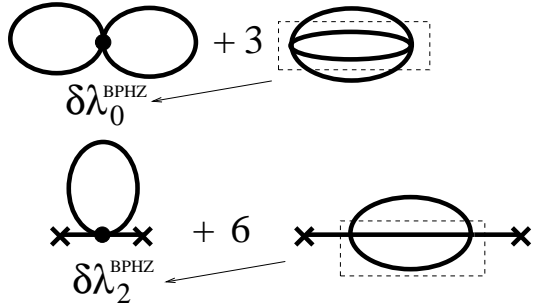}
 \caption{\small Diagrammatic analysis of four-point singularities. 
 The factors arise from the different possibilities 
 to draw a four-point box.\label{fig:4point}}
\end{center}
\end{figure}

We now turn to four-point singularities. If a given loop 
diagram is included in the truncation, all topologies generated by the 
BPHZ procedure described above must be included as well. Note that the latter 
only generates topologies with lower number of loops. For instance, if the 
three-loop (basket-ball) diagram is included, one needs to include the 
two-loop (eight) diagram as well. Similarly, the renormalization of the 
two-loop diagram with two external fields (setting-sun) requires the presence 
of the one-loop diagram with two external fields (tadpole) in the truncation. 
This is illustrated in Fig.~\ref{fig:4point}. We note that this
procedure does not mix diagrams with different number of external fields and
can therefore be applied to each of the $\Gamma_\inter^{(2n)}$ separately.
It turns out that this simple analysis gives the relevant 
subtractions needed to renormalize all 2PI kernels
$\delta^{m+2n}\Gamma_\inter^R/\delta G_R^{m}\delta\phi_R^{2n}|_{\bG_R}=
\delta^m\Gamma_\inter^{(2n)}/\delta G_R^m|_{\bG_R}$
having a number of legs $2(n+m)\ge 4$.\footnote{In practice it is
sufficient to renormalize the kernels with $2(n+m)=4$ as well as the functions
$\Gamma_\inter^{(2n)}(G_R)$ with $n\ge6$. All the other kernels with more than
four legs can be obtained from the latter by taking derivatives with respect to
$G_R$ and are, therefore, automatically finite.}

\begin{figure}
\begin{center}
\includegraphics[width=12cm]{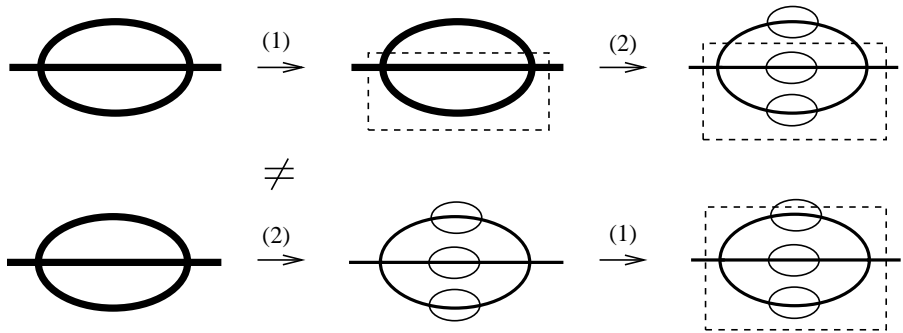}
\caption{\small \label{fig:missing_sing}
The BPHZ procedure applied to diagrams with resummed propagators misses singularities. 
This is simply because the two operations ``drawing boxes'' (1) and ``iterating diagrams'' 
(2) do not commute as illustrated in the figure for the case of the self-energy 
\eqn{eq:2kernel0}. The first line shows the coupling singularities which are accounted 
for by the BPHZ analysis regardless of the content of the propagator. The second line 
shows what happens if one first iterates the resummed propagator and then apply the 
BPHZ analysis. There are clearly more divergent topologies in the second case.}
\end{center}
\end{figure}

\begin{figure}
\begin{center}
\includegraphics[width=10cm]{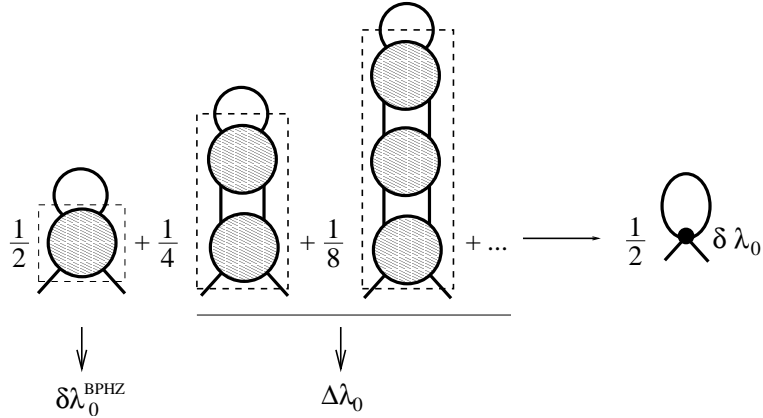}
 \caption{\small The coupling divergences in the two-point kernel $\delta\Gamma_\inter/\delta G$ are 
 those of the function $\bV$. The total counterterm $\dl_0=\dl_0^{\rm BPHZ}+\Dl_0$ accounts for 
 the divergences of the four-point kernel  $\delta^2\Gamma_\inter^R/\delta G_R^2|_{\bG_R}$ 
 ($\dl_0^{\rm BPHZ}$) and the divergences generated by superposing kernels through 
 the integral Bethe-Salpeter--type equation for $\bV_R$ ($\Dl_0$).
 \label{fig:coupling_sig3_bis}}
\end{center}
\end{figure}

The case of two-point kernels given in 
Eqs.\ (\ref{eq:se}) and (\ref{eq:2kernel2res}) is, however, more subtle
and requires special attention. Indeed, drawing boxes on 2PI diagrams 
misses an infinite number of coupling sub-divergences which arise for
$G_R = \bG_R$ at the stationary point of the effective action as illustrated in Fig. \ref{fig:missing_sing}. 
The case of the two-point kernel 
\beq
\label{eq:2kernel0}
 \bSigma_R = 2i\frac{\delta\Gamma_\inter^{(0)}}{\delta G_R}\Big|_{\bG_R}\,,
\eeq
has been analyzed in detail in Refs.~\cite{vanHees:2001ik,Blaizot:2003br}. There, it
has been shown that these coupling singularities precisely correspond to those 
of the Bethe-Salpeter--type equation (\ref{eq:BSsymb}) and that they can be 
absorbed in a shift of the tree-level contribution to the four-point kernel
\eqn{eq:BSkernel}. Denoting this local shift by $\Dl_0$, we write:\footnote{Here, 
we employ the notation introduced in Sec.~\ref{sec:explicit} and omit the explicit space-time 
dependence. A local contribution to a four-point function, such as the shift $\Dl_0$, 
is understood as a product of delta-functions, i.e.\ $\delta_{12}\delta_{13}\delta_{14}$, 
in real space, or, equivalently, as a constant in momentum space.} 
\beq
\label{eq:4kernel0}
 4\frac{\delta^2\Gamma_\inter^{(0)}}{\delta G_R^2}\Big|_{\bG_R}
 \equiv Z^2\bLambda=\bLambda_R-\Dl_0\,.
\eeq
where $Z$ is the wave-function renormalization introduced in \Eqn{wfr}.
The function $\bLambda_R$ is made finite by the previous 2PI BPHZ analysis.
The local shift $\Dl_0$ arises from a single graph in the 2PI expansion of the 
zero-field contribution $\Gamma_\inter^{(0)}[G_R]$ in \Eqn{eq:fieldexp}, namely 
the two-loop (eight) diagram. It contributes to the corresponding counterterm, 
which we denote by $\dl_0$, 
and adds to the contribution $\dl_0^{\rm BPHZ}$ arising from 
the BPHZ analysis described previously (see Fig.~\ref{fig:4point}). 
One has:\footnote{We stress that the present 
splitting of the counterterm $\dl_0$ associated to the eight diagram is not essential 
and is introduced for purely pedagogical purposes, in order to emphasize the origin of 
the various contributions as well as their role in the cancellation of divergences. 
In practice, the counterterm $\dl_0$ is computed from a single renormalization condition,
see below.} 
\beq
\label{dl0}
 \dl_0=\dl_0^{\rm BPHZ}+\Dl_0\,.
\eeq
Successive iterations of the kernel \eqn{eq:4kernel0} through the Bethe-Salpeter--type
equation \eqn{eq:BSsymb} define a finite four-point function 
$\bV_R=Z^2\bV$. This is actually sufficient to show 
that the coupling divergences hidden in the kernel $\bSigma_R$\footnote{Note that the 
two-point singularities generated by the perturbative iterations have already been taken 
into account in the previous BPHZ analysis. Indeed, the perturbative expansion of the first 
graph of Fig.~\ref{fig:2point} precisely generates an appropriate counterterm for 
each of these divergences.} have been removed by the 
subtraction employed in 
\Eqn{eq:4kernel0} \cite{vanHees:2001ik,Blaizot:2003br}. 
An illustration of this is given in Fig.~\ref{fig:coupling_sig3_bis}.

\begin{figure}
\begin{center}
\includegraphics[width=10cm]{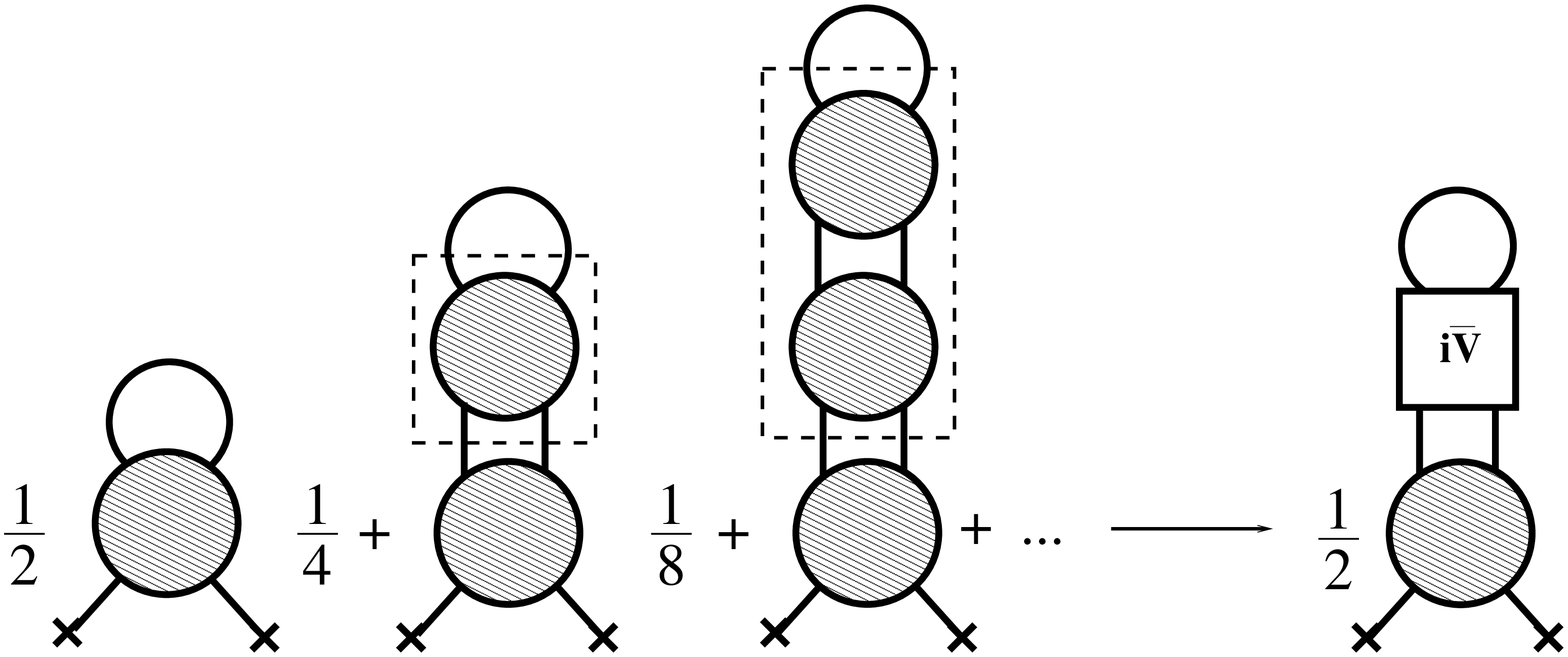}
 \caption{\small Illustration of the part of the four-point singularities in 
   $\Gamma_\inter^{(2)}(G_R)$, which is already taken into 
   account in the renormalization of $\bV_R$. 
   \label{fig:coupling_sig}}
\end{center}
\end{figure}

\begin{figure}
\begin{center}
\includegraphics[width=10cm]{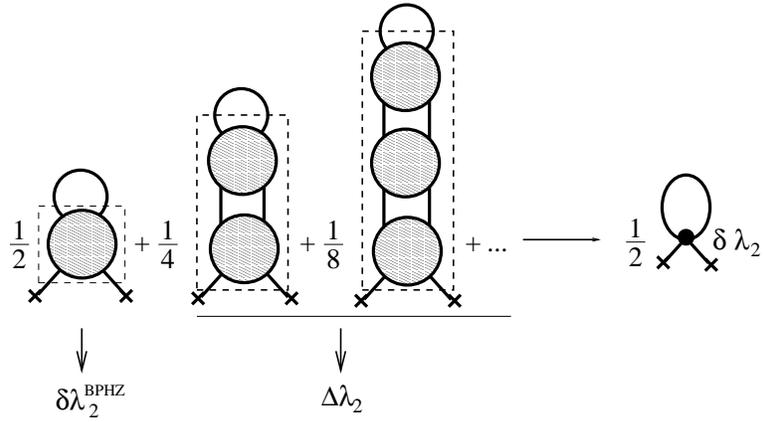}
 \caption{\small Illustration of the fact that 
   the remaining singularities can all be accounted for by the counterterm
   $\delta\lambda_2=\dl_2^{\rm BPHZ}+\Dl_2$. The latter arises from the tadpole diagram in 
   the 2PI effective action.
   \label{fig:coupling_sig2_bis}}
\end{center}
\end{figure}

A similar analysis can be applied to the two-point kernel (\ref{eq:2kernel2res})
which is obtained from the two-field part $\Gamma_\inter^{(2)}(G_R)$ in the decomposition 
\eqn{eq:fieldexp} as
\beq
\label{eq:2kernel2}
 \Sigma_R = i\Gamma_\inter^{(2)}(\bG_R)\,.
\eeq
As before, the perturbative expansion of this equation generates coupling 
singularities which were not taken into account by the previous BPHZ analysis
applied to resummed diagrams. Expanding the resummed propagator in terms of
perturbative contributions in \Eqn{eq:2kernel2}, 
one observes that these singularities actually correspond to those generated
by the integral equation \eqn{eq:intVsymb}. By drawing all possible four-point 
boxes,\footnote{It is important to realize that there can be no four-point box
entering inside the 2PI kernels because they arise from two-particle
irreducible diagrams (see also Appendix~\ref{2PItop}).} 
one sees that part of these singularities actually correspond to those discussed 
previously for the renormalization of the function $\bV_R$ and are, therefore, 
absorbed in the counterterm $\dl_0$. This is illustrated in Fig.~\ref{fig:coupling_sig}.
The remaining singularities all have the topology of the tadpole contribution to 
$\Gamma_\inter^{(2)}(G_R)$, as illustrated in Fig.~\ref{fig:coupling_sig2_bis}.
They may therefore be absorbed in the corresponding counterterm $\delta\lambda_2$.
We denote the corresponding contribution by $\Dl_2$. In complete analogy 
with the previous case, this corresponds to a redefinition of the tree-level 
contribution to the four-point kernel (\ref{eq:4kernel2res}):
\beq
\label{eq:4kernel2}
 2\frac{\delta\Gamma_\inter^{(2)}}{\delta G_R}\Big|_{\bG_R}
 \equiv Z^2\Lambda=\Lambda_R-\Dl_2\,.
\eeq
where the function $\Lambda_R$ is finite thanks to the 2PI BPHZ analysis.
Similarly as before, the shift $\Dl_2$ gives a contribution to the counterterm 
$\dl_2$ associated with the tadpole contribution to $\Gamma_\inter^{(2)}(G_R)$, which 
adds to the contribution $\dl_2^{\rm BPHZ}$ arising from the BPHZ analysis applied
to resummed diagrams:
\beq
\label{dl2}
 \dl_2=\dl_2^{\rm BPHZ}+\Dl_2\,.
\eeq
In the following, we show that the kernels \eqn{eq:4kernel0} and \eqn{eq:4kernel2} 
define a finite four-point function $V_R=Z^2V$ through the integral equation 
\eqn{eq:intVsymb} and that all coupling sub-divergences of the two-point kernel 
\eqn{eq:2kernel2} can indeed be absorbed in the renormalization of the functions
$\bV_R$ and $V_R$. The remaining overall divergences are absorbed in the mass and 
field-strength counterterms $\dmsq_2$ and $\delta Z_2$.

It is remarkable that the previous analysis is almost all what is needed 
for the renormalization for the 2PI-resummed effective action (\ref{eq:effac}). 
For instance, it is already clear from \Eqn{eq:phi4_2pf1} that the second 
derivative or two-point function is finite. However, one observes that the loop 
integrals in Eq.~(\ref{eq:4ptsymb}) together with the successive iterations 
of Eq.~(\ref{eq:intVsymb}) bring {\it a priori} infinitely many new divergences. 
The same holds for all higher $n$-point functions. We will show below that all 
potential sub-divergences can in fact be absorbed in the renormalization of 2PI
kernels with two, six, or more than six legs, and of the four-point functions $\bV_R$ 
and $V_R$. The only remaining singularity is a global divergence of the four-point vertex 
function (\ref{eq:4ptsymb}), which can be absorbed in a contribution 
$\Dl_4$ to the counterterm $\dl_4$ associated with the tree-level contribution 
to the four-point kernel\footnote{We mention that for approximations where 
\eqn{simpler} is valid (see Appendix~\ref{sec:appcorrel}), the shift $\Dl_4$ is 
trivially obtained as: $\Dl_4=3\Dl_2$ (see also \cite{Berges:2004hn}).}
\beq
\label{eq:4kernel4}
 \frac{\delta\Gamma_\inter^R}{\delta\phi_R^4}\Big|_{\bG_R}=\Gamma_\inter^{(4)}(\bG_R)\,.
\eeq

\begin{figure}
\begin{center}
\includegraphics[width=6cm]{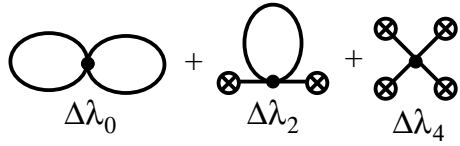}
\caption{\small \label{fig:tadpoles} Diagrammatic representation
of the counterterms needed to renormalize the divergences 
missed by the BPHZ analysis applied to diagrams with resummed propagators.}
\end{center}
\end{figure}

To summarize, we have seen that a first condition for renormalizability is that
for each 2PI graph included in the approximation all topologies generated by 
the BPHZ procedure described above (i.e., drawing all possible four-point 
boxes and replacing them by a point) must be included as well. 
In addition, one must include an infinite shift of the tree-level contribution 
to the four-point kernels \eqn{eq:4kernel0}, \eqn{eq:4kernel2} and \eqn{eq:4kernel4}. 
At the level of the 2PI functional, this corresponds to an infinite shift of the 
contributions from each local, mass-dimension four operator. To make this explicit, 
we write:
\bea
 \Gamma_\inter^R[\phi_R,G_R]&=&\gamma_\inter[\phi_R,G_R]
 -\int_x\,\left\{\frac{\Dl_0}{8}\,G_R^2(x,x)
 +\frac{\Dl_2}{4}\,G_R(x,x)\phi_R^2(x)\right.\nonumber\\
\label{eq:shift}
 &&\hspace*{3.2cm}\left.+\frac{\Dl_4}{4!}\,\phi_R^4(x)\right\}\,.
\eea
The diagrams corresponding to the shifted part of the 2PI 
functional are shown in Fig.~\ref{fig:tadpoles}. 
Equation (\ref{eq:shift}) defines the functional $\gamma_\inter[\phi_R,G_R]$.
Employing an equivalent expansion in terms of powers of the field $\phi_R$
as in (\ref{eq:fieldexp}), the four-point kernels $\delta^2\gamma_\inter^{(0)}/
\delta G_R^2|_{\bG_R}$, $\delta\gamma_\inter^{(2)}/\delta G_R|_{\bG_R}$ 
and $\gamma_\inter^{(4)}|_{\bG_R}$, as well as all higher kernels 
$\delta^m\gamma_\inter^{(2n)}/\delta G_R^m|_{\bG_R}$ with $n+m>2$ are made 
finite by the BPHZ subtraction procedure applied to graphs with resummed propagator.
For later discussions it is useful to extract the mass and field-strength
counterterms explicitely (cf.~\Eqn{eq:gammaint_ren}). Accordingly, we write:
\bea
 \gamma_\inter[\phi_R,G_R]&=&\tilde\gamma_\inter[\phi_R,G_R]
 -\frac{1}{2}\int_x\,\Big\{\phi_R(x)\left[\dZ_0\square_x+\dmsq_0\right]\phi_R(x)\nn
\label{eq:mfsct}
 &&+\left[\dZ_2\square_x+\dmsq_2\right]G_R(x,y)|_{y=x}\Big\}\,.
\eea
 
Section \ref{sec:finite} below is devoted to a direct proof that
the three counterterms $\Dl_0$, $\Dl_2$ and $\Dl_4$ are actually sufficient 
to absorb the remaining divergences missed by the 2PI BPHZ procedure and, thereby, 
to obtain renormalized $n$-point functions from the 2PI-resummed effective action. 
We show how to compute these counterterms explicitly.
We emphasize that, as far as the 2PI BPHZ analysis is concerned, the various 
$\Gamma_\inter^{(2n)}$ can be treated independently from each other. This is only 
slightly modified by the complete analysis. In particular, the 
$\gamma_\inter^{(2n)}$ introduced above remain independent. 

%%%%%%%%%%%%%%%%%%%%%%%%%%%%%%%%%%%%%%%%%%%%%%%%%%%%%%%%%%%%%%%%%%%%%%%%%%%%%%%%%%%%%%%%%%%%%%%%%%%%%%%%%%%%%%%%%%%%%%%%%%%%%%%%%%%%%%%%%%%%%%%%%%%%%%   Renormalization conditions  %%%%%%%%%%%%%%%%%%%%%%%%%%%%%%%%%%%%%%%%%%%%%%%%%%%%%%%%%%%%%%%%%%%%%%%%%%%%%%%%%%%%%%%%%%%%%%%%%%%%%%%%%%%%%%%%%%%%%%%%%%%%%%%%%%%%%%%%%%%%%%%%%%%%%%%%%%%%%%%%%%%%%%%%%%%%%%%%
\subsection{Counterterms and renormalization conditions}
\label{sec:ren_cond}

We have seen that, in order to renormalize the functions $\bV_R$,
$V_R$ and $\Gamma_R^{(4)}$, we need three {\it a priori} different coupling 
counterterms: $\Dl_0$, $\Dl_2$ and $\Dl_4$ respectively. It is important 
to realize that, although these functions may be different at a given order 
of approximation, they are in fact not independent in the exact theory, where 
one has $\bV_R=V_R=\Gamma_R^{(4)}$ (cf.~Appendix \ref{sec:appcorrel}). 
To guarantee
that a given approximation scheme converges to the correct theory, it is
therefore important to maintain this relation when stating renormalization 
conditions at each approximation order.\footnote{For the universality class of 
the $\phi^4$ theory there are
only two independent input parameters, here $m_R$ and $\lambda_R$, as employed  
in Eqs.~(\ref{eq:renormcondl}) and (\ref{eq:renormcondm}).} 
For instance, the renormalized coupling $\lambda_R$ may be defined at a given 
renormalization point $\{p_i=\tp_i\,,\,i=1,\ldots,4\}$ in Fourier space 
by:
\beq
\label{eq:renormcondl}
 \Gamma_R^{(4)} (\tp_i)=\bV_R (\tp_i)=V_R (\tp_i)
 =-\lambda_R\,.
\eeq
In the following, we will choose the renormalization point in Euclidean 
momentum space such that $\tp_1=\tp_2=\ldots=\tp$, with $\tp^2=\mu^2$.
We stress that \Eqn{eq:renormcondl} expresses a single renormalization condition for 
the three functions $\bV_R$, $V_R$ and $\Gamma_R^{(4)}$ and uniquely determines the 
three counterterms $\Dl_0$, $\Dl_2$ and $\Dl_4$. Similarly, the two
{\it a priori} different sets of field-strength and mass counterterms
$(\dZ_0,\,\dmsq_0)$ and $(\dZ_2,\,\dmsq_2)$ are uniquely determined by
imposing the same renormalization 
conditions for the functions $i\bG_R^{-1}$ and $\Gamma_R^{(2)}$, respectively. The
latter are equal in the exact theory. For instance, one has: 
\bea
\label{eq:renormcondm}
 &&\Gamma_R^{(2)}(p=\tp)=i\bG_R^{-1}(p=\tp)\,\,=\, - m_R^2\\
\label{eq:renormcondZ}
 &&\frac{d\Gamma_R^{(2)}}{dp^2}\Big|_{p=\tp}
 =i\frac{d\bG_R^{-1}}{dp^2}\Big|_{p=\tp}\,=\, - 1\,.
\eea

Finally, we would like to comment on the BPHZ subtraction procedure described
previously. It is clear that the latter is equivalent to adding a different
counterterm at each vertex appearing in the diagrammatic expansion of the 2PI
functional. We stress that these, together with the $\Dl$'s discussed above
actually represent different approximations to one and the same counterterm $\dl$
introduced in Sec.~\ref{sec:ct}. Similarly the field-strength and 
mass counterterms $(\dZ_0,\dmsq_0)$ and $(\dZ_2,\dmsq_2)$ are two different 
approximations to $(\dZ,\dmsq)$ introduced before.\footnote{We point out that 
for approximations where the relation \eqn{eq:relation} is satisfied, one has 
$\dZ_0=\dZ_2$ and $\dmsq_0=\dmsq_2$ since, in that case, $\Gamma_R^{(2)}=i\bG_R^{-1}$
at $\phi_R=0$.} This 
is similar to standard perturbation theory, where one would expand the 
counterterm appearing in each graph to different orders, depending on the order 
of the graph itself. For practical purposes, this means different counterterms 
are associated to different graphs. In terms of the BPHZ 
subtraction scheme, it is crucial that the subtractions are always performed
at the same subtraction point. In the present context, the latter should 
coincide with the renormalization point employed in 
Eqs.~(\ref{eq:renormcondl})--(\ref{eq:renormcondZ}).

%%%%%%%%%%%%%%%%%%%%%%%%%%%%%%%%%%%%%%%%%%%%%%%%%%%%%%%%%%%%%%%%%%%%%%%%%%%%%%%%%%%%%%%%%%%%%%%%%%%%%%%%%%%%%%%%%%%%%%%%%%%%%%%%%%%%%%%%%%%%%%%%%%%%%%   Renormalization   %%%%%%%%%%%%%%%%%%%%%%%%%%%%%%%%%%%%%%%%%%%%%%%%%%%%%%%%%%%%%%%%%%%%%%%%%%%%%%%%%%%%%%%%%%%%%%%%%%%%%%%%%%%%%%%%%%%%%%%%%%%%%%%%%%%%%%%%%%%%%%%%%%%%%%%%%%%%%%%%%%%%%%%%%%%%%%%%%%%%%%%%%%
\section{Proof}\label{sec:finite}

Without loss of generality, the detailed analysis of divergences 
is most conveniently done in Euclidean Fourier space. Accordingly, 
we introduce the Euclidean four-momentum $q^\mu\equiv(iq_0,\vec q)$. 
We denote the Euclidean integration measure by: 
$\int_q\equiv \int\frac{d^4q}{(2\pi)^4}$. 
%%%%%%%%%%%%%%%%%%%%%%%%%%%%%%%%%%%%%%%%%%%%%%%%%%%%%%%%%%%%%%%%%%%%%%%%%%%%%%%%%%%%%%%%%%%%%%%%%%%%%%%%%%%%%%%%%%%%%%%%%%%%%%%%%%%%%%%%%%%%%%%%%%%%%%   2PI kernels  %%%%%%%%%%%%%%%%%%%%%%%%%%%%%%%%%%%%%%%%%%%%%%%%%%%%%%%%%%%%%%%%%%%%%%%%%%%%%%%%%%%%%%%%%%%%%%%%%%%%%%%%%%%%%%%%%%%%%%%%%%%%%%%%%%%%%%%%%%%%%%%%%%%%%%%%%%%%%%%%%%%%%%%%%%%%%%%%%%%%%%%%%%%%%%%
\subsection{Renormalization of 2PI kernels}

\begin{figure}[tb]
\begin{center}
\includegraphics[width=8cm]{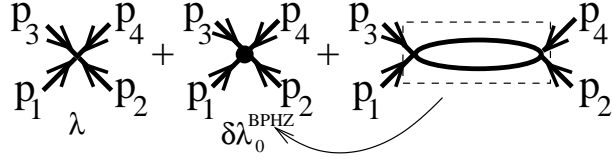}
\caption{\small The one-loop approximation to the function $\bLambda_R$, obtained
from the 2PI effective action to three loop after cutting two lines, that is 
after taking two derivatives with respect to $G$. The first two terms arise
from the two-loop (eight) diagram whereas the third one comes from the basket-ball
diagram. The logarithmic divergence of the third term can be absorbed in the
counterterm $\dl_0^{\rm BPHZ}$.\label{fig:BPHZex}}
\end{center}
\end{figure}

As described above, the renormalization of kernels with more than four 
external legs is based on the BPHZ procedure applied to diagrams with 
resummed propagator. To illustrate the procedure, we consider the example 
of the function $\bLambda_R$ as obtained from the three-loop
approximation of the 2PI effective action. After opening two propagator 
lines to obtain the function $\bLambda_R$, this corresponds to the one-loop 
expression represented in Fig.~\ref{fig:BPHZex}. In
momentum space this reads:\footnote{Here and in the following, 
we factor out the  
momentum conservation term $(2\pi)^4\delta^{(4)}(p_1+\cdots+p_4)$.}
\beq
\label{eq:BPHZex}
 \bLambda_R(p_1,\ldots,p_4)=-\Big(\lambda_R+\delta\lambda_0^{\rm BPHZ}\Big)
 +\lambda_R^2\int_q\,\bG_R(q)\bG_R(p_1+p_3+q)\,.
\eeq
The 2PI BPHZ procedure amounts here to choosing $\delta\lambda_0^{\rm BPHZ}$ in 
order to renormalize the one-loop integral in \Eqn{eq:BPHZex}. 
With our convention for the renormalization point (cf.~Sec.~\ref{sec:ren_cond} 
above) we choose:
\begin{equation}
 \delta\lambda_0^{\rm BPHZ}=
 \lambda_R^2\int_q\,\bG_R(q)\bG_R(p+q)|_{p^2=4\mu^2}\,,
\end{equation}
so that the function
\begin{equation}
\label{eq:BPHZex2}
 \bLambda_R(p_1,\ldots,p_4)=-\lambda_R
 +\lambda_R^2\int_q\,\bG_R(q)\,\Big(\bG_R(p_1+p_3+q)-\bG_R(p+q)|_{p^2=4\mu^2}\Big)
\end{equation}
is finite. A similar analysis holds for the function $\Lambda_R$ as well as
for all 2PI kernels with more than four external legs.\footnote{In general, such
an analysis does not only involve overall subtractions but also the elimination 
of possible sub-divergences, e.g. following an iterative procedure.}

As explained in the previous subsection, in order to renormalize the two-point kernels, 
one first needs to renormalize the vertices $\bV_R$ and $V_R$, obtained from the 
integral equations \eqn{eq:BSsymb} and \eqn{eq:intVsymb} respectively. This requires 
additional contributions $\Dl_0$ and $\Dl_2$ to the counterterms $\dl_0$ and $\dl_2$ 
respectively, see Eqs.~\eqn{dl0} and \eqn{dl2}. We note that due to these additional 
infinite contributions, the kernels \eqn{eq:4kernel0} and \eqn{eq:4kernel2} are not 
finite, in contrast to the functions $\bLambda_R$ and $\Lambda_R$. However, this poses
no problem for the renormalization program, since these kernels can always be traded for 
the finite vertices $\bV_R$ and $V_R$. We make this explicit in the next subsections.

%%%%%%%%%%%%%%%%%%%%%%%%%%%%%%%%%%%%%%%%%%%%%%%%%%%%%%%%%%%%%%%%%%%%%%%%%%%%%%%%%%%%%%%%%%%%%%%%%%%%%%%%%%%%%%%%%%%%%%%%%%%%%%%%%%%%%%%%%%%%%%%%%%%%%%   BS equation  %%%%%%%%%%%%%%%%%%%%%%%%%%%%%%%%%%%%%%%%%%%%%%%%%%%%%%%%%%%%%%%%%%%%%%%%%%%%%%%%%%%%%%%%%%%%%%%%%%%%%%%%%%%%%%%%%%%%%%%%%%%%%%%%%%%%%%%%%%%%%%%%%%%%%%%%%%%%%%%%%%%%%%%%%%%%%%%%%%%%%%%%%%%%%%%
\subsubsection{The first vertex-equation}\label{sec:renBS}

\begin{figure}[tb]
\begin{center}
\includegraphics[width=12cm]{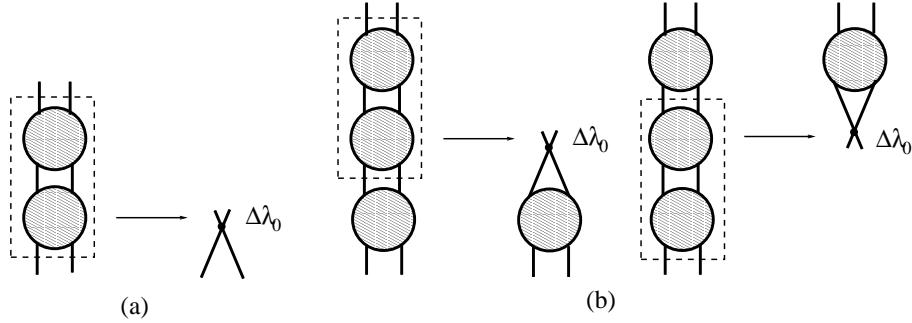}
\caption{\small The figures (a) and (b) show the procedure by which the coupling 
divergences in the vertex equation are absorbed in the shift $\Delta\lambda_0$. 
The overall divergence absorbed at a given iteration is used to remove all 
the sub-divergences at the next iteration.\label{fig:BS_diag}}
\end{center}
\end{figure}

The integral equation (\ref{eq:BSsymb}) resums the infinite series
of ladder diagrams with rungs given by the kernel (\ref{eq:BSkernel}).
In terms of renormalized quantities, each iteration brings a new finite rung 
$\bLambda_R=4\delta^2\gamma_\inter^{(0)}/\delta G_R^2|_{\bG_R}$ together with 
a counterterm $-\Dl_0$. The new logarithmic 
divergences generated at each iteration can be absorbed in the counterterms 
appearing in the previous iterations. This is illustrated 
in Fig. \ref{fig:BS_diag}.
\begin{figure}[tb]
\begin{center}
\includegraphics[width=1.2cm]{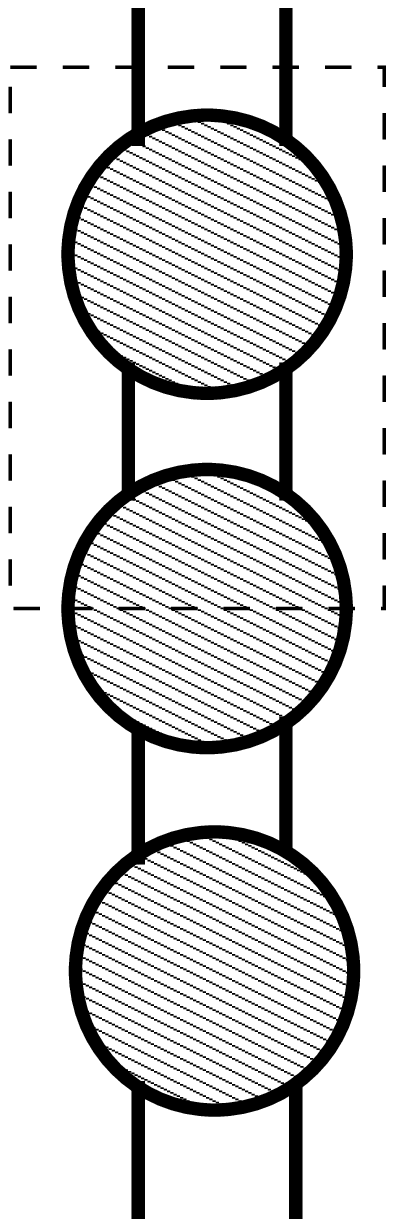}
\caption{\small The 2PI character of the kernels plays a crucial role in the 
renormalization of the vertex $\bV_R$: There can be no four-point boxes such 
as the one depicted in the figure. For the same reason, there is no counterterm 
to absorb such structures.\label{fig:BS_diag3}}
\end{center}
\end{figure}
The 2PI character of the kernel (cf. the discussion in Appendix~\ref{2PItop}) 
prevents other potential divergences to appear 
(such as the one depicted in Fig.~\ref{fig:BS_diag3}). 
In fact, for precisely the same reason, there is no topology which could be used 
to cancel such a divergence. This diagrammatic property has its counterpart in the 
algebraic proof that we now recall \cite{vanHees:2001ik,Blaizot:2003br}. 
We introduce the notation:
\beq
\label{eq:skernel0}
 \bLambda_R(p,k) \equiv \bLambda_R(p,-p,-k,k)\,,
\eeq
where $\bLambda_R(p_1,\ldots,p_4)$ is the Fourier transform of the 
renormalized function $\bLambda_R$, as defined from (\ref{eq:BSkernel}). 
One has $\bLambda_R(p,k)\sim\ln k$ at large $k$ and fixed $p$. Furthermore, 
it follows from Weinberg's theorem and from the two-particle irreducible character 
of the function $\bLambda_R$, that \cite{vanHees:2001ik,Blaizot:2003br}:
\beq
 \bLambda_R(p,k)-\bLambda_R(q,k)\sim \frac{1}{k}\,,
\label{eq:asymptotic}
\eeq 
at large $k$ and fixed $p$ and $q$. We first consider the vertex equation in 
the $s$-chanel, namely for the function $\bV_R(p,k)\equiv\bV_R(p,-p,-k,k)$:
\bea
\label{eq:BSmom}
 \bV_R(p,k)&=&\bLambda_R(p,k)-\Dl_0
 +\frac{1}{2}\,\int_q\,\bV_R(p,q)\,\bG_R^2(q)\,\Big(\bLambda_R(q,k)-\Dl_0\Big)\nonumber\\
 &=&\bLambda_R(p,k)-\Dl_0+
 \frac{1}{2}\,\int_q\,\Big(\bLambda_R(p,q)-\Dl_0\Big)\,\bG_R^2(q)\,\bV_R(q,k)\,,\nn
\eea
where we used the fact that $\bV_R(p,k)=\bV_R(k,p)$, and similarly for $\bLambda_R(p,k)$. 
As for $\bLambda_R(p,k)$, one has $\bV_R(p,k)\sim\ln k$ at large $k$ and fixed 
$p$.\footnote{However, unlike $\bLambda_R$, the function $\bV_R$ is not 2PI and 
$\bV_R(p,k)-\bV_R(q,k)$ is only $\sim\ln k$ at large $k$ and fixed $p$ and $q$.}
This equation contains UV divergent loops but also counterterms. The input of 
renormalization is simply to state that the value of $\bV_R$ at a given 
renormalization point is finite. For the vertex equation to be self-consistently 
renormalizable, this has to be enough to obtain a finite equation for $\bV_R$. 
Imposing, as in \Eqn{eq:renormcondl}, that $\bV_R(\tp,\tp)=-\lambda_R$ is finite, 
we consider the difference:
\bea
\label{eq:BSren}
 \bV_R(p,k)-\bV_R(\tp,\tp) &=& \bLambda_R(p,k)-\bLambda_R(\tp,\tp)\nonumber\\
 &&+\frac{1}{2}\,\int_q\,\bV_R(p,q)\,\bG_R^2(q)\,
 \Big(\bLambda_R(q,k)-\bLambda_R(q,\tp)\Big)\nn
 &&+\frac{1}{2}\,\int_q\,\Big(\bLambda_R(p,q)-\bLambda_R(\tp,q)\Big)\,
 \bG_R^2(q)\,\bV_R(q,\tp)\,,\nn
\eea
where we used the symmetry property of the momentum integral in Eq.\ (\ref{eq:BSmom}) so 
that only differences of 2PI kernels appear. In this way, all counterterms -- which are
momentum independent -- disappear from this equation. Using the 
asymptotic behavior (\ref{eq:asymptotic}), one easily checks that the integrals
in Eq.\ (\ref{eq:BSren}) are finite.

We also need to consider the full momentum dependence of $\bV_R$ which 
satisfies the following equation, where $\bV_R(p,q,k)\equiv\bV_R(p,-q,-k,k+q-p)$ and 
similarly for $\bLambda_R$:
\bea
\label{eq:BSfull}
 \bV_R(p,q,k)&=&\bLambda_R(p,q,k)-\Dl_0\nn
 &&+\frac{1}{2}\,\int_r\,\Big(\bLambda_R(p,q,r)-\Dl_0\Big)\,
 \bG_R(r)\,\bG_R(l)\,\bV_R(r,l,k)\,,\nonumber\\
\eea
where $l=r+q-p$. The divergent part of the momentum integral in the equation above 
comes from the large $r$ contribution:
\bea
 \bV_R(p,q,k)|_{\rm div}&=&\bLambda_R(p,q,k)-\Dl_0\nn
 &&+\frac{1}{2}\,\int_r\,\Big(\bLambda_R(p,q,r)-\Dl_0\Big)\,
 \bG_R^2(r)\,\bV_R(r,r,k)\,.\nonumber\\
\eea
The latter precisely coincides with the divergences of $\bV_R(p,k)=\bV_R(p,p,k)$
discussed above. Indeed, one has:
\bea
 &&\bV_R(p,q,k)|_{\rm div}-\bV_R(p,p,k)=
 \bLambda_R(p,q,k)-\bLambda_R(p,p,k)\nn
\label{eq:fullmom}
 &&\qquad\qquad\qquad\qquad\quad
 +\frac{1}{2}\,\int_r\,\Big(\bLambda_R(p,q,r)-\bLambda_R(p,p,r)\Big)\,
 \bG_R^2(r)\,\bV_R(r,r,k)\,,\nn
\eea
which is finite due to the fact that, as a consequence of two-particle-irreducibility, 
$\bLambda_R(p,q,r)-\bLambda_R(p,p,r)\sim 1/r$ at large $r$ and fixed $p$ and $q$.

%%%%%%%%%%%%%%%%%%%%%%%%%%%%%%%%%%%%%%%%%%%%%%%%%%%%%%%%%%%%%%%%%%%%%%%%%%%%%%%%%%%%%%%%%%%%%%%%%%%%%%%%%%%%%%%%%%%%%%%%%%%%%%%%%%%%%%%%%%%%%%%%%%%%%%   First two-point kernel  %%%%%%%%%%%%%%%%%%%%%%%%%%%%%%%%%%%%%%%%%%%%%%%%%%%%%%%%%%%%%%%%%%%%%%%%%%%%%%%%%%%%%%%%%%%%%%%%%%%%%%%%%%%%%%%%%%%%%%%%%%%%%%%%%%%%%%%%%%%%%%%%%%%%%%%%%%%%%%%%%%%%%%%%%%%%%%%%%%%%%%
\subsubsection{Renormalization of $\delta\Gamma_\inter^R/\delta G_R|_{\bG_R}$}
\label{sec:ren2pt1}

The renormalization of the kernel $\delta\Gamma_\inter^R/\delta G_R|_{\bG_R}$, 
or equivalently, of the auxiliary propagator $\bG_R$ has been
extensively discussed in Refs.\ \cite{vanHees:2001ik,Blaizot:2003br}. 
Here, we briefly recall the main arguments of the analysis. This only 
involves the zero-field part 
$\Gamma_\inter^{(0)}[G_R]$ of the 2PI effective action. The 2PI BPHZ procedure fixes 
the counterterms $\dZ_0$, $\dmsq_0$, and a set of $\dl_0^{\rm BPHZ}$'s for each 
topology appearing in the truncation (cf.~Sec.~\ref{sec:ren_cond}), adjusted so that 
the four-point kernel $\delta^2\gamma_\inter^{(0)}/\delta G_R^2|_{\bG_R}$ is finite. 
The remaining divergences 
in the two-point kernel $\delta\Gamma_\inter^R/\delta G_R|_{\bG_R}$ 
are made explicit by performing an asymptotic expansion of $\bG_R$ in its defining 
equation\footnote{Extracting the contribution from
the counterterms $\Delta \lambda_0$, $\delta Z_0$ and $\delta m_0^2$, one has 
explicitly (cf. \Eqn{eq:mfsct})
$$
 \bSigma_R(p)=2i\frac{\delta\tilde\gamma_\inter^{(0)}}{\delta G_R}\Big|_{\bG_R}(p)
 -i(\dZ_0\,p^2+\dmsq_0)-\frac{\Dl_0}{2}\int_q\,\bG_R(q)\,.
$$}
\beq
\label{eq:gapeq}
 \bG_R^{-1}=G_{0R}^{-1}-\bSigma_R(\bG_R)\, ,
\eeq
with
\beq\label{eq:kernelgap}
\bSigma_R(\bG_R)=2i\frac{\delta\Gamma_\inter^R}{\delta G_R}\Big|_{\bG_R} \, .
\eeq

Following Ref.~\cite{Blaizot:2003br}, we separate 
the leading asymptotic behavior\footnote{Here and 
in the following, the notation 
`$\sim$' includes possible powers of $\ln p$.} $\bSigma_2(p)\sim p^2$ of the 
self-energy $\bSigma_R(p)$ at large momentum $p$:
\beq\label{eq:bob3}
 \bSigma_R(p)=\bSigma_2 (p)+\bSigma_0 (p)\,,
\eeq
where the function $\bSigma_0(p)$ grows at most as powers of $\ln p$ at large~$p$. 
Similarly, we extract the leading asymptotic behavior $\bG_2(p)\sim 1/p^2$ of the 
propagator\footnote{We do not pay attention to possible infrared singularities 
arising from the expansion around $\bG_2$, since we are interested in UV singularities. 
A more careful analysis can be found in Ref.~\cite{Blaizot:2003br}, 
where it is shown 
how to deal with the infrared sector in a safe way.} $\bG_R(p)$:
\beq\label{eq:propexp}
 \bG_R(p)=\bG_2 (p)+ \delta\bG(p)\,,
\eeq
where the function $\delta\bG(p)\sim 1/p^4$ at most. One has, explicitly:
\beq
 \bG_R^{-1}(p)=i(p^2+m_R^2)-\bSigma_2(p)-\bSigma_0(p)
\eeq
and
\beq
 \bG_2^{-1}(p)=ip^2-\bSigma_2(p)\,,
\eeq
from which follows
\beq
\label{eq:deltaG}
 \delta\bG (p)=\bG_2(p)\left[-im_R^2+\bSigma_0(p)\right]\bG_2(p)+\bG_r(p)\,,
\eeq
where $\bG_r(p)\sim 1/p^6$ at most. 
We now expand the right hand side of Eq. (\ref{eq:kernelgap}) according to 
(\ref{eq:propexp}). Using\footnote{This equation arises from 
the fact that the leading piece $\bSigma_2$ is obtained from the gap equation by 
removing all the masses. The masses are already absent from the propagators $\bG_2$. 
The role of the term $i\delta m_0^2$ is to remove the mass counterterm since it does 
not enter in the definition of $\bSigma_2$. This is indeed true in 
dimensional regularization. With an explicit cut-off $\Lambda_c$ there are mass 
divergences proportional to $\Lambda_c^2$. These are removed by a convenient shift 
of the mass counterterm $i\delta m_{\Lambda_c}^2$ which does enter the definition of 
$\bSigma_2$. In that case $i\delta m_0^2$ in Eq. (\ref{eq:bob3}) only represents 
divergences logarithmic in $\Lambda_c$. Finally, in order to renormalize $\bSigma_2$, 
one only needs the coupling counterterms $\delta\lambda_0^{\rm BPHZ}$ given by 
the 2PI
BPHZ subtraction procedure, together with the field-strength counterterm $\delta Z_0$ 
which is implemented as an overall subtraction (see \cite{vanHees:2001ik,Blaizot:2003br}). 
There is no need at this level to use $\delta m_0^2$ or $\Delta \lambda_0$ as there is 
no explicit scale but the renormalization scale $\mu$ in the momentum integrals 
(cf.~also the discussion in Ref.~\cite{Blaizot:2003br}).}
\beq
\label{eq:lasigma}
\bSigma_2 \equiv\bSigma_R(\bG_2)+i\dmsq_0\,,
\eeq
the leading term of the expansion cancels with $\bSigma_2$ and one obtains, 
for the sub-leading part
\beq
\label{eq:eom}
 \bSigma_0(p)=-i\dmsq_0+\frac{1}{2}\,\int_q\Big(\bLambda_2(p,q)-\Dl_0\Big)\,
 \delta\bG(q)+\bSigma_r(p)\,,
\eeq
where $\bLambda_2\equiv\bLambda_R(\bG_2)$, see \Eqn{eq:skernel0}, and $\bSigma_r(p)$ 
is a finite function decreasing at least as $1/p$ at large $p$.
It is possible to absorb all the non-localities in a redefinition of $\bSigma_r(k)$
($p=\tp$ denotes the renormalization point):
\beq
\label{eq:eom2}
 \bSigma_0(p)=-i\dmsq_0+\frac{1}{2}\,\int_q\Big(\bLambda_2(\tp,q)-\Dl_0\Big)\,
 \delta\bG(q)+\bSigma'_r(p)\,,
\eeq
with
\beq
\label{eq:redef}
 \bSigma'_r(p)=\bSigma_r(p)
 +\frac{1}{2}\,\int_q\Big(\bLambda_2(p,q)-\bLambda_2(\tp,q)\Big)\,\delta\bG(q)\,.
\eeq 
Indeed, using Weinberg's theorem and the asymptotic behavior 
(\ref{eq:asymptotic}), one can show that the integral on the 
RHS of (\ref{eq:redef}) is finite and decreases at least as $1/p$ at large $p$.
We now use the vertex equation \eqn{eq:BSmom} with propagator $\bG_2$
-- which solution we denote by $\bV_2(p,k)$ -- to trade $\bLambda_2(\tp,q)$ 
for $\bV_2(\tp,q)$ in (\ref{eq:eom2}):
\bea
 \bSigma_0(p)&=&-i\dmsq_0+\bSigma'_r(p)
 +\frac{1}{2}\,\int_q\,\bV_2(\tp,q)\,\delta\bG(q)\nonumber\\
\label{eq:eom3}
 &&-\frac{1}{4}\,\int_q\int_k\,\bV_2(\tp,q)\,\bG_2^2(q)\,
 \Big(\bLambda_2(q,k)-\Delta\lambda_0\Big)\,\delta\bG(k)\,.
\eea
Using (\ref{eq:eom}), one can rewrite the integral over $k$ in the last term:
\bea
 \bSigma_0(p)&=&-i\dmsq_0+\bSigma'_r(p)
 +\frac{1}{2}\,\int_q\,\bV_2(\tp,q)\,\delta\bG(q)\nonumber\\
 &&-\frac{1}{2}\,\int_q\,\bV_2(\tp,q)\,\bG_2^2(q)\,
 \Big\{\bSigma_0(q)-\bSigma_r(q)+i\dmsq_0\Big\}\,.
\eea
Using (\ref{eq:deltaG}), one sees that the potentially divergent terms which 
depend on $\bSigma_0$ cancel out.\footnote{These correspond to coupling 
sub-divergences and would lead, at finite temperature, to temperature dependent 
singularities.} One finally obtains
\bea
\label{eq:bsigma0}
 \bSigma_0(p)&=&\frac{1}{2}\,\int_q\,\bV_2(\tp,q)\,\Big\{\bG_r(q)
 +\bG_2^2(q)\,\bSigma_r(q)\Big\}+\bSigma'_r(p)\nonumber\\
 &&-i\dmsq_0-\frac{i}{2}(m_R^2+\dmsq_0)\,\int_q\,\bV_2(\tp,q)\,\bG_2^2(q)\,.
\eea
The first line is finite by power counting. The logarithmic divergence in 
the second line of Eq.\ (\ref{eq:bsigma0}) is independent of $\bSigma_0$ and 
can be absorbed in $\dmsq_0$.

%%%%%%%%%%%%%%%%%%%%%%%%%%%%%%%%%%%%%%%%%%%%%%%%%%%%%%%%%%%%%%%%%%%%%%%%%%%%%%%%%%%%%%%%%%%%%%%%%%%%%%%%%%%%%%%%%%%%%%%%%%%%%%%%%%%%%%%%%%%%%%%%%%%%%%   Modified BS equation  %%%%%%%%%%%%%%%%%%%%%%%%%%%%%%%%%%%%%%%%%%%%%%%%%%%%%%%%%%%%%%%%%%%%%%%%%%%%%%%%%%%%%%%%%%%%%%%%%%%%%%%%%%%%%%%%%%%%%%%%%%%%%%%%%%%%%%%%%%%%%%%%%%%%%%%%%%%%%%%%%%%%%%%%%%%%%%%%%%%%%%
\subsubsection{The second vertex-equation}

In terms of renormalized quantities, the second vertex-equation in momentum 
space reads (cf.~\Eqn{eq:intVsymb}):
\beq
\label{eq:mbs}
 V_R(p,k)=\Lambda_R(p,k)-\Dl_2+\frac{1}{2}\,\int_q V_R(p,q)\,
 \bG_R^2(q)\,\Big(\bLambda_R(q,k)-\Dl_0\Big)\,,
\eeq
using a similar notation as above. Its solution in terms of $\bV_R$ reads
\beq
\label{eq:mbss}
 V_R(p,k)=\Lambda_R(p,k)-\Dl_2+\frac{1}{2}\,\int_q\Big(\Lambda_R(p,q)-\Dl_2\Big)\,
 \bG_R^2(q)\,\bV_R(q,k)\,.
\eeq
It follows from these two equations that
\beq
\label{eq:diff}
 \int_q V_R(p,q)\,\bG_R^2(q)\,\Big(\bLambda_R(q,k)-\Dl_0\Big)=
 \int_q\Big(\Lambda_R(p,q)-\Dl_2\Big)\,\bG_R^2(q)\,\bV_R(q,k)\,.
\eeq
One has $\Lambda_R(p,k)\sim\ln k$ at large $k$ and fixed $p$, and similarly
for $V_R(p,k)$.
To show that Eq.\ (\ref{eq:mbs}) is made finite by a single renormalization
condition, we subtract the value $V_R(\tp,\tp)=-\lambda_R$ and write:
\bea
\label{eq:mbs2}
 V_R(p,k)-V_R(\tp,\tp)&=&\Lambda_R(p,k)-\Lambda_R(\tp,\tp)\nonumber\\
 &&+\frac{1}{2}\,\int_q\,V_R(p,q)\,\bG_R^2(q)\,\Big(\bLambda_R(q,k)-\Dl_0\Big)\nonumber\\
 &&-\frac{1}{2}\,\int_q\,\Big(\Lambda_R(\tp,q)-\Dl_2\Big)\,\bG_R^2(q)\,\bV_R(q,\tp)\,.\nn
\eea
Using Eq.\ (\ref{eq:diff}), one can rewrite this equation in such a way that
only differences of 2PI kernels appear:
\bea
 V_R(p,k)-V_R(\tp,\tp)&=&\Lambda_R(p,k)-\Lambda_R(\tp,\tp)\nn
 &&+\frac{1}{2}\,\int_q\,V_R(p,q)\,\bG_R^2(q)\,\Big(\bLambda_R(q,k)-\bLambda_R(q,\tp)\Big)\nn
\label{eq:Vfinite}
 &&+\frac{1}{2}\,\int_q\,\Big(\Lambda_R(p,q)-\Lambda_R(\tp,q)\Big)\,\bG_R^2(q)\,\bV_R(q,\tp)
 \,.\nn
\eea
All counterterms disappear from this equation. Finally, exploiting the 2PI
character of the kernel $\Lambda_R$, one can show that, similarly to 
(\ref{eq:asymptotic}):
\beq
\label{eq:asymptotic2}
 \Lambda_R(p,k)-\Lambda_R(q,k)\sim\frac{1}{k}\,,
\eeq
at large $k$ and fixed $p$ and $q$. It follows that the momentum integrals on the RHS
of Eq.\ (\ref{eq:Vfinite}) are finite. As for the case of the first vertex equation, 
it can be shown that the function $V_R(p,q,k)$ does not contain new divergences.

%%%%%%%%%%%%%%%%%%%%%%%%%%%%%%%%%%%%%%%%%%%%%%%%%%%%%%%%%%%%%%%%%%%%%%%%%%%%%%%%%%%%%%%%%%%%%%%%%%%%%%%%%%%%%%%%%%%%%%%%%%%%%%%%%%%%%%%%%%%%%%%%%%%%%%   Second two-point kernel  %%%%%%%%%%%%%%%%%%%%%%%%%%%%%%%%%%%%%%%%%%%%%%%%%%%%%%%%%%%%%%%%%%%%%%%%%%%%%%%%%%%%%%%%%%%%%%%%%%%%%%%%%%%%%%%%%%%%%%%%%%%%%%%%%%%%%%%%%%%%%%%%%%%%%%%%%%%%%%%%%%%%%%%%%%%%%%%%%%%%%
\subsubsection{Renormalization of $\delta^2\Gamma_\inter^R/\delta\phi_R^2|_{\bG_R}$}
\label{sec:ren2pt2}

We now come to the renormalization of the second two-point 2PI kernel, 
namely:\footnote{Extracting the field-strength and mass counterterms as before,
one has, explicitly:
$$
 \Sigma_R(p)=2\tilde\gamma_\inter^{(2)}(p;\bG_R)
 -i(\dZ_2\,p^2+\dmsq_2)-\frac{\Dl_2}{2}\int_q\,\bG_R(q)\,.
$$}
\beq
 \Sigma_R(\bG_R)=i\frac{\delta^2\Gamma_\inter^R}{\delta\phi_R^2}\Big|_{\bG_R}\,.
 \eeq
Following the previous analysis, we write
\beq
 \Sigma_R(p)=\Sigma_2(p)+\Sigma_0(p)\,,
\eeq
where\footnote{As before, $\delta Z_2$ and the $\dl_2^{\rm BPHZ}$ are used to 
define a finite $\Sigma_2$.}
\beq
 \Sigma_2\equiv\Sigma_R(\bG_2)+i\dmsq_2\,.
\eeq
Repeating the steps leading to Eq.\ (\ref{eq:eom2}), one can write:
\beq
 \Sigma_0(p)=-i\dmsq_2+\frac{1}{2}\,\int_q\Big(\Lambda_2(\tp,q)-\Dl_2\Big)\,
 \delta\bG(q)+\Sigma'_r(p)\,,
\eeq
where $\Sigma'_r(p)$ is a finite function decreasing at least as $1/p$ at large $p$
and where $\Lambda_2\equiv\Lambda_R(\bG_2)$. Similarly to the previous case, the 
kernel $\Lambda_2(\tp,q)$ can be traded for the finite function $V_2(\tp,q)$,
which is the solution
of the integral equation (\ref{eq:mbs}) with propagator $\bG_2$. Repeating 
the same steps as for the kernel $\delta\Gamma_\inter^R/\delta G_R|_{\bG_R}$,
one can check that the potentially divergent integrals involving the function 
$\bSigma_0(p)$ cancel out. One finally obtains 
(compare to Eq.\ (\ref{eq:bsigma0})):
\bea
 \Sigma_0(p)&=&\frac{1}{2}\,\int_q\,V_2(\tp,q)\,\Big\{\bG_r(q)
 +\bG_2^2(q)\,\bSigma_r(q)\Big\}+\Sigma'_r(p)\nonumber\\
 && -i\dmsq_2-\frac{i}{2}\left(m_R^2+\dmsq_0\right)
   \int_q\,V_2(\tp,q)\,\bG_2^2(q)\,.
\label{eq:bsigma2}
\eea
As in the previous case, the first line is finite by power counting and the remaining
logarithmic divergence in the second line can be absorbed in $\dmsq_2$.

%%%%%%%%%%%%%%%%%%%%%%%%%%%%%%%%%%%%%%%%%%%%%%%%%%%%%%%%%%%%%%%%%%%%%%%%%%%%%%%%%%%%%%%%%%%%%%%%%%%%%%%%%%%%%%%%%%%%%%%%%%%%%%%%%%%%%%%%%%%%%%%%%%%%%%   $n$-point functions  %%%%%%%%%%%%%%%%%%%%%%%%%%%%%%%%%%%%%%%%%%%%%%%%%%%%%%%%%%%%%%%%%%%%%%%%%%%%%%%%%%%%%%%%%%%%%%%%%%%%%%%%%%%%%%%%%%%%%%%%%%%%%%%%%%%%%%%%%%%%%%%%%%%%%%%%%%%%%%%%%%%%%%%%%%%%%%%%%%%%%
\subsection{The $n$-point functions}
Now that all the 2PI kernels have been renormalized together with the 
vertex functions $\bV_R$ and $V_R$, we can discuss the renormalization of the
$n$-point functions derived from the 2PI-resummed effective action. 
We first consider the symmetric phase.

\subsubsection{Two-point function}
The two-point function in the symmetric phase is nothing but the kernel 
$\delta^2\Gamma_\inter^R/\delta\phi_R^2|_{\bG_R}$, see \Eqn{eq:2ptsymb}. Its 
renormalization has been dealt with in the previous section.

\subsubsection{Four-point function}
\begin{figure}
\begin{center}
\includegraphics[width=8cm]{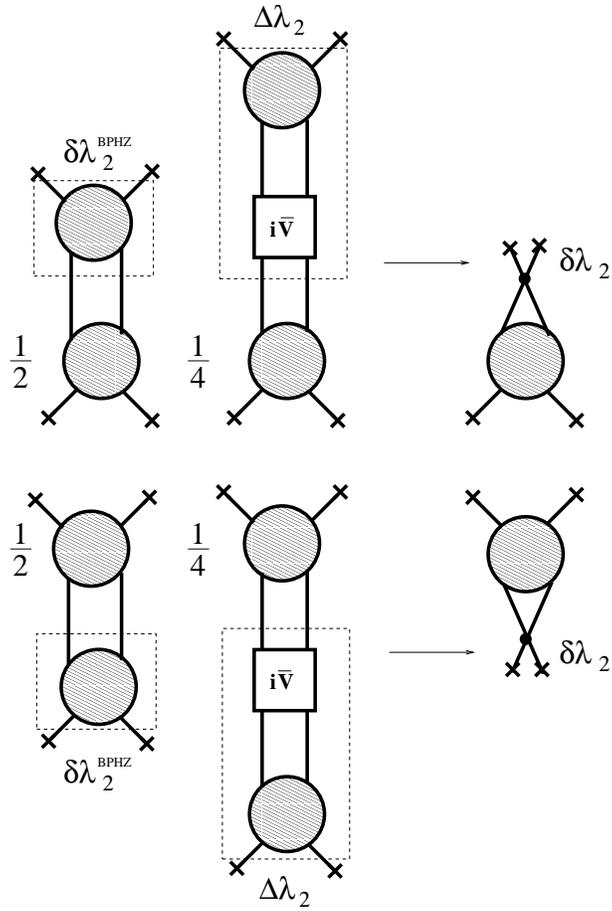}
 \caption{\small The sub-divergences in the four-point function are nothing but the divergences 
   in $V_R$.\label{fig:coupling_sing42}}
\end{center}
\end{figure}

\begin{figure}[tb]
\begin{center}
\includegraphics[width=8cm]{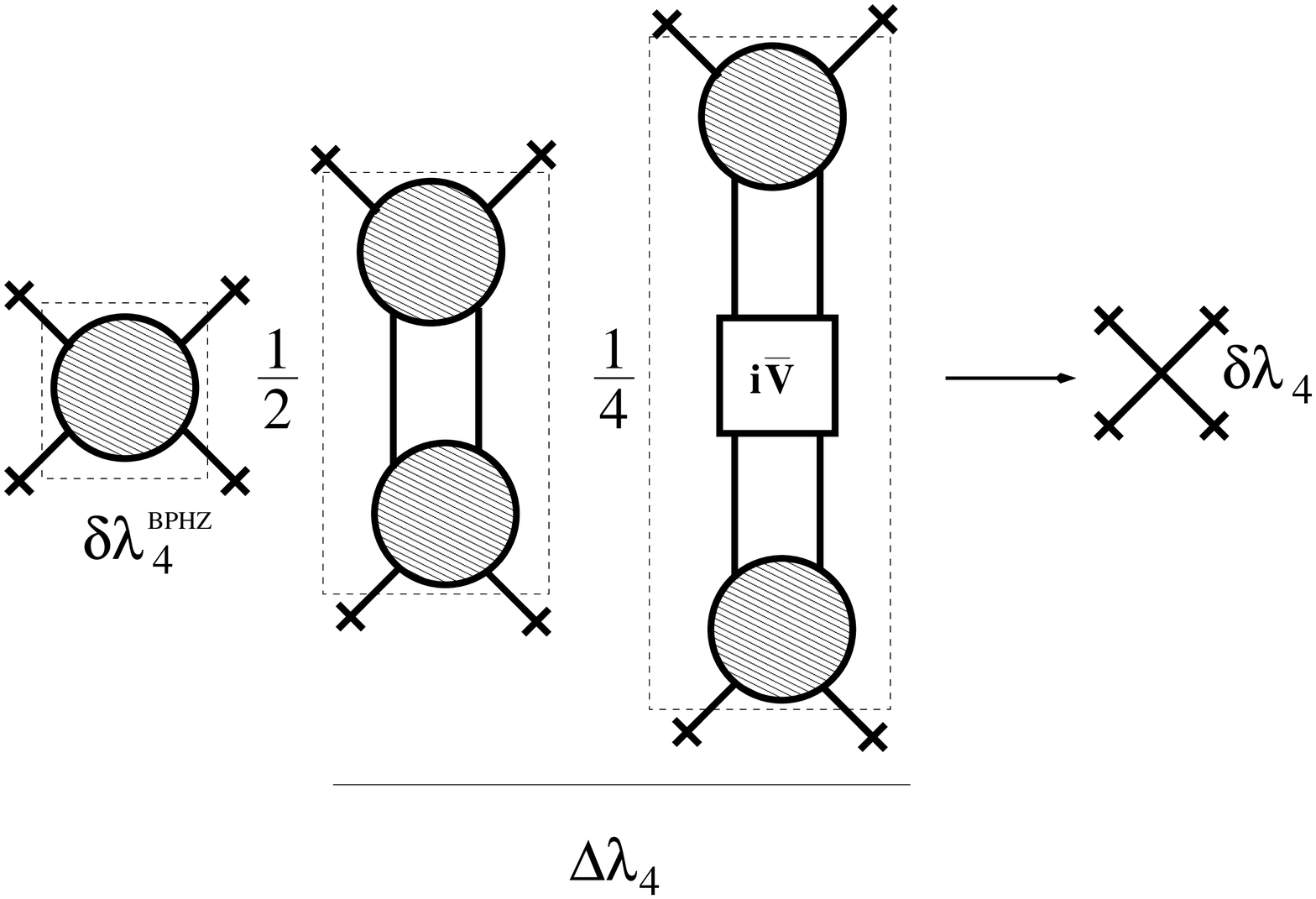}
 \caption{\small The remaining divergences are only overall divergences. They are thus local 
   and are absorbed in a redefinition of the tree-level contribution to
   $\delta^4\Gamma_\inter^R/\delta\phi_R^4|_{\bG_R}$. Part of them ($\dl^{\rm BPHZ}_4$) 
   have already been taken into account in the renormalization of the four-point kernel 
   $\delta^4\Gamma_\inter/\delta\phi^4|_{\bG_R}$. The rest can be absorbed in a shift 
   $\Delta\lambda_4$ of the tree-level contribution to this kernel.\label{fig:coupling_sing422}}
\end{center}
\end{figure}

For a diagrammatic analysis of divergences of 
the four-point function we use \Eqn{eq:4pt2symb} 
(cf.~also Fig.~\ref{fig:fourth_der2}) expressed in terms of renormalized 
quantities. One sees that
all possible coupling sub-divergences in this expression have been absorbed in the 
renormalization of the vertex $V_R$. This is illustrated in 
Fig.~\ref{fig:coupling_sing42}.
Therefore, only overall, local divergences remain which can be absorbed altogether 
in a contribution $\Dl_4$ to the counterterm $\dl_4$ associated with the classical interaction 
term in the effective action, as shown in Fig.~\ref{fig:coupling_sing422}. This adds to the
contribution $\dl^{\rm BPHZ}_4$ arising from the renormalization of the kernel 
$\gamma_\inter^{(4)}(\bG_R)$ through the 2PI BPHZ analysis. 
It corresponds to a local 
shift of the tree-level contribution to 
the kernel $\Gamma_\inter^{(4)}(\bG_R)$.

For an algebraic proof, we use \Eqn{eq:4ptsymb}. As for the functions $\bV_R$ and $V_R$ above, 
we use the notations $\Gamma_R^{(4)}(p,q,k)\equiv\Gamma_R^{(4)}(p,-q,-k,k+q-p)$ and 
$\Gamma_R^{(4)}(p,k)\equiv\Gamma_R^{(4)}(p,p,k)$, where $\Gamma_R^{(4)}(p_1,\ldots,p_4)$
is the four-dimensional Fourier transform of the renormalized four-point function.
According to our discussion above \Eqn{eq:shift}, we write the kernel 
\beq
 \frac{\delta^4\Gamma_\inter^R}{\delta\phi_R^4}\Big|_{\bG_R}=\Gamma_\inter^{(4)}(\bG_R)=
 \gamma_\inter^{(4)}(\bG_R)-\Dl_4\,,
\eeq
where the function $\gamma_\inter^{(4)}(\bG_R)$ has been made finite by means of the BPHZ 
analysis applied to resummed 2PI diagrams, as described previously. The three channels of \Eqn{eq:4ptsymb}
contribute the same for what concerns UV-divergences so one can write: 
\bea
 \Gamma_R^{(4)}(p,k)&=&\gamma_\inter^{(4)}(p,k)-\Dl_4+\frac{3}{2}\,\int_q 
 V_R(p,q)\,\bG_R^2(q)\,\Big(\Lambda^\dagger_R(q,k)-\Dl_2\Big)\nn
&+& \mbox{finite}\nn
 &=&\gamma_\inter^{(4)}(p,k)-\Dl_4+\frac{3}{2}\,\int_q 
 \Big(\Lambda_R(p,q)-\Dl_2\Big)\,\bG_R^2(q)\,V^\dagger_R(q,k)\nn
&+& \mbox{finite}\,,
\eea
where we used the fact that $\Gamma_R^{(4)}(p,k)=\Gamma_R^{(4)}(k,p)$ -- and similarly
for $\gamma_\inter^{(4)}(p,k)$ -- to write the second line, and where $\Lambda_R^\dagger(q,k)=\Lambda_R(k,q)$ 
and $V_R^\dagger(q,k)=V_R(k,q)$. The two equations above imply 
that:
\bea
 \int_q V_R(p,q)\,\bG_R^2(q)\,\Big(\Lambda_R^\dagger(q,k)-\Dl_2\Big)&=&
 \int_q \Big(\Lambda_R(p,q)-\Dl_2\Big)\,\bG_R^2(q)\,V^\dagger_R(q,k)\nn
&+& \mbox{finite} \,.
\eea
From this, and using similar manipulations as for 
the discussions of the functions $\bV_R$ and
$V_R$, one obtains the following finite equation:
\bea
 \Gamma_R^{(4)}(p,k)-\Gamma_R^{(4)}(\tp,\tp)&=&
 \gamma_\inter^{(4)}(p,k)-\gamma_\inter^{(4)}(\tp,\tp)\nn
 &&+\frac{3}{2}\,\int_q V_R(p,q)\,\bG_R^2(q)\,
 \Big\{\Lambda^\dagger_R(q,k)-\Lambda^\dagger_R(q,\tp)\Big\}\nn
 &&+\frac{3}{2}\,\int_q \Big\{\Lambda_R(p,q)-\Lambda_R(\tp,q)\Big\}\,
 \bG_R^2(q)\, V^\dagger_R(q,\tp)\nn
&&+ \mbox{finite}\,.
\eea
The contribution $\Dl_4$ to the counterterm $\dl_4$ has been traded for the finite
number $\Gamma_R^{(4)}(\tp,\tp)=-\lambda_R$. Using similar arguments as before, see 
\Eqn{eq:fullmom}, it can be shown that the function $\Gamma_R^{(4)}(p,q,k)$ does not 
contain new divergences.

\subsubsection{Higher $n$-point functions}\label{sec:higher}
\begin{figure}[tb]
\begin{center}
\includegraphics[width=12cm]{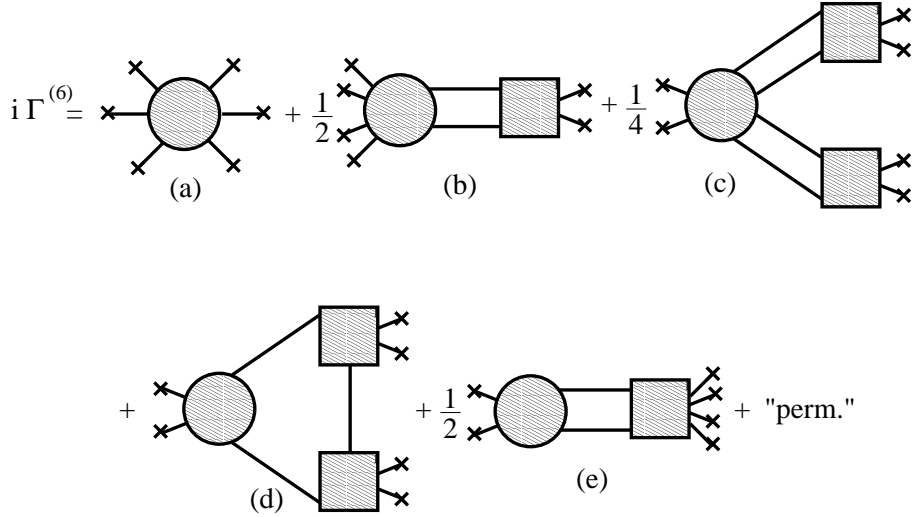}
 \caption{\small Topologies entering the six-point function. The combinatorial factors are shown. It is understood
   that each topology comes with the relevant permutations of external legs needed
   by symmetry.\label{fig:sixth_der}}
\end{center}
\end{figure}

\begin{figure}[tb]
\begin{center}
\includegraphics[width=10cm]{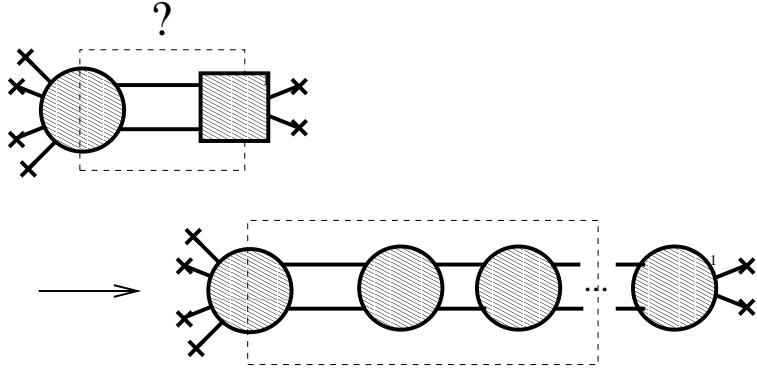}
\caption{\small The only possible origin for a four-point sub-divergence in the
  contribution (b) to the six-point function, see Fig.~\ref{fig:sixth_der}.\label{fig:kernels}}
\end{center}
\end{figure}

We now show that higher $n$-point functions are automatically finite once 
the two- and
four-point functions have been renormalized. To this aim, we discuss the case of 
the six-point function in detail. The argument generalizes to arbitrary $n$-point 
functions. The various topologies appearing in the diagrammatic representation of 
the six-point function are shown in Fig.~\ref{fig:sixth_der}. 
There, we show the combinatorial factors associated to each topology, arising from 
the various functional derivatives with the rules described earlier in Sec.~\ref{sec:npoint}. 
We do not specify explicitly the various possible orderings of the external legs
as this plays no role for the present argument. It is understood that each topology 
comes with all the permutations of its external legs needed to symmetrize it in a 
proper way, that is according to the symmetry property of the six-point 
function.\footnote{An explicit example of the relevant permutations is given in 
Fig.~\ref{fig:fourth_der} for the case of the four-point function.}
Most of the diagrams in Fig. \ref{fig:sixth_der} are explicitly finite. Diagram (a) 
is a six-point 2PI kernel, namely $\delta^6\Gamma_\inter^R/\delta\phi_R^6|_{\bG_R}=
\Gamma_\inter^{(6)}(\bG_R)$, and is thus renormalized by the 2PI BPHZ procedure described 
previously. Diagrams (b) and (c) both contain loop integrals involving the six-point kernels 
$\delta^6\Gamma_\inter^R/\delta\phi_R^4\delta G_R|_{\bG_R}=\delta\Gamma_\inter^{(4)}/
\delta G_R|_{\bG_R}$ and $\delta^6\Gamma_{\rm int}^R/\delta\phi_R^2\delta
G_R^2|_{{\bar G}_R}=\delta^2\Gamma_{\rm int}^{(2)}/\delta G_R^2|_{{\bar G}_R}$, 
the propagator $\bG_R$ and the box with two external legs 
$\delta^2\bSigma_R/\delta\phi_R^2\equiv iV_R$. Each of these objects is finite. Thus 
potential divergences in diagrams (b) and (c) can only arise from the loop integrals. 
By power counting there is no overall divergence thus one has to look for potential 
sub-divergences. The only possible candidates are depicted in Fig. \ref{fig:kernels}
in the case of diagram (b).
Clearly, the square box being a 1PI structure,\footnote{This follows from the fact
that the self-energy $\bSigma_R$ itself is 1PI.} one needs to cut at least two lines
of it in order to draw a four-point box -- corresponding to a coupling sub-divergence.
This implies that one needs to cut not more than two lines in the six-point kernel
involved in this diagram. This is, however, not possible due 
to the 2PI character of the latter (cf. Appendix~\ref{2PItop}). 
We conclude that there are no sub-divergences in diagram (b), 
which is thus finite. 
The same argument applies to diagram (c) as well. Diagrams (d) and (e) are more subtle 
since they contain the four-point kernel $\Lambda_R-\Delta\Lambda_2$, which is not finite. 
If one would replace this kernel by its finite part $\Lambda_R$, diagram (d) would be 
finite by the same argument as the one used above for diagrams (b) and (c). However, 
there is an extra contribution arising from $\Dl_2$, which is infinite. 
In fact, as we now show, this contribution is crucial to remove divergences which are 
present in diagram (e). To see this explicitly, we first have to discuss the renormalization
of the function $\delta^4\bSigma_R/\delta\phi_R^4$, involved in diagram (e).

\begin{figure}[tb]
\begin{center}
\includegraphics[width=12cm]{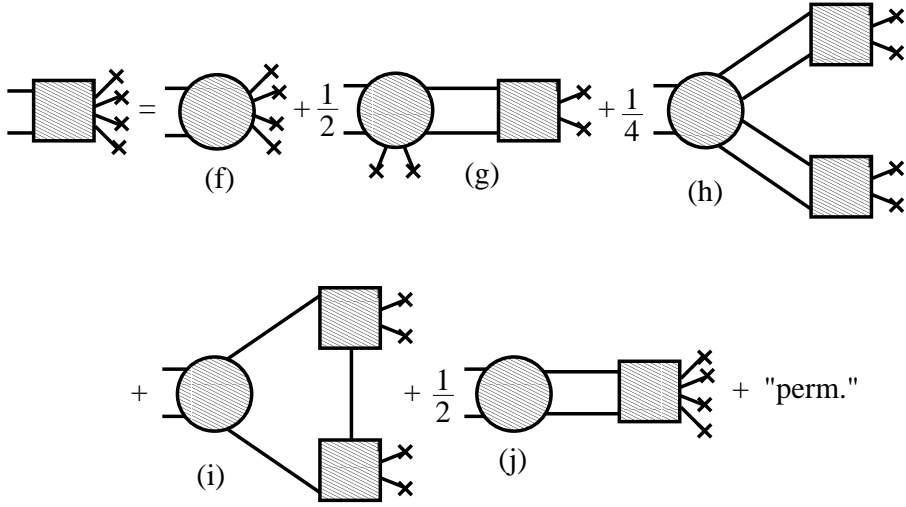}
 \caption{\small Topologies appearing in the self-consistent equation for 
   $\delta^4\bSigma_R/\delta\phi_R^4$. The combinatorial factors are shown. It is understood
   that each topology comes with the relevant permutations of external legs needed
   by symmetry.\label{fig:Sigma_der6}}
\end{center}
\end{figure}

\begin{figure}
\begin{center}
\includegraphics[width=14cm]{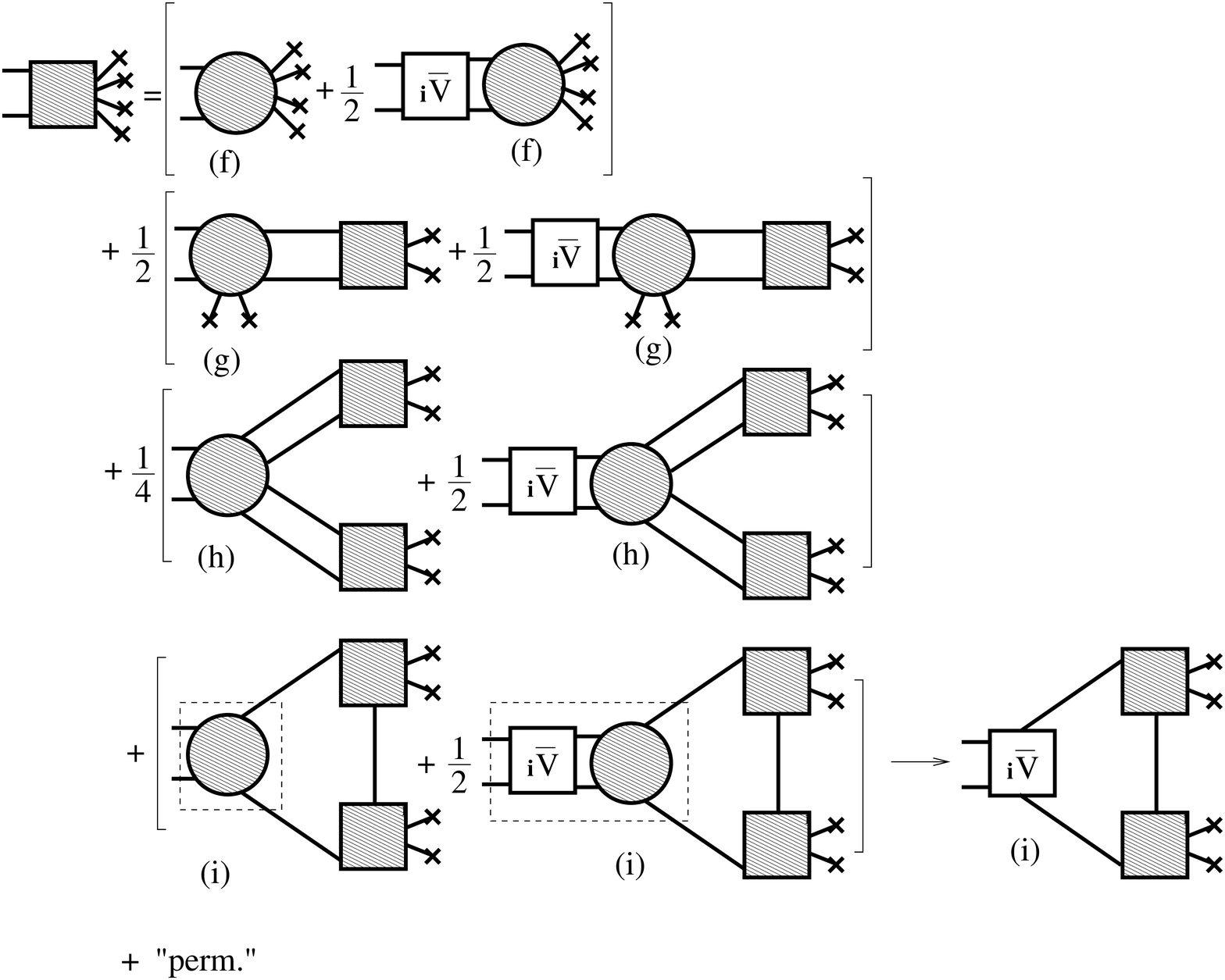}
 \caption{\small Solution and finiteness of $\delta^4\bSigma_R/\delta\phi_R^4$: The only 
   two potential divergent diagrams combine to a finite term (last line). \label{fig:Sigma_der6sol}}
\end{center}
\end{figure}

As discussed earlier in Sec.~\ref{sec:npoint}, this function satisfies a linear integral 
equation. In Fig.~\ref{fig:Sigma_der6}, we show the relevant topologies appearing in this 
equation together with the appropriate combinatorial factors. As before, it is understood
that each topology comes with the relevant permutations of external legs needed to ensure
the correct symmetry properties according to the LHS.
Using the same argument as for the diagram (a)-(c) in Fig.~\ref{fig:sixth_der}, we conclude
that the contributions (f)-(h) are finite, whereas diagrams (i) and (j) are infinite.
Diagram (j) involves the unknown function $\delta^4\bSigma_R/\delta\phi_R^4$ itself. To 
pursue the discussion, it is thus more appropriate to solve for $\delta^4\bSigma_R/
\delta\phi_R^4$ using the vertex function $\bV_R$, as explained in Sec.~\ref{sec:npoint}. 
The topologies appearing in the solution are depicted in Fig.~\ref{fig:Sigma_der6sol} with
the relevant combinatorial factors.
As expressed in \Eqn{eq:resumsol}, one obtains all the previous diagrams of 
Fig.~\ref{fig:Sigma_der6} but (j), plus the same diagrams convoluted with the 
function $\bV_R$. Using a similar reasoning as above, exploiting the 2PI character
of the kernels, one finds that all the diagrams 
appearing in the first three lines of Fig.~\ref{fig:Sigma_der6sol} -- 
that is those 
containing diagrams (f)-(h) as a sub-diagram -- are explicitly finite. The only 
divergent contributions are those containing diagrams (i), shown on the last line of 
the figure. These two divergent diagrams combine to a finite term
by means of \Eqn{eq:BSsymb}, as represented 
in Fig.~\ref{fig:Sigma_der6sol}. This shows 
that the function $\delta^4\bSigma_R/\delta\phi_R^4$ is finite.

\begin{figure}
\begin{center}
\includegraphics[width=14cm]{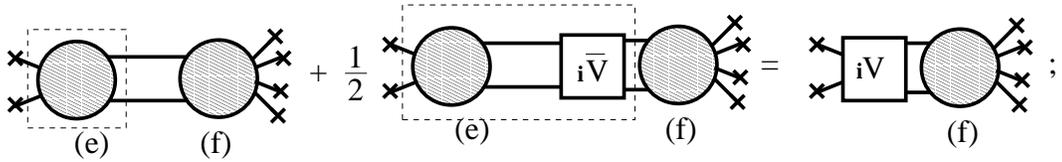}
 \caption{\small Structures appearing in the six-point function after inserting 
the expression 
for $\delta^4\bSigma_R/\delta\phi_R^4$ represented on Fig.~\ref{fig:Sigma_der6sol} 
   into the diagram (e) of Fig.~\ref{fig:sixth_der}. Each contribution on the LHS
   is potentially divergent, but their sum is finite.
  \label{fig:exemplestruct}}
\end{center}
\end{figure}

\begin{figure}
\begin{center}
\includegraphics[width=13.5cm]{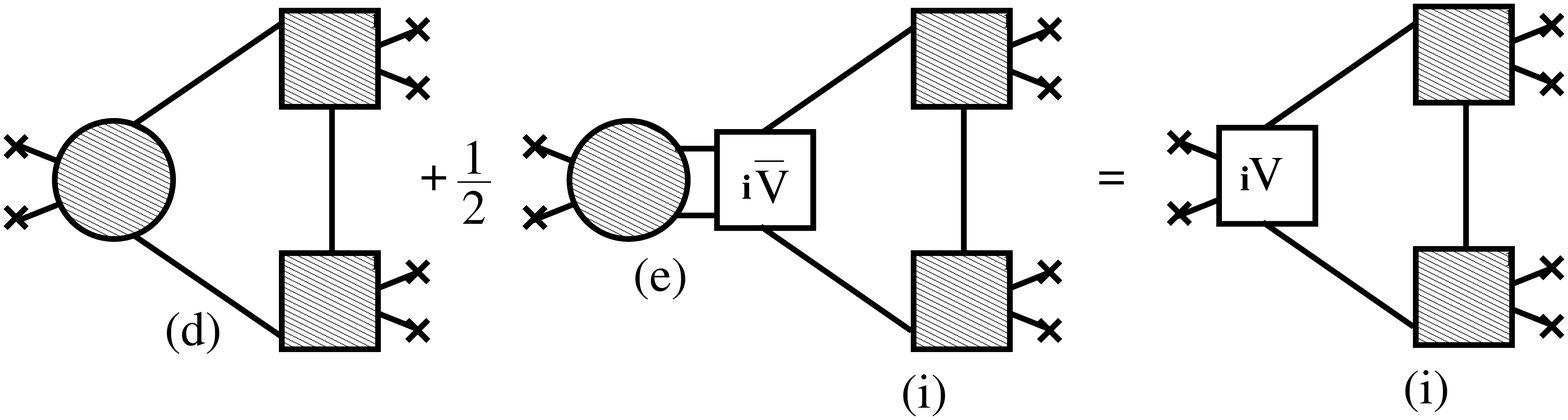}
 \caption{\small Contributions to the six-point function. The structure involving sub-diagram
   (i) arises after plugging the expression of $\delta^4\bSigma_R/\delta\phi_R^4$ represented 
   on Fig.~\ref{fig:Sigma_der6sol} into the diagram (e) of Fig.~\ref{fig:sixth_der}. 
   The latter combines with the previous contribution (d) into a finite structure by
   means of \Eqn{eq:intVsymb}.\label{fig:exemplestruct2}}
\end{center}
\end{figure}

We now return to the six-point function. 
We insert the diagrams of Fig.~\ref{fig:Sigma_der6sol} 
into diagram (e) of Fig.~\ref{fig:sixth_der}. This generates structures such as the one 
depicted in Fig.~\ref{fig:exemplestruct}, which shows the convolution of diagram (e)
with the first line of Fig.~\ref{fig:Sigma_der6sol}, that is with the diagrams containing
(f) as a sub-structure. 
Each separate contribution containing the sub-diagram (f) contains potential 
sub-divergences. However, using \Eqn{eq:intVsymb} one sees that their sum combines
in a simpler structure involving the finite function $V_R$, as shown in 
Fig.~\ref{fig:exemplestruct}. Exploiting the 2PI character of the kernel
$\delta\Gamma_\inter^{(4)}/\delta G_R|_{\bG_R}$ -- which corresponds to the
diagram (f) --
one concludes that the resulting structure is finite. A similar reasoning shows that 
the insertion of the second and the third lines of Fig.~\ref{fig:Sigma_der6sol}, i.e.\ those
involving the sub-diagrams (g) and (h) respectively, into diagram (e) of 
Fig.~\ref{fig:sixth_der} also leads to finite structures involving the function $V_R$. 
In contrast, insertion of the last diagram of Fig.~\ref{fig:Sigma_der6sol} -- involving
the sub-diagram (i) -- into (e) is not enough to generate a finite structure. However,
the resulting contribution combines with diagram (d) of Fig.~\ref{fig:sixth_der} into
a structure involving the finite function $V_R$, as shown in Fig.~\ref{fig:exemplestruct2}.
The latter contribution is finite by similar arguments as before.
Thus we conclude that the six-point function is finite, as announced.

The present analysis straightforwardly generalizes to the case of higher $n$-point
functions. Although the number of terms to consider increases rapidly, 
the relevant structures needed for the cancellation of divergences by
means of Eqs.~\eqn{eq:BSsymb} and \eqn{eq:intVsymb} always appear. In particular,
using a similar reasoning as above, one can show by recurrence that all the functions 
$\delta^{2n}\bSigma_R/\delta\phi_R^{2n}$ as well as all $n$-point functions $\Gamma_R^{2n}$ 
are finite. Finally, we mention that the zero-point function, namely $\Gamma_\inter^{(0)}[\bG_R]$,
can be shown to be finite up to an irrelevant, 
field and temperature independent infinite constant~\cite{vanHees:2001ik}.

%%%%%%%%%%%%%%%%%%%%%%%%%%%%%%%%%%%%%%%%%%%%%%%%%%%%%%%%%%%%%%%%%%%%%%%%%%%%%%%%%%%%%%%%%%%%%%%%%%%%%%%%%%%%%%%%%%%%%%%%%%%%%%%%%%%%%%   Broken phase   %%%%%%%%%%%%%%%%%%%%%%%%%%%%%%%%%%%%%%%%%%%%%%%%%%%%%%%%%%%%%%%%%%%%%%%%%%%%%%%%%%%%%%%%%%%%%%%%%%%%%%%%%%%%%%%%%%%%%%%%%%%%%%%%%%%%%%%%%%%%%%%%%%%%%%%%%%%%%%%%%%%%%%%%%%%%%%%%%%%%%
\section{Renormalization at non-vanishing field}
\label{sec:finite2}

It has been shown above 
that the resummed effective action is finite. For this we have employed 
derivatives of the effective action taken at $\phi_R=0$. 
As far as UV singularities are concerned,
it thus follows that derivatives at $\phi_R\neq 0$ 
are also finite. The latter are exactly the $n$-point 
functions that one has to consider 
in the broken phase. It is therefore 
sufficient to renormalize the resummed effective 
action in the symmetric phase. Divergences are simply reshuffled among the various
$n$-point functions as compared to the zero-field case. 

In this section, we illustrate these general arguments for a number of
relevant examples. We first 
discuss the divergences appearing in the equation for $\bG_R(\phi_R)$ 
for arbitrary $\phi_R\neq0$. We show that to make the latter finite, one does not only
need the counterterms determined from the renormalization of $\bG_R(\phi_R=0)$, but also
the counterterms used to renormalize the two-point function 
$\Gamma^{(2)}_R(\phi_R=0)$. 
We also discuss the UV singularities of the two-point function $\Gamma_R^{(2)}(\phi_R)$. 
The renormalization of the latter for arbitrary $\phi_R\neq0$ is connected to the 
renormalization of all higher $n$-point functions in the symmetric 
phase.

\subsection{Renormalization of $\bG_R(\phi_R)$}\label{sec:GRphi}
A diagrammatic analysis reveals that the divergences appearing in the function 
$\bG_R(\phi_R)$ for non-zero field are precisely those of the functions $\bG_R(\phi_R=0)$ 
and part of those in $\Gamma^{(2)}_R(\phi_R=0)$. The renormalization of $\bG_R(\phi_R)$, therefore, does not 
only involve the counterterms $\dZ_0$, $\dmsq_0$, $\dl_0$, etc., needed to renormalize the 
former, but also the coupling counterterm $\dl_2$, determined from the renormalization of 
the latter.\footnote{We point out, in particular, that the renormalization 
of $\bG_R(\phi_R=0)$ is actually enough to renormalize $\bG_R(\phi_R)$ for approximations 
where these counterterms are equal. This is, for instance, the case of the approximation 
discussed in Ref.~\cite{Berges:2004hn}, or in the 2PI $1/N$-expansion 
\cite{Berges:2001fi,Aarts:2002dj}, where one has, in particular $\dl_0=\dl_2$ 
(see Sec.~\ref{sec:sym} below and Ref.~\cite{Cooper:2004rs}).} Here, we present 
an algebraic discussion of these aspects. Using the field-expansion 
\eqn{eq:fieldexp}, the self-energy $\bSigma_R(\phi_R)$ may be written as
\beq
\label{eq:sigmafieldexp}
 \bSigma_R(\phi_R)=2i\left\{\frac{\delta\Gamma_\inter^{(0)}}{\delta G_R}\Big|_{\bG_R}
 +\frac{1}{2}\phi_R^2\frac{\delta\Gamma_\inter^{(2)}}{\delta G_R}\Big|_{\bG_R}+
 \sum_{n\geq 2}\frac{1}{(2n)!}\phi_R^{2n}\frac{\delta\Gamma_\inter^{(2n)}}
     {\delta G_R}\Big|_{\bG_R}\right\}\,.
\eeq
where $\bG_R\equiv\bG_R(\phi_R)$. Applying the same technique as in the previous 
section, and generalizing the notations of Secs.~\ref{sec:ren2pt1} and
\ref{sec:ren2pt2} to the case with non-vanishing field, we obtain
\bea
 \bSigma_0(p)&=&-i\delta m_0^2
 +\frac{1}{2}\,\int_q\Big(\bLambda_2(p,q)-\Dl_0\Big)\,\delta \bG(q)\nn
\label{eq:toto4}
 &&+\frac{i}{2}\,\phi_R^2\Big(\Lambda_2(0,p)-\Dl_2\Big)+\bSigma_r(p)\,,
\eea
where the function $\delta \bG(p)$ can be written as in \Eqn{eq:deltaG}.
The above equation can be rewritten as ($\tp$ being the renormalization
point):
\bea
 \bSigma_0(p)&=&-i\delta m_0^2
 +\frac{1}{2}\,\int_q\Big(\bLambda_2(\tp,q)-\Dl_0\Big)\,\delta \bG(q)\nn
\label{eq:toto3}
 &&+\frac{i}{2}\,\phi_R^2\Big(\Lambda_2(0,p)-\Dl_2\Big)+\bSigma'_r(p)\,,
\eea
where the functions $\bSigma_r(p)$ and $\bSigma'_r(p)$ are finite and $\sim 1/p$ at large 
$p$. They are related to each other as in \Eqn{eq:redef}. Here, we have used the fact that 
possible subdivergences in the terms higher than quadratic in the field in \Eqn{eq:sigmafieldexp} 
have been removed by means of the previously described BPHZ analysis. 
Moreover, these contributions decrease at least as $\sim 1/p$ at large $p$, which follows 
from power counting arguments and the 2PI character of the kernels 
$\delta\Gamma_\inter^{(2n)}/\delta G_R|_{\bG_R}$. Using the vertex equation for 
$\bV_R\equiv\bV_R(\phi_R)$, with propagators replaced by $\bG_2$, we get:
\bea
 \bSigma_0(p) & = & -i\delta m_0^2+\bSigma'_r(p)
 +\frac{i}{2}\,\phi_R^2\Big(\Lambda_2(0,p)-\Dl_2\Big)
 +\frac{1}{2}\,\int_q\,\bV_2(\tp,q)\,\delta \bG(q)\nonumber\\
 &&-\frac{1}{4}\,\int_q\int_k\,\bV_2(\tp,q)\,\bG_2^2(q)\,
 \Big(\bLambda_2(q,k)-\Dl_0\Big)\,\delta \bG(k)\,.
\eea
In the third term, the integral over $k$ is known from \Eqn{eq:toto4}:
\bea
 \bSigma_0(p) & = & -i\delta m_0^2+\bSigma'_r(p)
 +\frac{1}{2}\,\int_q\,\bV_2(\tp,q)\,\delta \bG(q)\nn
 &-&\frac{1}{2}\,\int_q\,\bV_2(\tp,q)\,\bG_2^2(q)\,
 \Big\{\bSigma_0(q)-\bSigma_r(q)+i\delta m_0^2\Big\}\nn
 &+&\frac{i}{2}\phi_R^2\Big\{\Lambda_2(0,p)-\Dl_2+\frac{1}{2}
 \int_q\bV_2(\tp,q)\,\bG_2^2(q)\,\Big(\Lambda_2(0,q)-\Dl_2\Big)\Big\}.\nn
\eea
Using the expression of the function $\delta\bG(p)$ (cf. \Eqn{eq:deltaG}), one observes
that potentially divergent terms depending on $\bSigma_0$ vanish and 
we are left with
\bea
 \bSigma_0(p)&=&\frac{1}{2}\,\int_q\,\bV_2(\tp,q)\,
 \Big\{\bG_r(q)+\bG_2^2(q)\,\bSigma_r(q)\Big\}+\bSigma'_r(p)\nn
 &&-i\delta m_0^2-\frac{i}{2}(m_R^2+\dmsq_0)\,\int_q\,\bV_2(\tp,q)\,\bG_2^2(q)\nn
 &&+\frac{i}{2}\,\phi_R^2\Big\{\Lambda_2(0,p)-\Lambda_2(0,\tp)+V_2(0,\tp)\Big\}\,,
\eea
where we have used the defining equation for the finite number $V_2(0,\tp)$ (cf.
\Eqn{eq:mbss}) to absorb the counterterm $\Dl_2$. The above equation has the same 
structure as the corresponding \Eqn{eq:bsigma0} for the case of a vanishing field, 
except for the last line, which is finite due to the counterterm $\dl_2$. 

\subsection{The two-point function}
For the purpose of discussing the structure of UV divergences due to the presence
of a non-vanishing field, it is sufficient to assume
the following field-expansion of the two-point function:\footnote{This can be
thought of as the field-expansion of the UV 
divergent part of the two-point function.}
\beq
\label{eq:fieldexp2pt}
 \Gamma_R^{(2)}(\phi_R)=\Gamma_R^{(2)}(0)+\frac{1}{2}\phi_R^2\Gamma_R^{(4)}(0)+
 \frac{1}{4!}\phi_R^4\Gamma_R^{(6)}(0)+\ldots\,.
\eeq
The latter is finite for arbitrary field, provided
the theory has been properly renormalized at $\phi_R=0$, that is all $n$-point
functions have been made finite in the symmetric phase. Equation \eqn{eq:fieldexp2pt}
also shows how the divergences appearing in the symmetric phase are reshuffled 
in the presence of a non-vanishing field expectation value. In particular, unlike
in the symmetric case, all the counterterms of the theory are needed to renormalize 
the two-point function. 

\begin{figure}[tb]
\begin{center}
\includegraphics[width=13.5cm]{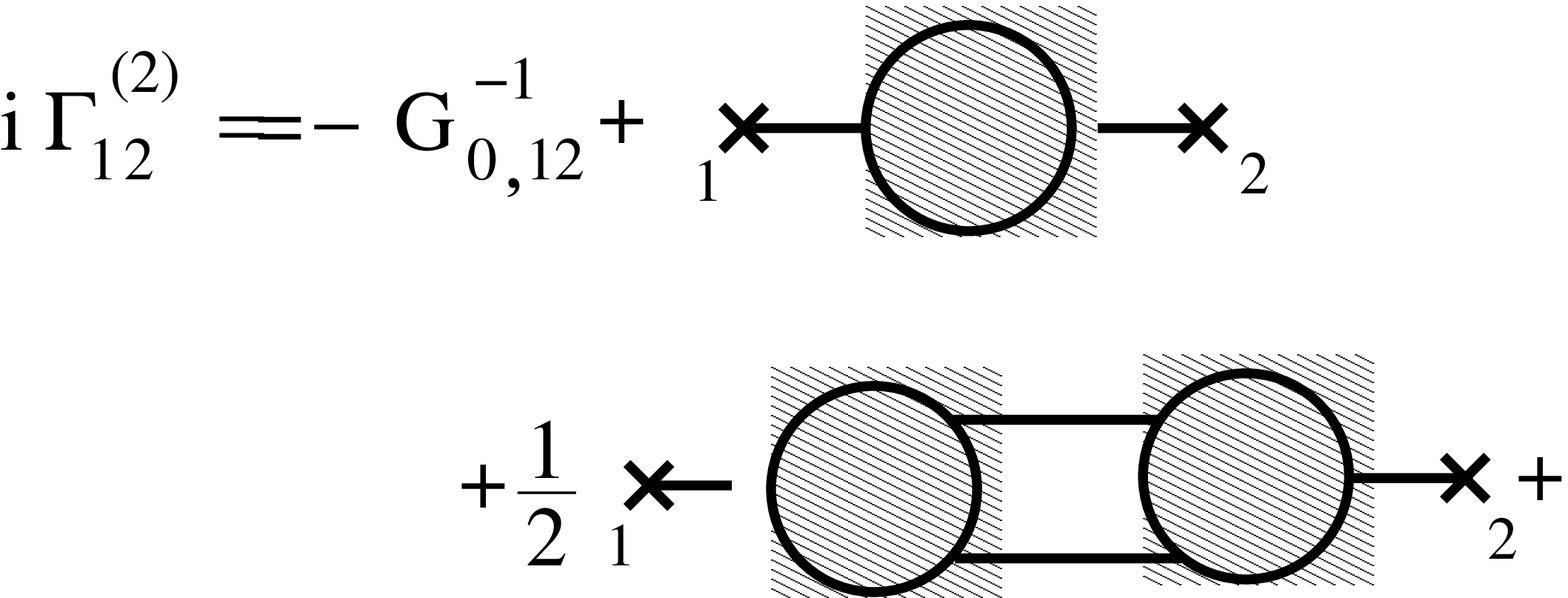}
 \caption{\small The two-point function in the broken phase $\Gamma^{(2)}_R(\phi_R)$, in terms of 
 the vertex function $\bV_R$.\label{fig:2pointsol}}
\end{center}
\end{figure}

It is instructive to see how \Eqn{eq:fieldexp2pt} arises from
the expression of the two-point function in terms of 2PI kernels. Here, we
illustrate this for the example of the term quadratic in the field, which
involves the four-point function at $\phi_R=0$. In the broken phase, 
the non-trivial part of the two-point function is no longer simply given by the 
kernel $\delta^2\Gamma_\inter^R/\delta\phi_R^2|_{\bG_R}$, but receives a new 
contribution which involves the function $\delta\bSigma_R/\delta\phi_R$, as 
depicted in Fig.~\ref{fig:Sigma} (cf.~also \Eqn{eq:2pt}). In the present 
$Z_2$-symmetric theory, the latter vanishes in the 
symmetric phase. In the general case, it satisfies the integral equation 
\eqn{eq:resum1}, represented in Fig.~\ref{fig:Sigma}, expressed in terms of 
renormalized kernels. The latter has the general form of \Eqn{eq:resumgen} and 
can be solved in terms of the finite vertex function $\bV_R$, as in \Eqn{eq:resumsol}. 
Plugging the resulting expression into the expression for the two-point function
(cf. \Eqn{eq:2pt}) one finds
\bea
 \Gamma^{(2)}_R(\phi_R)&=&iG_{0R}^{-1}+\frac{\delta^2\Gamma_\inter^R}{\delta\phi_R^2}\Big|_{\bG_R}
 +2i\frac{\delta^2\Gamma_\inter^R}{\delta\phi_R \delta G_R}\Big|_{\bG_R}\bG_R^2
 \frac{\delta^2\Gamma_\inter^R}{\delta G_R\delta\phi_R}\Big|_{\bG_R}\nn
\label{eq:BP2pt}
 &&-\frac{\delta^2\Gamma_\inter^R}{\delta\phi_R \delta G_R}\Big|_{\bG_R}
 \bG_R^2\bV_R\bG_R^2\frac{\delta^2\Gamma_\inter^R}{\delta G_R\delta\phi_R}\Big|_{\bG_R}\,,
\eea
where $\bG_R\equiv\bG_R(\phi_R)$ and $\bV_R\equiv\bV_R(\phi_R)$. Equation \eqn{eq:BP2pt}
is depicted in Fig.~\ref{fig:2pointsol}. Using again power counting and the 2PI character
of the kernels, one finds that possible sub-divergences in the last 
two terms of \Eqn{eq:BP2pt} can only arise from the quadratic field-dependence of the 
latter. Using the field-expansion\footnote{Notice that, unlike in the symmetric 
case, the various 2PI kernels are not related to only one term in the field 
expansion \eqn{eq:fieldexp}.}
\beq
 \frac{\delta^2\Gamma_\inter^R}{\delta\phi\delta G_R}\Big|_{\bG_R}
 =\phi_R\frac{\delta\Gamma_\inter^{(2)}}{\delta G_R}\Big|_{\bG_R}
 +\frac{1}{3!}\phi_R^3\frac{\delta\Gamma_\inter^{(4)}}{\delta G_R}\Big|_{\bG_R}+\dots\,,
\eeq
one finds that the quadratic field-dependence of the last two terms of \Eqn{eq:BP2pt}
is given by
\bea
 &&2i\phi_R\,\frac{\delta\Gamma_\inter^{(2)}}{\delta G_R}\Big|_{\bG_R(0)}
 \Big\{\bG_R^2(0)+\frac{i}{2}\bG_R^2(0)\,\bV_R(0)\,\bG_R^2(0)\Big\}
 \frac{\delta\Gamma_\inter^{(2)}}{\delta G_R}\Big|_{\bG_R(0)}\phi_R\nn
\label{eq:quadBP}
 &&=i\phi_R\,V_R(0)\,\bG_R^2(0)\,\frac{\delta\Gamma_\inter^{(2)}}
      {\delta G_R}\Big|_{\bG_R(0)}\phi_R\,,
\eea
where we used Eqs.~\eqn{eq:4kernel2res} and \eqn{eq:intVsymb} to write the result in a 
compact form involving the function $V_R(\phi_R=0)$.
We see that the divergences contained in this contribution have
a similar structure as that of the four-point function in the symmetric phase
(cf.~Eqs.~\eqn{eq:4ptsymb} and \eqn{eq:4pt2symb}). 

All the remaining divergences in the two-point function
at non-vanishing field in fact come from the second term on the RHS of \Eqn{eq:BP2pt},
which was already present in the symmetric phase. The explicit field dependence
of the latter is given by
\beq
\label{eq:explicit_fieldexp}
 \frac{\delta^2\Gamma_\inter^R}{\delta\phi_R^2}\Big|_{\bG_R}=\Gamma_\inter^{(2)}(\bG_R)
 +\frac{1}{2}\phi_R^2\Gamma_\inter^{(4)}(\bG_R)+\dots\,.
\eeq
One observes that the explicit quadratic term, involving $\Gamma_\inter^{(4)}[\bG_R]$, almost
combines with the quadratic term \eqn{eq:quadBP} to give the four-point function, as 
in \Eqn{eq:4ptsymb}. However, one of the three possible channels appearing in this
equation is missing. In fact, the latter arise from the implicit field-dependence 
of the kernels, that is through the field-dependence of $\bG_R(\phi_R)$. Indeed,
writing
\beq
 \bG_R(\phi_R)=\bG_R(0)+\frac{i}{2}\phi_R^2\,V_R(0)\,\bG_R^2(0)+\ldots\,,
\eeq
where we used the definition of the vertex function $V_R(0)$, cf. \Eqn{eq:V},
one obtains:
\beq
\label{eq:quad3}
 \Gamma_\inter^{(2)}(\bG_R)=\Gamma_\inter^{(2)}(\bG_R(0))+
 \frac{i}{2}\phi_R^2\,V_R(0)\,\bG_R^2(0)\,
 \frac{\delta\Gamma_\inter^{(2)}}{\delta G_R}\Big|_{\bG_R(0)}+\ldots\,.
\eeq
When inserted into the expression \eqn{eq:BP2pt} for the two-point function, the first term
on the RHS of \Eqn{eq:quad3} gives the first term in the field-expansion \eqn{eq:fieldexp2pt},
that is simply the two-point function in the symmetric phase. The term quadratic in the 
field in \Eqn{eq:quad3} gives the missing channel needed to reconstruct the four-point
function at vanishing field, as mentioned above. Indeed,
collecting all terms quadratic in the field in Eqs.~\eqn{eq:quadBP}, 
\eqn{eq:explicit_fieldexp} and \eqn{eq:quad3} and using the fact that 
$\delta\Gamma_\inter^{(2)}/\delta G_R|_{\bG_R(0)}=(\Lambda_R(0)-\Dl_2)/2$, 
one finds that the quadratic field-dependence of the two-point function \eqn{eq:BP2pt} 
is given by (see \Eqn{eq:4ptsymb}):
\beq
 \frac{1}{2}\phi_R^2\Big\{\Gamma_\inter^{(4)}(\bG_R(0))+\frac{3i}{2}(\Lambda_R(0)-\Dl_2)\,
\bG_R^2(0)\,V_R(0)\Big\}=\frac{1}{2}\phi_R^2\Gamma_R^{(4)}(0)\,,
\eeq
as expected. Higher order terms in the field-expansion \eqn{eq:fieldexp2pt} can be 
obtained along similar lines.

%%%%%%%%%%%%%%%%%%%%%%%%%%%%%%%%%%%%%%%%%%%%%%%%%%%%%%%%%%%%%%%%%%%%%%%%%%%%%%%%%%%%%%%%%%%%%%%%%%%%%%%%%%%%%%%%%%%%%%%%%%%%%%%%%%%%%%%%%%%%%%%%%%%%%%%%%%%%%%%%%%%%%%%%%%%%%%%%%%%%%%%%%%%%%%%%%%%%%%%%%%%%%%%%%%%%%%%   Multiple scalar fields   %%%%%%%%%%%%%%%%%%%%%%%%%%%%%%%%%%%%%%%%%%%%%%%%%%%%%%%%%%%%%%%%%%%%%%%%%%%%%%%%%%%%%%%%%%%%%%%%%%%%%%%%%%%%%%%%%%%%%%%%%%%%%%%%%%%%%%%%%%%%%%%%%%%%%%%%%%%%%%%%%%%%%%%%%%%%%%%%%%%%%%%%%%%%%%%%%%%%%%%%%%%%%%%

\section{Multiple scalar fields}\label{sec:sym}

So far, we have been concerned with the case of a single scalar field theory.
In this section, we show how the previous results generalize to 
theories with multiple fields. In particular, this will demonstrate 
that all the results established 
in the previous sections hold, provided all counterterms allowed by the 
symmetries of the theory are included. 

\subsection{Symmetries}
\label{sec:symmetries}

As an example, we consider the case of an $O(N)$-symmetric theory.
The 2PI functional is symmetric under simultaneous rotations of the
field $\phi_R^a(x)$ and of the propagator $G_R^{ab}(x,y)$:\footnote{In the 
present section, we use Latin letters to denote $O(N)$ indices and write 
space-time and/or momentum variables explicitly when needed.}
\bea
 \phi_R^a(x)&\to&{\mathcal R}^{ab}\phi_R^{b}(x)\\
  G_R^{ab}(x,y)&\to&{\mathcal R}^{ac}{\mathcal R}^{bd} G_R^{cd}(x,y)
\eea
where ${\mathcal R}$ denotes an arbitrary $O(N)$ rotation.
The generalization of \Eqn{eq:shift} to the $O(N)$ case, therefore, 
has the general structure:\footnote{For the $N$-component theory we
employ the classical bare interaction 
term $\lambda(\varphi_a\varphi_a)^2/4!N$.}
\bea
 \Gamma_\inter^R[\phi_R,G_R]&=&\gamma_\inter[\phi_R,G_R]\nn
 &-&\int_x\,\Big\{\frac{\Dla_0}{4!N}\,G_R^{aa}(x,x)\,G_R^{bb}(x,x)
 +\frac{\Dlb_0}{12N}\,G_R^{ab}(x,x)\,G_R^{ab}(x,x)\nn
 &+&\frac{\Dla_2}{12N}\,G_R^{aa}(x,x)\phi_R^b(x)\phi_R^b(x)
 +\frac{\Dlb_2}{6N}\,G_R^{ab}(x,x)\phi_R^a(x)\phi_R^b(x)\nn
 &+&\frac{\Dl_4}{4!N}\,\phi_R^a(x)\phi_R^a(x)\phi_R^b(x)\phi_R^b(x)\Big\}\,.\nn
\label{app:cct}
\eea
Extracting the mass and field counterterms as in \Eqn{eq:mfsct}, we write:
\bea
 \gamma_\inter[\phi_R,G_R]&=&\tilde\gamma_\inter[\phi_R,G_R]
 -\frac{1}{2}\int_x\,\Big\{\phi_R^a(x)\left[\dZ_0\square_x+\dmsq_0\right]\phi_R^a(x)\nn
\label{eq:mfsct2}
 &&+\left[\dZ_2\square_x+\dmsq_2\right]G_R^{aa}(x,y)|_{y=x}\Big\}\,.
\eea
The pairs of counterterms $\Dla_0$ and $\Dla_2$ as well as 
$\Dlb_0$ and $\Dlb_2$ are independent of each other and may have 
different expressions for a given approximation. As for the single scalar field, 
their values are determined by imposing renormalization conditions on the 
various components of the renormalized functions $\bV_R$ and 
$V_R$. The latter satisfy trivial generalizations of the integral 
equations \eqn{eq:BSsymb} and \eqn{eq:intVsymb}, respectively, where the subscript
$1,2,\dots$ now denote space-time variables as well as internal $O(N)$ indices. 

As an illustration, we consider the renormalization of $\bV_R$ 
at $\phi_R^a=0$. Exploiting the 
symmetries\footnote{The relevant symmetry relations are
$\bLambda_R^{ab;cd}(x,y;z,t)=\bLambda_R^{ba;cd}(y,x;z,t)=\bLambda_R^{ab;dc}
(x,y;t,z)$, and similarly for $\bV_R$, $\Lambda_R$ and $V_R$. The functions 
$\bLambda_R$ and $\bV_R$ possess the further property 
$\bLambda_R^{ab;cd}(x,y;z,t)=\bLambda_R^{cd;ab}(z,t;x,y)$, 
and similarly for $\bV_R$. This implies the following relations 
for the various components in \Eqn{components}: 
$\bLama_R(x,y;z,t)=\bLama_R(z,t;x,y)$, and similarly for $\bLamb_R$, 
as well as $\bVa_R$ and $\bVb_R$.} of the functions 
$\bLambda_R\equiv\bLambda_R^{ab;cd}(x,y;z,t)$ 
and $\bV_R\equiv\bV_R^{ab;cd}(x,y;z,t)$, we write
\bea
\label{components}
 \bLambda_R^{ab;cd}(x,y;z,t)&=&\delta^{ab}\delta^{cd}\,\bLama_R(x,y;z,t)\nn
 &+&\delta^{ac}\delta^{bd}\,\bLamb_R(x,y;z,t)+\delta^{ad}\delta^{bc}\,\bLamb_R(x,y;t,z)\,,
\eea
with $\bLama_R(x,y;z,t)=\bLama_R(y,x;z,t)=\bLama_R(x,y;t,z)$
and $ \bLamb_R(x,y;z,t)=\bLamb_R(y,x;t,z)$,
and similarly for $\bV_R^{ab;cd}(x,y;z,t)$. One also has
\beq
\label{Gcomp}
 \bG_R^{ab}(x,y)=\delta^{ab}\,\bG_R(x,y)\,,
\eeq 
with $\bG_R(x,y)=\bG_R(y,x)$. Using Eqs.~\eqn{components} and \eqn{Gcomp}, one 
obtains the integral equations satisfied by the various components 
$\bVa_R$ and
$\bVb_R$ from the generalization of \Eqn{eq:BSsymb} to the $O(N)$ case. 
For notational convenience, we write the latter in terms of bare quantities
\bea
\label{app:Va}
 \bVa&=&\bLama+\frac{iN}{2}\bVa\bG^2\bLama+i\bVa\bG^2\bLamb+i\bVb\bG^2\bLama\nn
 &=&\bLama+\frac{iN}{2}\bLama\bG^2\bVa+i\bLamb\bG^2\bVa+i\bLama\bG^2\bVb\,,
\eea
and
\beq
\label{app:Vb}
 \bVb=\bLamb+i\bVb\bG^2\bLamb=\bLamb+i\bLamb\bG^2\bVb\,.
\eeq
The corresponding equations for renormalized quantities are obtained using the
relations $G=ZG_R$ as well as $Z^2\bLama=\bLama_R-\Dla_0$ and $Z^2\bVa=\bVa_R$, 
and similarly for $\bLamb$ and $\bVb$. Equation~\eqn{app:Vb} has the same structure 
as \Eqn{eq:BSsymb} for the case $N=1$ discussed in this paper. It can, therefore, be 
made finite by the single shift $\Dlb_0$. Equation \eqn{app:Va} mixes the various kernels 
$\bLama_R$ and $\bLamb_R$ and the associated counterterms $\Dla_0$ and $\Dlb_0$. Iterating
this integral equation in powers of the kernels $\bLama_R$ and $\bLamb_R$
and exploiting, as before, the two-particle-irreducibility of the latter, one
finds that, once $\Dlb_0$ has been adjusted to renormalize $\bVb_R$, all the 
remaining divergences can be absorbed in the shift $\Dla_0$. 
Alternatively, employing the same steps as for the case $N=1$ 
(cf.~\Eqn{eq:BSren}), one 
obtains the following finite equation for 
$\bVa_R(p,k)\equiv\bVa_R(p,-p,-k,k)$ in Euclidean momentum space:
\bea
 &&\hspace*{-0.5cm}\bVa_R(p,k)-\bVa_R(\tp,\tp) = \bLama_R(p,k)-\bLama_R(\tp,\tp)\nn
 &&\quad+\frac{N}{2}\,\int_q\,\Big\{\bVa_R(p,q)\,\bG_R^2(q)\,\Delta\bLama_R(k,q)
 +\Delta\bLama_R(p,q)\,\bG_R^2(q)\,\bVa_R(\tp,q)\Big\}\nn
 &&\qquad+\int_q\,\Big\{\bVa_R(p,q)\,\bG_R^2(q)\,\Delta\bLamb_R(k,q)
 +\Delta\bLama_R(p,q)\,\bG_R^2(q)\,\bVb_R(\tp,q)\Big\}\nn
 &&\qquad+\int_q\,\Big\{\bVb_R(p,q)\,\bG_R^2(q)\,\Delta\bLama_R(k,q)
 +\Delta\bLamb_R(p,q)\,\bG_R^2(q)\,\bVa_R(\tp,q)\Big\},\nn
\eea
where $\Delta\bLama_R(p,q)\equiv\bLama_R(p,q)-\bLama_R(\tp,q)$ and similarly for
$\Delta\bLamb_R(p,q)$. The counterterm $\Dla_0$
is replaced by the finite number $\bVa_R(\tp,\tp)$. The renormalization of the function 
$V_R^{ab;cd}(x,y;z,t)$ can be treated along similar lines, using the same steps as those 
leading to \Eqn{eq:Vfinite} for the case $N=1$. Once the kernels $\Lama_R$ and $\Lamb_R$ 
have been made finite by the previous BPHZ analysis, the functions $\Va_R$ and $\Vb_R$ 
are renormalized through the shifts $\Dla_2$ and $\Dlb_2$. 

In the exact theory, one has $\bVa_R=\bVb_R=\Va_R=\Vb_R$. These 
relations must be maintained when imposing renormalization conditions,
which is necessary 
for the employed approximation scheme to converge to the correct theory. For 
instance, one may demand that the four-point functions at a given 
renormalization point $p_1=p_2=\dots=\tp$, be fixed by the following 
condition:\footnote{We note that for the 
$1/N$-expansion to leading order (see below), 
one has $\bLamb=\Lamb=0$ and, 
therefore, $\bVb=\Vb=0$ and $\Dlb_0=\Dlb_2=0$.}
\beq
\label{app:rencond}
 \Gamma^{(4)}_R(\tp_i)=\bVa_R(\tp_i)=\bVb_R(\tp_i)
 =\Va_R(\tp_i)=\Vb_R(\tp_i)=-\frac{\lambda_R}{3N}\,,
\eeq
where $\Gamma^{(4)}_R$ denotes the $O(N)$-invariant component of the 
renormalized four-point vertex function at vanishing field, defined as:
\beq
\label{4pton}
 \Gamma^{(4)\,abcd}_R(p_1,\ldots,p_4)=\Big(\delta^{ab}\delta^{cd}
 +\delta^{ac}\delta^{bd}+\delta^{ad}\delta^{bc}\Big)\,\Gamma^{(4)}_R(p_1,\ldots,p_4)\,.
\eeq
The latter is related to 2PI kernels and the functions $\bV_R$ by 
the generalization of \Eqn{eq:4pt2symb} to the $O(N)$ case. 
As for the case $N=1$, 
once the kernel $\delta^4\gamma_\inter^R/\delta\phi_R^4|_{\bG_R}$ 
has been made finite by means of the BPHZ analysis applied to diagrams 
with resummed 
propagators, the renormalization of the four-point vertex \eqn{4pton} 
is achieved by the shift $\Dl_4$. 

Finally, we mention that, similarly to the case $N=1$, the four-point vertex 
$\Gamma_R^{(4)}$ can be given a particularly simple expression for approximations 
where\footnote{For instance, this is the case of the 2PI coupling and $1/N$-expansions 
described in the next subsections. We emphasize that this relation is satisfied in the 
exact theory, as shown in Appendix~\ref{sec:appcorrel}.}  
\beq
\label{relationker}
 \frac{\delta^2\Gamma^R_\inter}{\delta\phi_R^a(x)\delta\phi_R^b(y)}\Big|_{\phi_R=0}
 = 2\frac{\delta \Gamma^R_\inter}{\delta G_R^{ab}(x,y)}\Big|_{\phi_R=0}\,.
\eeq
In this case, one has $\Gamma_R^{(2)}=i\bG_R^{-1}$ and $\bV_R=V_R$, as shown in 
Appendix~\ref{sec:appcorrel}. In particular, this implies that $\dZ_0=\dZ_2$ and
$\dmsq_0=\dmsq_2$ as well as $\Dla_0=\Dla_2$ and $\Dlb_0=\Dlb_2$.
Moreover, the four-point vertex reads, in real space:
\bea
\label{eq:sanslabel}
 \Gamma_R^{(4)}(x,y;z,t)&=&\gamma_\inter^{(4)}(x,y;z,t)-\Dl_4\nn
 &+&\Big(\bVa_R(x,y;z,t)-\bLama_R(x,y;z,t)+\Dla_0\Big)\nn
 &+&\Big(\bVb_R(x,z;y,t)-\bLamb_R(x,z;y,t)+\Dlb_0\Big)\nn
 &+&\Big(\bVb_R(x,t;z,y)-\bLamb_R(x,t;z,y)+\Dlb_0\Big)\,,
\eea
where $\gamma_\inter^{(4)}$ denotes the $O(N)$-invariant component of the 
kernel $\delta^4\gamma_\inter/\delta\phi_R^4|_{\bG_R}$, as in \eqn{4pton}. 
In that case, the renormalization of $\Gamma_R^{(4)}$ is trivial. 
In particular, one immediately concludes that $\Dl_4=\Dla_0+2\Dlb_0$.

\subsection{The renormalized 2PI coupling-expansion to order $\lambda_R$:
Hartree approximation}

In the previous subsection, we have emphasized that renormalization
requires taking into account
all counterterms consistent with the symmetries of the theory. As an illustration,
we consider here the simplest non-trivial approximation where this plays a role,
namely the 2PI coupling expansion to order $\lambda_R$. This corresponds to the 
so-called Hartree approximation and has been extensively discussed 
in the literature \cite{Hartree}.

We define the renormalized 2PI coupling-expansion as in
Eqs.~\eqn{app:cct}-\eqn{eq:mfsct2}, with
\beq
 \tilde\gamma_\inter[\phi_R,G_R]=\tilde\gamma_\inter^{(\lambda_R)}[\phi_R,G_R] 
 +\tilde\gamma_\inter^{(\lambda_R^2)}[\phi_R,G_R]+\ldots\,,
\eeq
where $\tilde\gamma_\inter^{(\lambda_R^n)}[\phi_R,G_R]$ denotes the ${\cal O}(\lambda_R^n)$
contribution. The first non-trivial contribution is given by
\bea
 \tilde\gamma_\inter^{(\lambda_R)}[\phi_R,G_R]&=&-\frac{\lambda_R}{4!N}\int_x\,
 \Big\{G_R^{aa}(x,x)G_R^{bb}(x,x)+2G_R^{ab}(x,x)G_R^{ab}(x,x)\nn
 &&\quad+2G_R^{aa}(x,x)\phi_R^b(x)\phi_R^b(x)+4G_R^{ab}(x,x)\phi_R^a(x)\phi_R^b(x)\nn
 &&\quad+\phi_R^a(x)\phi_R^a(x)\phi_R^b(x)\phi_R^b(x)\Big\}\,.\nn
\label{lambdaR}
\eea
We note that it is enough to choose $\Dla_0=\Dla_2\equiv\Dla$ 
and $\Dlb_0=\Dlb_2\equiv\Dlb$ as well as 
$\dZ_0=\dZ_2$ and $\dmsq_0=\dmsq_2$ in Eqs.~\eqn{app:cct}-\eqn{eq:mfsct2} 
since, in that case, the relation \eqn{relationker} is satisfied.
As explained above (cf. \Eqn{eq:sanslabel}), this also implies that $\Dl_4=\Dla+2\Dlb$. 
Notice also that at the present order of approximation, there is no need for 
BPHZ contributions to the various counterterms in \Eqn{lambdaR}. Therefore, 
the relevant counterterms are entirely given by the shifts in \Eqn{app:cct}: 
$\Dla$, $\Dlb$ and $\Dl_4$.

At this level of approximation, the kernel $\bLambda_R$ is simply given by the 
renormalized tree-level four-point vertex. One has, in Euclidean momentum space:
\beq
 \bLama_R(p_1,\ldots,p_4)=\bLamb_R(p_1,\ldots,p_4)=-\frac{\lambda_R}{3N}\,.
\eeq
As a consequence, the various components of the resummed four-point function 
$\bV_R$ have the form
\beq
 \bVa_R(p_1,\ldots,p_4)\equiv\bva_R(p_1+p_2)\,,
\eeq
and similarly for $\bVb_R$. The integral equations~\eqn{app:Va}
and \eqn{app:Vb} reduce to 
the following algebraic equations for the functions $\bva_R$ and $\bvb_R$:
\bea
 \bva_R(p)&=&Z^2\bLama-iZ^2\bLama\bPi(p)\bva_R(p)\nn
\label{bva}
 &&-\frac{2i}{N}Z^2\bLamb\bPi(p)\bva_R(p)
 -\frac{2i}{N}Z^2\bLama\bPi(p)\bvb_R(p)\,,
\eea
and
\beq
\label{bvb}
 \bvb_R(p)=Z^2\bLamb-\frac{2i}{N}Z^2\bLamb\bPi(p)\bvb_R(p)\,,
\eeq
where we defined the one-loop bubble as (note that 
one factor of $i$ is absorbed in the Euclidean momentum
integral)
\beq 
\label{eq:oneloopbubble}
 i\bPi(p)=-\frac{N}{2}\int_q \bG_R(q)\bG_R(q+p)\,.
\eeq
Here, we used the relations $Z^2\bLama=\bLama_R-\Dla$ and $Z^2\bLamb=\bLamb_R-\Dlb$ 
between renormalized and bare quantities to simplify notations. In terms of the function
$\bva_R$ an $\bvb_R$, the renormalization conditions \eqn{app:rencond} read
\beq
 \bva_R(p=2\tp)=\bvb_R(p=2\tp)=-\frac{\lambda_R}{3N}\,.
\eeq
Evaluating Eqs.~\eqn{bva} and \eqn{bvb} at $p=2\tp$, one can solve for 
the counterterm $\Dla$ and $\Dlb$, or, equivalently, for the bare kernels:
\beq
\label{ctb}
 \frac{1}{Z^2\bLamb}=-\frac{3N}{\lambda_R}-\frac{2i}{N}\bPi(2\tp)\,,
\eeq
and
\beq
\label{cta}
 \frac{\bLamb}{\bLama}=1+i\frac{\lambda_R}{3N}\frac{N+2}{N}\,\bPi(2\tp)\,.
\eeq
In particular, we note that $\bLama\neq\bLamb$, or, equivalently, $\Dla\neq\Dlb$.
Writing \Eqn{bvb} as
\beq
 \frac{1}{\bvb_R(p)}=\frac{1}{Z^2\bLamb}+\frac{2i}{N}\bPi(p)\,,
\eeq
and using \Eqn{ctb}, one obtains the following finite equation:
\beq
 \frac{1}{\bvb_R(p)}=-\frac{3N}{\lambda_R}+\frac{2i}{N}\Delta\bPi(p,2\tp). 
\eeq
where we introduced the finite function
\beq
\label{eq:finitepi}
 i\Delta\bPi(p,k)=i\bPi(p)-i\bPi(k)
 =-\frac{N}{2}\int_q\bG_R(q)\,\Big\{\bG_R(q+p)-\bG_R(q+k)\Big\}\,,
\eeq
which grows as $\ln p$ at large $p$ and fixed $k$. 
Similarly, using \Eqn{bvb}, \Eqn{bva} can be written as:
\beq
 \frac{\bvb_R(p)}{\bva_R(p)}=\frac{\bLamb}{\bLama}+iZ^2\bLamb
 \left(1+\frac{2}{N}\frac{\bLamb}{\bLama}\right)\bPi(p)\,.
\eeq
Using the expressions \eqn{ctb} and \eqn{cta} for the bare kernels, one
finally obtains the following equation: 
\beq
 \frac{\bvb_R(p)}{\bva_R(p)}=1-i\frac{\lambda_R}{3N}\frac{N+2}{N}\,\Delta\bPi(p,2\tp)\,,
\eeq
which is finite, as desired. This simple example illustrates the importance
of allowing for all possible counterterms consistent with symmetries. In particular,
the present Hartree approximation cannot be made finite by imposing $\Dla=\Dlb$.

\subsection{The renormalized 2PI $1/N$-expansion}
\label{sec:appN}

In the previous subsections, we have seen that the $\Dl$--counterterms 
allowed by symmetries 
and power counting are sufficient to absorb the coupling sub-divergences generated by
iterations of the vertex equations for the various components of the four-point
functions $\bV_R$, $V_R$ and $\Gamma_R^{(4)}$. This assumes that the functions $\bLambda_R$,
$\Lambda_R$ and $\gamma_\inter^{(4)}(\bG_R)$ have been made finite by means of the 2PI BPHZ
procedure described previously. Here, we show in detail 
how the latter proceeds in practice
for the case of the 2PI $1/N$-expansion \cite{Berges:2001fi,Aarts:2002dj}. In particular,
we show that, in the next-to-leading order (NLO) approximation, the BPHZ procedure can be explicitely 
reformulated as a single renormalization condition.

The renormalized 2PI $1/N$-expansion can be 
written as in Eqs.~\eqn{app:cct}-\eqn{eq:mfsct2}, with
\beq
\label{define1}
 \tilde\gamma_\inter[\phi_R,G_R]=\tilde\gamma_\inter^{\rm LO}[\phi_R,G_R] 
 +\tilde\gamma_\inter^{\rm NLO}[\phi_R,G_R]+\ldots
\eeq
where $\gamma_\inter^{\rm LO}$ is the leading-order (${\mathcal O}(N)$) contribution,
$\gamma_\inter^{\rm NLO}$ the next-to-leading--order (${\mathcal O}(1)$) contribution, etc. 
The first two contributions read 
explicitly \cite{Cornwall:1974vz,Berges:2001fi,Aarts:2002dj}:
\bea
 &&\tilde\gamma_\inter^{\rm LO}[\phi_R,G_R]=-\frac{\lambda_R}{4!N}\int_x\,
 \phi_R^a(x)\phi_R^a(x)\phi_R^b(x)\phi_R^b(x)\nn
 &&\qquad\quad-\frac{\lambda_R+\dl^{\rm LO}}{4!N}\int_x\,
 \Big\{2G_R^{aa}(x,x)\phi_R^b(x)\phi_R^b(x)+G_R^{aa}(x,x)G_R^{bb}(x,x)\Big\}\,,\nn
\label{effacLO}
\eea
and
\beq 
 \tilde\gamma_\inter^{\rm NLO}[\phi_R,G_R]=\frac{i}{2}\Tr{\rm Ln}\,{\bf B}(G_R)
 +\frac{i}{2}\int_{xy}{\bf D}_0(x,y)\,\phi_R^a(x)G_R^{ab}(x,y)\phi_R^b (y)\,.
\eeq
where
\beq
 {\bf B}(x,y) = \delta^{(4)}(x-y)-i\frac{\lambda_R+\dl^{\rm NLO}}{3N}\Pi(x,y)\,,
\eeq
with the one-loop bubble
\beq
\label{bubble}
 \Pi(x,y) = -\frac{1}{2} G_R^{ab}(x,y)G_R^{ab}(x,y)\,.
\eeq
The function ${\bf D}_0$ is defined as:\footnote{In terms of the (bare) function 
${\bf D}(\phi)$ introduced in Ref.~\cite{Aarts:2002dj} 
one has ${\bf D}_0=Z^2{\bf D}(\phi=0)$.}
\beq
\label{Dfunction}
 {\bf D}_0(x,y)
 =i\frac{\lambda_R+\dl^{\rm NLO}}{3N}{\bf B}^{-1}(x,y)\,,
\eeq
and satisfies the following integral equation:
\beq
\label{Deq}
 {\bf D}_0(x,y)=i\frac{\lambda_R+\dl^{\rm NLO}}{3N}\delta^{(4)}(x-y)
 +i\frac{\lambda_R+\dl^{\rm NLO}}{3N}\int_z\Pi(x,z){\bf D}_0(z,y)\,.
\eeq
In other words, ${\bf D}_0$ resums the infinite chain of bubbles given by $\Pi$
(cf.~\Eqn{bubble}).
Equations \eqn{define1} to \eqn{Dfunction} above, together with \Eqn{app:cct}, 
define the renormalized 2PI $1/N$-expansion at NLO. As for the case
of the Hartree approximation discussed previously, it is enough to choose 
$\Dla_0=\Dla_2$ and $\Dlb_0=\Dlb_2$ as well as $\dZ_0=\dZ_2$ and 
$\dmsq_0=\dmsq_2$ in Eqs.~\eqn{app:cct}-\eqn{eq:mfsct2} since, in that case, the
relation \eqn{relationker} between 2PI kernels is satisfied in the present 
approximation as well.
We note also that at NLO, there is no other quartic contribution in the 
field $\phi_R$ beyond the interaction term in the classical action. Consequently, 
there is no need for a BPHZ contribution to the corresponding counterterm in the 
first term on the RHS of \Eqn{effacLO}. Therefore, one has, for the counterterm 
associated with the classical contribution to $\Gamma_\inter^R[\phi_R,G_R]$:
$\dl_4=\Dl_4=\Dla_0+2\Dlb_0$, where the last equality follows from \Eqn{relationker},
as discussed above..

Here, we discuss the renormalization of the four-point kernel $\bLambda_R$ at $\phi_R=0$, 
defined as:
\beq
 \bLambda_R^{ab;cd}(x,y;z,t)=4\frac{\delta^2\gamma_\inter[\phi_R=0,G_R]}
	{\delta G_R^{ab}(x,y)\delta G_R^{cd}(z,t)}\Big|_{\bG_R}\,.
\eeq
Once the latter is finite, the renormalization of the four-point functions $\bV_R$
and $\Gamma^{(4)}_R$ goes along the lines described in the previous section and is
achieved through the shifts $\Dla_0$ and $\Dlb_0$ for the former and $\Dl_4$ for
the latter. It is sufficient to discuss renormalization at $\phi_R^a=0$, where
one can write $\bG_R^{ab}(x,y)=\delta^{ab}\,\bG_R(x,y)$. Using the expansion 
\eqn{define1}, the renormalized kernel can be written as
\beq
\label{NLOkernel}
 \bLambda_R=\bLambda^{\rm LO}+\bLambda^{\rm NLO}+\ldots\,.
\eeq
where the LO and NLO contributions read explicitly, in Euclidean momentum space:
\beq
\label{LOker}
 \bLambda^{\rm LO}_{ab;cd}(p_1,\ldots,p_4)=
 -\frac{\lambda_R+\dl^{\rm LO}}{3N}\delta^{ab}\delta^{cd}
\eeq
and
\bea
 \bLambda^{\rm NLO}_{ab;cd}(p_1,\ldots,p_4)&=&
 2\delta^{ab}\delta^{cd}\int_q \bbD_0(q)\bG(q+p_1)\bbD_0(q+p_1+p_2)\bG(q-p_3)\nn
\label{BSkernelNLO}
 &&+i\delta^{ac}\delta^{bd}\bbD_0(p_1+p_3)+i\delta^{ad}\delta^{bc}\bbD_0(p_1+p_4)
\eea
where we introduced the notation $\bbD_0\equiv{\bf D}_0(G_R=\bG_R)$. The NLO
contribution contains explicitly divergent momentum integrals, as well as
implicit ones in the function $\bbD_0$. To make the kernel \eqn{NLOkernel} 
finite at NLO, we first discuss the renormalization of the latter.

From \Eqn{Deq}, written at $G_R=\bG_R$, one gets
\beq
\label{bubbleresum}
 \frac{1}{i\bbD_0(p)}=-\frac{3N}{\lambda_R+\dl^{\rm NLO}}+i\bPi(p). 
\eeq
where $i\bPi(p)$ is the one-loop bubble defined in \Eqn{eq:oneloopbubble}.
The counterterm $\delta\lamNLO$ is easily eliminated with
\beq
 \frac{1}{i\bbD(p)}=\frac{1}{i\bbD(\tilde p)}+i\Delta\bPi(p,2\tp). 
\eeq
where the function $\Delta\bPi(p,k)$ has been defined in \Eqn{eq:finitepi}. It is finite
and grows as $\ln p$ at large $p$ and fixed $k$. Therefore, the function $\bbD_0$ is made
finite by the counterterm $\dl^{\rm NLO}$. 
The latter is determined by imposing a suitable renormalization condition, e.g.\ by requiring
that the number $\bbD(p=2\tp)$ be finite. The relevant renormalization
condition is
\beq
\label{renD}
 i\bbD(p=2\tp)=-\frac{\lambda_R}{3N}\,,
\eeq
Indeed, using the LO expression \Eqn{LOker} for the kernel $\bLambda_R$, 
one finds that at 
LO in the $1/N$-expansion the four-point function $\bV_R$ satisfies the very same
integral equation as the function $i\bbD_0$ (cf.~\Eqn{Deq}), with propagators replaced
by the solution of the gap equation at LO: $\bG_R\to\bG_R^{\rm (LO)}$. In fact, one 
can write, at LO: $\bV_R^{\rm (LO)}(p_1,\ldots,p_4)=i\bbD_0^{\rm (LO)}(p_1+p_2)$
where $\bbD_0^{\rm (LO)}\equiv{\bf D}_0(G_R=\bG_R^{\rm (LO)})$. 
Therefore, \Eqn{renD} 
corresponds to the requirement that the renormalization condition for the four-point 
function $\bV_R$ be satisfied order by order in the 2PI $1/N$-expansion.

Once the function $\bbD_0$ has been made finite, there remains only one global 
divergence in the expression of the kernel $\bLambda_R$, due to the momentum 
integral in \Eqn{BSkernelNLO}. The latter is trivially absorbed in the counterterm
$\dl^{\rm LO}$ appearing in the LO contribution \eqn{LOker}, thus making the kernel
$\bLambda_R$ finite.

%%%%%%%%%%%%%%%%%%%%%%%%%%%%%%%%%%%%%%%%%%%%%%%%%%%%%%%%%%%%%%%%%%%%%%%%%%%%%%%%%%%%%%%%%%%%%%%%%%%%%%%%%%%%%%%%%%%%%%%%%%%%%%%%%%%%%%%%%%%%%%%%%%%%%%   Conclusions   %%%%%%%%%%%%%%%%%%%%%%%%%%%%%%%%%%%%%%%%%%%%%%%%%%%%%%%%%%%%%%%%%%%%%%%%%%%%%%%%%%%%%%%%%%%%%%%%%%%%%%%%%%%%%%%%%%%%%%%%%%%%%%%%%%%%%%%%%%%%%%%%%%%%%%%%%%%%%%%%%%%%%%%%%%%%%%%%%%%%%
\section{Conclusions}
\label{Sec:summaryconclusions}

In this work the renormalization for the 2PI-resummed 
effective action $\Gamma[\phi_R]\equiv \Gamma_{\rm 2PI}[\phi_R,\bG_R(\phi_R)]$ 
has been presented. 
The remarkable result of the present analysis is that, as far as
renormalization is concerned, all the complications due to selective summations of 
an infinite set of perturbative diagrams in the 2PI scheme boil down to 
adjusting a finite set of coupling counterterms. 
These correspond to all local, mass-dimension four
operators allowed by the symmetries of the theory (cf.~\Eqn{eq:shift}
or \Eqn{app:cct}). These include the mass-dimension four 
operator that has been previously discussed in 
Refs.~\cite{vanHees:2001ik,Blaizot:2003br}.
As a consequence of two-particle irreducibility 
there is only a limited number (two) of places where mass and 
field-strength counterterms can appear. All these counterterms can be 
computed from the renormalization conditions 
\eqn{eq:renormcondl}-\eqn{eq:renormcondZ}. In this sense, we find
that renormalization for 2PI approximation schemes is remarkably similar to 
renormalization in standard perturbation theory. 

The 2PI-resummed effective action $\Gamma[\phi_R]$ is the generating functional for 
proper vertices, obtained by derivatives with respect to the field $\phi_R$. Its importance 
stems from the fact that the proper vertices respect all symmetry properties and, in 
particular, Goldstone's theorem in the phase with spontaneous symmetry breaking.
We emphasize that, since Ward identities may not be manifest at 
intermediate calculational steps of the 2PI resummation scheme, 
a consistent renormalization procedure for the 2PI-resummed 
effective action $\Gamma[\phi_R]$ is crucial in order to obtain 
the correct symmetry properties for physical results. To have achieved
this is an important step towards the renormalization
for 2PI-resummed effective actions in gauge theories. 
The application of our techniques to the latter 
will be presented in a separate work~\cite{BBRS}.

%%%%%%%%%%%%%%%%%%%%%%%%%%%%%%%%%%%%%%%%%%%%%%%%%%%%%%%%%%%%%%%%%%%%%%%%%%%%%%%%%%%%%%%%%%%%%%%%%%%%%%%%%%%%%%%%%%%%%%%%%%%%%%%%%%%%%%%%%%%%%%%%%%%%%%   Acknowledgements   %%%%%%%%%%%%%%%%%%%%%%%%%%%%%%%%%%%%%%%%%%%%%%%%%%%%%%%%%%%%%%%%%%%%%%%%%%%%%%%%%%%%%%%%%%%%%%%%%%%%%%%%%%%%%%%%%%%%%%%%%%%%%%%%%%%%%%%%%%%%%%%%%%%%%%%%%%%%%%%%%%%%%%%%%%%%%%%%%%%%%
\section*{Acknowledgments}

We thank J.-P. Blaizot, E. Iancu, A. Rebhan and N. Wschebor for fruitful
collaborations/discussions on these topics. U. Reinosa acknowledges support from
the Austrian Science Foundation FWF, project no. P16387-N08.
APC is unit\'e mixte de recherche UMR7164 (CNRS, Universit\'e Paris 7, CEA, 
Observatoire de Paris). LPT is UMR8627 (CNRS, Universtit\'e Paris-Sud~11).

\appendix

\section{Correlation functions from the 2PI effective action}
\applabel{sec:appcorrel}

The generating functional $W[J,K]$ for connected Green's function in the
presence of source terms linear and bilinear in the fields is given in terms
of the functional integral
\beq 
 \e^{iW[J,K]}=\int{\mathcal D}\varphi\,\e^{i({\mathcal S}[\varphi]+J_a\varphi_a+
   \frac{1}{2}K_{ab}\varphi_a\varphi_b)}\,.
\eeq
The connected one- and two-point functions, $\phi$ and $G$ 
in presence of sources are defined through
\bea
 \frac{\delta W}{\delta J_1}&=&\phi_1\\
 \frac{\delta W}{\delta K_{12}}&=&\frac{1}{2}\left(G_{12}+\phi_1\phi_2\right)\,,
\eea
In the exact theory, one has the obvious relation
\beq
\label{eq:W_JJ}
 \frac{\delta^2 W}{\delta J_1\delta J_2}=iG_{12}\,.
\eeq

The 2PI effective action is defined as the double Legendre transform of the functional
$W[J,K]$:
\beq
 \Gammatpi[\phi,G]=W[J,K]-J_a\frac{\delta W}{\delta J_a}
 -K_{ab}\frac{\delta W}{\delta K_{ab}}\,.
\eeq
One has, in particular
\bea
 \frac{\delta \Gammatpi}{\delta \phi_1}&=&-J_1-K_{1a}\phi_a\\
 \frac{\delta \Gammatpi}{\delta G_{12}}&=&-\frac{1}{2}K_{12}\,.
\eea
In the following we concentrate
on a $Z_2$-symmetric scalar theory. 
In that case, all derivatives of the 2PI effective action with respect to an
odd number of fields vanish at
$\phi=0$. Expressing the fact that the Jacobian of the double Legendre 
transformation is unity, one obtains the following relations at $\phi=0$:
\bea
\label{eq:W_JJ_inv}
 \frac{\delta^2W}{\delta J_1\delta J_a}
 \left(\frac{\delta^2\Gammatpi}{\delta\phi_a\delta\phi_2}
 -2\frac{\delta \Gammatpi}{\delta G_{a2}}\right)\Big|_{\phi=0}&=&-\delta_{12}\, 
,\\
\label{eq:W_KK_inv}
 \left(2\frac{\delta^2W}{\delta K_{12}\delta K_{ab}}\right)
 \left(2\frac{\delta^2\Gammatpi}{\delta G_{ab}\delta G_{34}}\right)\Big|_{\phi=0}
 &=&-\frac{\delta_{13}\delta_{24}+\delta_{14}\delta_{23}}{2}\,.
\eea
From the parametrization \eqn{eq:2PI} of the 2PI effective action, one has
\beq
 \frac{\delta^2\Gammatpi}{\delta\phi_1\delta\phi_2}
 -2\frac{\delta \Gammatpi}{\delta G_{12}}=iG^{-1}_{12}
 +\frac{\delta^2\Gamma_\inter}{\delta\phi_1\delta\phi_2}
 -2\frac{\delta \Gamma_\inter}{\delta G_{12}}\,.
\eeq
It follows, using Eqs.~\eqn{eq:W_JJ} and \eqn{eq:W_JJ_inv}, that:
\beq
\label{eq:2pt2PI}
  \frac{\delta^2\Gamma_\inter}{\delta\phi_1\delta\phi_2}\Big|_{\phi=0}=
  2\frac{\delta \Gamma_\inter}{\delta G_{12}}\Big|_{\phi=0}\,.
\eeq
Writing the above equation for vanishing source $K=0$, that is at $G=\bG$, 
one obtains the following exact relation between the two-point 2PI kernels 
\eqn{eq:se} and \eqn{eq:2kernel2res}:
\beq
 \bSigma_{12}(\phi=0)=\Sigma_{12}(\phi=0)\,.
\eeq
We note that this is equivalent to the fact that the two-point
vertex function $\Gamma^{(2)}$, defined as the second derivative of the
1PI effective action (see e.g. \Eqn{eq:phi4_2pf1}), coincides with the 
inverse propagator:
\beq
 \Gamma^{(2)}_{12}(\phi=0)=i\bG_{12}^{-1}(\phi=0)\,.
\eeq
We stress that this remains true for any 2PI approximation where the 
relation \eqn{eq:2pt2PI} is satisfied.\footnote{Of course, in the exact
theory, the functions $\Gamma^{(2)}(\phi)$ and $i\bG^{-1}(\phi)$ also
coincide for arbitrary $\phi\neq0$. This is, however, not true for
generic approximations. For systematic approximation schemes, 
such as e.g.\ the 2PI $1/N$-expansion, $\Gamma^{(2)}(\phi)$ and 
$i\bG^{-1}(\phi)$ differ in general at higher orders than the truncation 
order~\cite{Aarts:2002dj,vanHees:2001ik,Cooper:2004rs}.}

Differentiating \Eqn{eq:2pt2PI} once with respect to $G$ gives
\beq
\label{eq:4pt2PI}
 \frac{\delta^3\Gamma_\inter}{\delta\phi_1\delta\phi_2\delta G_{34}}\Big|_{\phi=0}
 =2\frac{\delta \Gamma_\inter}{\delta G_{12}\delta G_{34}}\Big|_{\phi=0}\,,
\eeq
from which it follows that the four-point 2PI kernels \eqn{eq:BSkernel} and 
\eqn{eq:4kernel2res} are equal:
\beq
 \bLambda_{12,34}(\phi=0)=\Lambda_{12,34}(\phi=0)\,.
\eeq
This, in turn, implies that the four-point functions $\bV$ and 
$V$, which satisfy the integral equations \eqn{eq:BS} and \eqn{eq:intVsymb} 
respectively are also equal:
\beq
\label{app:bVequalsV}
 \bV_{12,34}(\phi=0)=V_{12,34}(\phi=0)\,.
\eeq

Finally, to exploit \Eqn{eq:W_KK_inv} in a similar way, we write:
\beq
\label{eq:KK}
 \frac{\delta^2W}{\delta K_{12}\delta K_{34}}\Big|_{\phi=0}=
 \frac{i}{4}\left(G_{13}G_{24}+G_{14}G_{23}+C^{(4)}_{1234}(\phi=0)\right)\,,
\eeq
where $C^{(4)}$ is the connected four-point function. It is related to 
the four-point irreducible vertex function $\Gamma^{(4)}$ 
(see e.g. \Eqn{eq:phi4_4pf1}) at $\phi=0$ through
\beq
\label{app:JJJJ}
 iC^{(4)}_{1234}(\phi=0)\equiv
 -\frac{\delta^4W}{\delta J_1\delta J_2\delta J_3\delta J_4}\Big|_{\phi=0}
 = - G_{1a}G_{2b}G_{3c}G_{4d}\Gamma^{(4)}_{abcd}(\phi=0)\,.
\eeq
From the parametrization \eqn{eq:2PI} of the 2PI effective action, one has
\beq
\label{eq:dGG}
 \frac{\delta^2\Gammatpi}{\delta G_{12}\delta G_{34}}=
 \frac{i}{4}\Big(G_{13}^{-1}G_{24}^{-1}+G_{14}^{-1}G_{23}^{-1}\Big)
 +\frac{\delta^2\Gamma_\inter}{\delta G_{12}\delta G_{34}}\,.
\eeq
It follows from Eqs.~\eqn{eq:W_KK_inv}, \eqn{eq:KK} and \eqn{eq:dGG}
that the four-point vertex function satisfies the following integral equation:
\bea
 \Gamma^{(4)}_{1234}(\phi=0)=\frac{4\delta^2\Gamma_\inter}
       {\delta G_{12}\delta G_{34}}\Big|_{\phi=0}
 +\frac{i}{2}\,\Gamma^{(4)}_{12ab}(\phi=0)\,G_{aa'}G_{bb'}\,
 \frac{4\delta^2\Gamma_\inter}{\delta G_{a'b'}\delta G_{34}}\Big|_{\phi=0}\,.\nn
\eea
The above equation, written for $G=\bG(\phi=0)$ 
coincides with the defining equation 
\eqn{eq:BSsymb} for the function $\bV$. Therefore, one has in the exact theory
\beq
\label{app:final}
 \Gamma^{(4)}_{1234}(\phi=0)=\bV_{12,34}(\phi=0)=V_{12,34}(\phi=0)\,.
\eeq
We note that, contrary to the 
relation \eqn{app:bVequalsV}, which is valid whenever 
\Eqn{eq:4pt2PI} is satisfied, this is not the case for the above 
equation for $\Gamma^{(4)}$. Indeed, in
deriving \Eqn{app:final}, we have assumed that the derivatives of the generating
functional $W[J,K]$ in Eqs.~\eqn{eq:KK} and \eqn{app:JJJJ} are related, 
which is typically 
not true for generic approximations.
\begin{figure}[tb]
\begin{center}
\includegraphics[width=10cm]{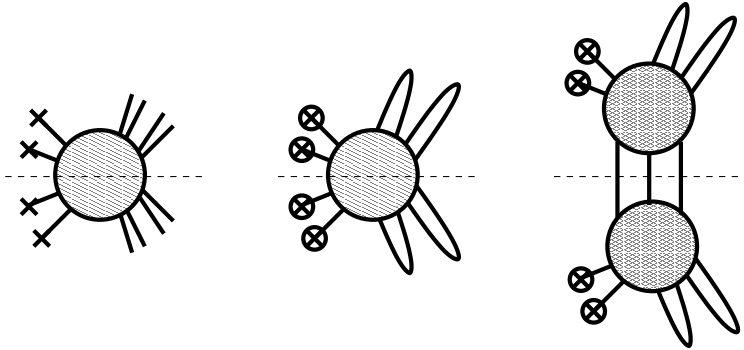}
\caption{\small Two-particle-irreducibility of 2PI kernel: If the 
cut does not separate legs originating from the same original pair, 
it must cut at least three internal lines in the original diagram, 
since the latter originates from a 2PI closed graph in the 2PI effective
action.\label{fig:2PIkernel}}
\end{center}
\end{figure}

\section{Two-particle irreducibility of kernels}
\applabel{2PItop}

Here, we consider a key topological property 
of the 2PI kernels, namely that they are two-particle irreducible in the following 
sense. Consider a given 2PI kernel $\delta^{m+n}\Gamma_\inter/\delta G^m\delta\phi^n|_\bG$
with $2m+n$ legs, as represented on the left of Fig.~\ref{fig:2PIkernel}. 
This kernel arises from a closed 2PI diagram in the 2PI effective action. 
In order to split the latter into two disconnected parts one has to cut 
at least three lines. A similar analysis can be done for the kernel. 
The kernel has $n$ ``single'' legs, corresponding
to derivatives with respect to the field $\phi$, and $m$ ``pairs'' of legs, arising
from derivatives with respect to $G$. After the graph has been cut, part of these 
legs are attached to one piece and the rest to the other piece. In particular, 
if the cut does not separate two legs from a single pair, it must cut at least
three internal lines in the diagram, as illustrated in
Fig.~\ref{fig:2PIkernel}.  This follows 
from the two-particle irreducibility of the diagrams contributing
to the 2PI effective action. As 
already emphasized in Refs.~\cite{vanHees:2001ik,Blaizot:2003br}, 
this property of 2PI 
kernels plays a crucial role in discussing renormalization (see also 
Sec.~\ref{sec:explicit}).


\begin{thebibliography}{10}

%\cite{Braaten:1989mz}
\bibitem{Braaten:1989mz}
  E.~Braaten and R.~D.~Pisarski,
  %``Soft Amplitudes In Hot Gauge Theories: A General Analysis,''
  Nucl.\ Phys.\ B {\bf 337} (1990) 569.
  %%CITATION = NUPHA,B337,569;%%

%\cite{Blaizot:2003tw}
\bibitem{Blaizot:2003tw}
  J.~P.~Blaizot, E.~Iancu and A.~Rebhan,
  %``Thermodynamics of the high-temperature quark gluon plasma,''
  in ``Quark Gluon Plasma 3'', eds.~R.C.\ Hwa and X.N.\ Wang, World Scientific,
  Singapore, 60 [arXiv:hep-ph/0303185].
  %%CITATION = HEP-PH 0303185;%%

%\cite{Blaizot:2000fc}
\bibitem{Blaizot:2000fc}
J.~P.~Blaizot, E.~Iancu and A.~Rebhan,
%``Approximately self-consistent resummations for the thermodynamics of  the
%quark-gluon plasma. I: Entropy and density,''
Phys.\ Rev.\ D {\bf 63} (2001) 065003
[arXiv:hep-ph/0005003];
%%CITATION = HEP-PH 0005003;%%
%\cite{Andersen:2004re}
%\bibitem{Andersen:2004re}
J.~O.~Andersen and M.~Strickland,
%``Three-loop Phi-derivable approximation in QED,''
Phys.\ Rev.\ D {\bf 71} (2005) 025011
[arXiv:hep-ph/0406163].
%%CITATION = HEP-PH 0406163;%%

%\cite{Berges:2004hn}
\bibitem{Berges:2004hn}
  J.~Berges, S.~Bors\'anyi, U.~Reinosa and J.~Serreau,
  %``Renormalized thermodynamics from the 2PI effective action,''
  Phys.\ Rev.\ D {\bf 71} (2005) 105004 
  [arXiv:hep-ph/0409123].
  %%CITATION = HEP-PH 0409123;%%

%\cite{Berges:2003pc}
\bibitem{Berges:2003pc}
  J.~Berges and J.~Serreau,
  ``Progress in nonequilibrium quantum field theory,''
  in SEWM02, ed.\ M.G.~Schmidt (World Scientific, 2003) [arXiv:hep-ph/0302210];
  %%CITATION = HEP-PH 0302210;%%
  %\cite{Berges:2004vw}
%\bibitem{Berges:2004vw}
  %J.~Berges and J.~Serreau,
  %``Progress in nonequilibrium quantum field theory. II,''
  {\it ibid.} SEWM04, eds.\ K.J.~Eskola, K.~Kainulainen,
  K.~Kajantje and K.~Rummukainen (World Scientific) [arXiv:hep-ph/0410330].
  %%CITATION = HEP-PH 0410330;%%

%\cite{Berges:2004yj}
\bibitem{Berges:2004yj}
  J.~Berges,
  ``Introduction to nonequilibrium quantum field theory'',
  AIP Conf.\ Proc.\  {\bf 739} (2005) 3
  [arXiv:hep-ph/0409233].
  %%CITATION = HEP-PH 0409233;%%

%\cite{Berges:2000ur}
\bibitem{Berges:2000ur}
  J.~Berges and J.~Cox,
  %``Thermalization of quantum fields from time-reversal invariant evolution  
  %equations,''
  Phys.\ Lett.\ B {\bf 517} (2001) 369
  [arXiv:hep-ph/0006160].
  %%CITATION = HEP-PH 0006160;%%

%\cite{Berges:2001fi}
\bibitem{Berges:2001fi}
  J.~Berges,
  %``Controlled nonperturbative dynamics of quantum fields out of equilibrium,''
  Nucl.\ Phys.\ A {\bf 699} (2002) 847 [arXiv:hep-ph/0105311].
  %%CITATION = HEP-PH 0105311;%%

%\cite{Cooper:2002qd}
\bibitem{Cooper:2002qd}
  F.~Cooper, J.~F.~Dawson and B.~Mihaila,
   %``Quantum dynamics of phase transitions in broken symmetry lambda phi**4 field
  %theory,''
  Phys.\ Rev.\ D {\bf 67} (2003) 056003
  [arXiv:hep-ph/0209051];
  %%CITATION = HEP-PH 0209051;%%
  %\cite{Juchem:2003bi}
%\bibitem{Juchem:2003bi}
  S.~Juchem, W.~Cassing and C.~Greiner,
  %``Quantum dynamics and thermalization for out-of-equilibrium phi**4-theory,''
  Phys.\ Rev.\ D {\bf 69} (2004) 025006 [arXiv:hep-ph/0307353].

%\cite{Berges:2002wr}
\bibitem{Berges:2002wr}
  J.~Berges, S.~Bors\'anyi and J.~Serreau,
  %``Thermalization of fermionic quantum fields,''
  Nucl.\ Phys.\ B {\bf 660} (2003) 51
  [arXiv:hep-ph/0212404];
  %%CITATION = HEP-PH 0212404;%%
%\cite{Berges:2004ce}
%\bibitem{Berges:2004ce}
  J.~Berges, S.~Bors\'anyi and C.~Wetterich,
  %``Prethermalization,''
  Phys.\ Rev.\ Lett.\  {\bf 93} (2004) 142002
  [arXiv:hep-ph/0403234].
  %%CITATION = HEP-PH 0403234;%%
%\cite{Berges:2002cz}

\bibitem{Berges:2002cz}
  J.~Berges and J.~Serreau,
  %``Parametric resonance in quantum field theory,''
  Phys.\ Rev.\ Lett.\  {\bf 91} (2003) 111601
  [arXiv:hep-ph/0208070].
  %%CITATION = HEP-PH 0208070;%%

%\cite{Aarts:2001yn}
\bibitem{Aarts:2001yn}
  G.~Aarts and J.~Berges,
  %``Classical aspects of quantum fields far from equilibrium,''
  Phys.\ Rev.\ Lett.\  {\bf 88} (2002) 041603
  [arXiv:hep-ph/0107129];
  %%CITATION = HEP-PH 0107129;%%
%\cite{Arrizabalaga:2004iw}
%\bibitem{Arrizabalaga:2004iw}
  A.~Arrizabalaga, J.~Smit and A.~Tranberg,
  %``Tachyonic preheating using 2PI - 1/N dynamics and the classical
  %approximation,''
  JHEP {\bf 0410} (2004) 017
  [arXiv:hep-ph/0409177].
  %%CITATION = HEP-PH 0409177;%%

%\cite{Alford:2004jj}
\bibitem{Alford:2004jj}
  M.~Alford, J.~Berges and J.~M.~Cheyne,
  %``Critical phenomena from the two-particle irreducible 1/N expansion,''
  Phys.\ Rev.\ D {\bf 70} (2004) 125002
  [arXiv:hep-ph/0404059].
  %%CITATION = HEP-PH 0404059;%%

\bibitem{Baym}
  J.~M.~Luttinger and J.~C.~Ward, Phys.\ Rev.\ {\bf 118} (1960) 1417.
  G.~Baym, Phys.\ Rev.\ {\bf 127} (1962) 1391.

%\cite{Knoll:2001jx}
\bibitem{Knoll:2001jx}
  J.~Knoll, Y.~B.~Ivanov and D.~N.~Voskresensky,
  %``Exact Conservation Laws of the Gradient Expanded Kadanoff-Baym Equations,''
  Annals Phys.\  {\bf 293} (2001) 126
  [arXiv:nucl-th/0102044].
  %%CITATION = NUCL-TH 0102044;%%

\bibitem{vanHees:2001ik}
  H.~van Hees and J.~Knoll,
  %``Renormalization in self-consistent approximations schemes at finite
  %temperature. I: Theory,''
  Phys.\ Rev.\ D {\bf 65} (2002) 025010
  [arXiv:hep-ph/0107200];
%\bibitem{VanHees:2001pf}
  %H.~Van Hees and J.~Knoll,
  %``Renormalization of self-consistent approximation schemes. II:  Applications
  %to the sunset diagram,''
  Phys.\ Rev.\ D {\bf 65} (2002) 105005
  [arXiv:hep-ph/0111193];
%\bibitem{vanHees:2002bv}
  %H.~van Hees and J.~Knoll,
  %``Renormalization in self-consistent approximation schemes at finite
  %temperature. III: Global symmetries,''
  Phys.\ Rev.\ D {\bf 66} (2002) 025028
  [arXiv:hep-ph/0203008].
  %%CITATION = HEP-PH 0203008;%% 

\bibitem{Blaizot:2003br}
  J.~P.~Blaizot, E.~Iancu and U.~Reinosa,
  %``Renormalizability of Phi-derivable approximations in scalar phi**4 theory,''
  Phys.\ Lett.\ B {\bf 568} (2003) 160
  [arXiv:hep-ph/0301201];
%\bibitem{Blaizot:2003an}
  %J.~P.~Blaizot, E.~Iancu and U.~Reinosa,
  %``Renormalization of phi-derivable approximations in scalar field theories,''
  Nucl.\ Phys.\ A {\bf 736} (2004) 149
  [arXiv:hep-ph/0312085]. 

%\cite{Cornwall:1974vz}
\bibitem{Cornwall:1974vz}
  J.~M.~Cornwall, R.~Jackiw and E.~Tomboulis,
  %``Effective Action For Composite Operators,''
  Phys.\ Rev.\ D {\bf 10} (1974) 2428.
  %%CITATION = PHRVA,D10,2428;%%

\bibitem{Hartree}
%\cite{Coleman:1974jh}
%\bibitem{Coleman:1974jh}
  S.~R.~Coleman, R.~Jackiw and H.~D.~Politzer,
  %``Spontaneous Symmetry Breaking In The O(N) Model For Large N*,''
  Phys.\ Rev.\ D {\bf 10} (1974) 2491;
  %%CITATION = PHRVA,D10,2491;%%
%\cite{Bardeen:1983st}
%\bibitem{Bardeen:1983st}
  W.~A.~Bardeen and M.~Moshe,
  %``Phase Structure Of The O(N) Vector Model,''
  Phys.\ Rev.\ D {\bf 28} (1983) 1372;
  %%CITATION = PHRVA,D28,1372;%%
%\cite{Amelino-Camelia:1992nc}
%\bibitem{Amelino-Camelia:1992nc}
  G.~Amelino-Camelia and S.~Y.~Pi,
  %``Selfconsistent improvement of the finite temperature effective potential,''
  Phys.\ Rev.\ D {\bf 47} (1993) 2356
  [arXiv:hep-ph/9211211].
  %%CITATION = HEP-PH 9211211;%%

%\cite{Cooper:2004rs}
\bibitem{Cooper:2004rs}
  F.~Cooper, B.~Mihaila and J.~F.~Dawson,
  %``Renormalizing the Schwinger-Dyson equations in the auxiliary field
  %formulation of lambda phi**4 field theory,''
  Phys.\ Rev.\ D {\bf 70} (2004) 105008
  [arXiv:hep-ph/0407119];
  %%CITATION = HEP-PH 0407119;%%
%\cite{Cooper:2005vw}
%\bibitem{Cooper:2005vw}
  F.~Cooper, J.~F.~Dawson and B.~Mihaila,
  %``Renormalized broken-symmetry Schwinger-Dyson equations and the 2PI-1/N
  %expansion for the O(N) model,''
  arXiv:hep-ph/0502040.
  %%CITATION = HEP-PH 0502040;%%

%\cite{Aarts:2002dj}
\bibitem{Aarts:2002dj}
  G.~Aarts, D.~Ahrensmeier, R.~Baier, J.~Berges and J.~Serreau,
  %``Far-from-equilibrium dynamics with broken symmetries from the 2PI-1/N
  %expansion,''
  Phys.\ Rev.\ D {\bf 66} (2002) 045008
  [arXiv:hep-ph/0201308].
  %%CITATION = HEP-PH 0201308;%%

\bibitem{BBRS}
  J.~Berges, S.~Bors\'anyi, U.~Reinosa and J.~Serreau, in preparation.

\bibitem{Urko:SEWM04}
  U.~Reinosa, ``Renormalization and gauge symmetry for 
  2PI effective actions,''
  in SEWM04, eds.\ K.J.~Eskola, K.~Kainulainen,
  K.~Kajantje and K.~Rummukainen (World Scientific) [arXiv:hep-ph/0411255].
\bibitem{Dominicis}
 C.~De Dominicis and P.~C.~Martin, J.~Math.~Phys.~5 (1964) 14,
 {\it ibid} 31; R.E.~Norton and J.M.~Cornwall, Ann. Phys. (N.Y.) 91 (1975) 106; 
 H.~Kleinert, Fortschritte der Physik 30 (1982) 187;
 A.N.~Vasiliev, ``Functional Methods in Quantum Field Theory
 and Statistical Physics'', Gordon and Breach Science Pub.~(1998).

%\cite{Berges:2004pu}
\bibitem{Berges:2004pu}
  J.~Berges,
  %``n-PI effective action techniques for gauge theories,''
  Phys.\ Rev.\ D {\bf 70} (2004) 105010 [arXiv:hep-ph/0401172].
  %%CITATION = HEP-PH 0401172;%% 

%\cite{Schwinger:1960qe}
\bibitem{Schwinger:1960qe}
  J.~S.~Schwinger,
  %``Brownian Motion Of A Quantum Oscillator,''
  J.\ Math.\ Phys.\  {\bf 2} (1961) 407.
  %%CITATION = JMAPA,2,407;%%

\bibitem{Braaten:2001en}
  E.~Braaten and E.~Petitgirard,
  %``Solution to the 3-loop phi-derivable approximation for scalar
  %thermodynamics,''
  Phys.\ Rev.\ D {\bf 65} (2002) 041701
  [arXiv:hep-ph/0106045];
%\bibitem{Braaten:2001vr}
  %E.~Braaten and E.~Petitgirard,
  %``Solution to the 3-loop Phi-derivable approximation for massless scalar
  %thermodynamics,''
  Phys.\ Rev.\ D {\bf 65} (2002) 085039
  [arXiv:hep-ph/0107118].

%\cite{Aarts:2003bk}
\bibitem{Aarts:2003bk}
  G.~Aarts and J.~M.~Martinez Resco,
  %``Transport coefficients from the 2PI effective action,''
  Phys.\ Rev.\ D {\bf 68} (2003) 085009
  [arXiv:hep-ph/0303216];
  %%CITATION = HEP-PH 0303216;%%
  %\cite{Aarts:2004sd}
%\bibitem{Aarts:2004sd}
  %G.~Aarts and J.~M.~Martinez Resco,
  %``Shear viscosity in the O(N) model,''
  JHEP {\bf 0402} (2004) 061
  [arXiv:hep-ph/0402192].
  %%CITATION = HEP-PH 0402192;%%

%\cite{Serreau:2003wr}
\bibitem{Serreau:2003wr}
  J.~Serreau,
  %``Out-of-equilibrium electromagnetic radiation,''
  JHEP {\bf 0405} (2004) 078
  [arXiv:hep-ph/0310051].
  %%CITATION = HEP-PH 0310051;%%


\end{thebibliography}
\end{document}